\documentclass[12pt]{article}

\usepackage{tikz, adjustbox, float} 
\usepackage[colorlinks=true,
    linkcolor=red,citecolor=blue]{hyperref} 
\usepackage{amsmath, amsfonts, amssymb, amsthm, dsfont} 
\usepackage{mathrsfs} 
\usepackage{soul} 
\usepackage{bm} 
\usepackage{booktabs}  
\usepackage{enumitem} 
\usepackage{mathabx} 

\def\real{{\rm I\!R}}

\DeclareMathOperator*{\Int}{Int}

\def\0{{\bf 0}}

\def\bpi{{\bm{\pi}}}

\def\bOmega{{\bm{\Omega}}}

\def\X{{\bf X}}
\def\x{{\bf x}}
\def\I{{\bf I}}
\def\Y{\mathbf{Y}}

\def\A{{\bf A}}
\def\B{{\bf B}}

\DeclareMathOperator*{\var}{var}

\DeclareMathOperator*{\argzero}{argzero}

\usepackage{latexsym}
\usepackage{amsthm}
\usepackage{amsmath,amssymb,amsfonts,bm,amsbsy,bbm}
\usepackage{epsfig}
\usepackage{graphicx,rotating,lscape}
\usepackage{verbatim}
\usepackage{natbib}
\usepackage{multirow,color}
\usepackage{geometry}
\usepackage{setspace}
\usepackage{subfigure}
\usepackage{bbm}
\usepackage{ulem} 

\usepackage[para]{footmisc}

\def\X{{\bf X}}
\def\x{{\bf x}}
\def\U{{\bf U}}
\def\u{{\bf u}}
\def\Z{{\bf Z}}
\def\z{{\bf z}}

\def\s{{\bf s}}
\def\a{{\bf a}}
\def\b{{\bf b}}
\def\e{{\bf e}}
\def\c{{\bf c}}

\def\bt{{\boldsymbol\theta}}
\def\bT{{\boldsymbol\Theta}}
\def\bb{{\boldsymbol\beta}}

\def\boldeta{\boldsymbol\eta}
\def\bpsi{\boldsymbol\psi}

\def\0{{\bf 0}}
\def\trans{^{\rm T}}
\def\pr{\hbox{pr}}
\def\wh{\widehat}
\def\wt{\widetilde}
\def\wc{\widecheck}
\def\n{\nonumber}

\def\var{\hbox{var}}

\def\bse{\begin{eqnarray*}}
\def\ese{\end{eqnarray*}}
\def\be{\begin{eqnarray}}
\def\ee{\end{eqnarray}}
\def\bsq{\begin{equation*}}
\def\esq{\end{equation*}}
\def\bq{\begin{equation}}
\def\eq{\end{equation}}

\def\boxit#1{\vbox{\hrule\hbox{\vrule\kern6pt  \vbox{\kern6pt#1\kern6pt}\kern6pt\vrule}\hrule}}
\def\bse{\begin{eqnarray*}}
\def\ese{\end{eqnarray*}}
\def\be{\begin{eqnarray}}
\def\ee{\end{eqnarray}}
\def\bsq{\begin{equation*}}
\def\esq{\end{equation*}}
\def\bq{\begin{equation}}
\def\eq{\end{equation}}
\def\th{^{th}}
\def\var{\hbox{var}}

\def\wh{\widehat}
\def\wt{\widetilde}

\def\th{^{\rm th}}

\def\n{\nonumber}

\def\trans{^{\rm T}}

\def\bDel{{\boldsymbol\Delta}}

\def\A{{\bf A}}
\def\a{{\bf a}}
\def\B{{\bf B}}

\def\D{{\bf D}}

\def\g{{\bf g}}

\def\r{{\bf r}}

\def\b{{\bf b}}
\def\u{{\bf u}}
\def\I{{\bf I}}

\def\U{{\bf U}}
\def\u{{\bf u}}
\def\v{{\bf v}}

\def\w{{\bf w}}
\def\X{{\bf X}}

\def\R{{\bf R}}
\def\x{{\bf x}}

\def\Y{{\bf Y}}

\def\Z{{\bf Z}}
\def\z{{\bf z}}

\def\bSig{{\bf \Sigma}}

\def\log{\hbox{log}}

\def\squarebox#1{\hbox to #1{\hfill\vbox to #1{\vfill}}}

\def\bpi{{\boldsymbol \pi}}
\def\0{{\bf 0}}

\def\mB{\mathcal{B}}

\def\var{\hbox{var}}

\def\pr{\hbox{pr}}
\def\wh{\widehat}
\def\wt{\widetilde}

\def\log{\hbox{log}}

\newtheoremstyle{mytheoremstyle} 
    {0.3cm}                      
    {0cm}                        
    {}                   
    {}                           
    {\bf}                   
    {: }                          
    {0em}                       
    {}  

\theoremstyle{mytheoremstyle}
\newtheorem{Theorem}{Theorem}

\newtheorem*{Lemma*}{Lemma}
\newtheorem{Corollary}{Corollary}

\newtheoremstyle{myExampleRemarkstyle} 
    {0.3cm}                    
    {0cm}                           
    {}                   
    {}                           
    {\bf}                   
    {: }                          
    {0em}                       
    {}  

\theoremstyle{myExampleRemarkstyle}
 
\newtheorem{Remark}{Remark}
\newtheorem{Assumption}{Assumption}

\renewcommand{\theAssumption}{\Alph{Assumption}}

\newtheorem{innercustomgeneric}{\customgenericname}
\providecommand{\customgenericname}{}
\newcommand{\newcustomtheorem}[2]{%
  \newenvironment{#1}[1]
  {%
   \renewcommand\customgenericname{#2}%
   \renewcommand\theinnercustomgeneric{##1}%
   \innercustomgeneric
  }
  {\endinnercustomgeneric}
}

\newcustomtheorem{customthm}{Theorem}
\newcustomtheorem{customlemma}{Lemma}
\usepackage{mathalfa}
\usepackage{oubraces}

\let\refBKP\ref
\renewcommand{\ref}[1]{{\upshape\refBKP{#1}}}

\definecolor{pinegreen}{rgb}{0.0, 0.47, 0.44}

\usepackage{setspace}
\begin{document}


\begin{center}
\LARGE{Just Identified Indirect Inference Estimator: Accurate Inference through Bias Correction} 
\vspace{0.2cm}\\
\large{Yuming Zhang$^{1}$, Yanyuan Ma$^{2}$, Samuel Orso$^{1}$, Mucyo Karemera$^{1}$, Maria-Pia Victoria-Feser$^{1}$, St\'ephane Guerrier$^{1}$}
\vspace{0.2cm}\\
\large{$^{1}$University of Geneva, $^{2}$Pennsylvania State University}   
\end{center}




\begin{abstract}
An important challenge in statistical analysis lies in controlling the estimation bias when handling the ever-increasing data size and model complexity of modern data settings. In this paper, we propose a reliable estimation and inference approach for parametric models based on the Just Identified iNdirect Inference estimator (JINI). The key advantage of our approach is that it allows to construct a consistent estimator in a simple manner, while providing strong bias correction guarantees that lead to accurate inference. Our approach is particularly useful for complex parametric models, as it allows to bypass the analytical and computational difficulties (e.g., due to intractable estimating equation) typically encountered in standard procedures. The properties of JINI (including consistency, asymptotic normality, and its bias correction property) are also studied when the parameter dimension is allowed to diverge, which provide the theoretical foundation to explain the advantageous performance of JINI in increasing dimensional covariates settings. Our simulations and an alcohol consumption data analysis highlight the practical usefulness and excellent performance of JINI when data present features (e.g., misclassification, rounding) as well as in robust estimation. 
\end{abstract}




\vspace{0.2cm}


\section{Introduction}
\label{sec:intro}

Point estimates in parametric models are frequently encountered in practice. They are routinely used, for example, to predict outcomes, to compute mean squared errors, and so on. In particular, the construction of reliable Confidence Intervals (CIs) heavily depends on accurate point estimates. As an illustration, they can be used as plug-in values to obtain an estimated asymptotic variance in order to construct CIs based on their asymptotic distributions. They can also be used in various procedures that require to simulate data from an estimated model to obtain CIs. However, the ever-increasing model complexity  tends to impair the accuracy of point estimates based on classical estimation procedures. Despite being asymptotically unbiased under regularity conditions, the Maximum Likelihood Estimator (MLE) can suffer from severe finite sample bias, for example, when the number of parameters $p$ is relatively large compared to the sample size $n$ (see e.g., \citealp{sur2019modern}). Similarly, the asymptotically efficient moment-based estimators may exhibit substantial bias when applied to relatively small samples (see e.g., \citealp{imbens2002generalized}). A common consequence of biased point estimates is that they can lead to unreliable coverage (see e.g., \citealp{wang1999bias,mittelhammer2005empirical,kosmidis2014bias}).  

Moreover, in some complex situations, consistent estimators can become particularly demanding to obtain using classical estimation methods, due to various analytical and/or numerical difficulties. For example, data often exhibit features such as truncation, censoring or misclassification. In these cases, the data generating process can often be understood as being dependent on latent variables. In this context, a standard approach is to integrate out the latent variables from the likelihood, but it can easily render integrals with no closed-form expression. One can, for example, approximate integrals and hence the likelihood function (see e.g., \citealp{rabe2002reliable}), but crude numerical approximations often lead to biased or even inconsistent estimators. Alternatively, one can use iterative methods like the Expectation-Maximization (EM) algorithm of \cite{dempster1977maximum}, which can also suffer from some numerical drawbacks such as slow convergence. Another example is in robust estimation, which presents its own bottleneck due to the potentially intractable consistency correction terms in their estimating equations. Indeed, these correction terms generally have no closed-form expression, so they can be very difficult to compute even for simple regression models (see e.g., \citealp{HeCaCoVF:09} and the reference therein). Neglecting these correction terms from the estimating equations can significantly simplify the computation, but it leads to inconsistent estimators (see e.g., \citealp{moustaki2006bounded}).    

Considering the importance of accurate point estimates for reliable inference, as well as various challenges encountered by standard procedures when handling complex data settings, in this paper we construct the Just Identified iNdirect Inference estimator (JINI) that is applicable for a wide range of complex parametric models. The key advantage is that JINI allows to bypass the analytical and computational difficulties typically encountered in standard procedures, as it is constructed from an initial estimator that is chosen for its analytical and numerical simplicity. Among many properties, the bias correction property of JINI is particularly advantageous allowing, among others, to provide a reliable basis for accurate inference.

\subsection{A Motivating Example}
\label{sec:motivating_example}

To illustrate the advantages of JINI, we consider the simulation of a logistic regression with misclassified responses with 45 parameters and 395 observations. This simulation is based on a real public health dataset where the alcohol consumption among secondary school students is studied. More details are given in Section~\ref{sec:alcohol}. We compare the performance of MLE and JINI which is constructed from a Naive MLE (NMLE) neglecting the misclassification (i.e., MLE for a classical logistic regression). Compared to MLE, JINI is simpler to compute as it avoids the computation of the likelihood function that takes into account the misclassification, and only requires the computation of an inconsistent but readily available initial estimator (i.e., NMLE). We also construct 95\% CIs for both estimators based on their asymptotic normality to compare their inference performance. In Table~\ref{tab:emulation_subset_res}, we present the results on $\beta_1$ to $\beta_7$, which correspond to all the covariates that are found to have significant associations to students' alcohol consumption levels (i.e., whose corresponding CIs do not cover zero) by either MLE or JINI. We can see that, compared to MLE, JINI has considerably smaller biases, comparable standard errors, and more accurate empirical coverages. Moreover, the average CI lengths of both estimators are comparable, suggesting that the advantage of JINI in terms of coverage is a consequence of its reduced bias. Therefore, as illustrated in this example, JINI can be constructed in a simple manner based on a readily available initial estimator, and its bias correction property can lead to improvement in inference compared to MLE.  

\begin{table*}[tb]
    \centering
    \caption{Simulation results on $\beta_1$ to $\beta_7$ for the
      logistic regression with misclassified responses presented in
      Section~\ref{sec:alcohol:simu}. Each column corresponds
      respectively to the absolute bias, standard error, empirical coverage of 95\% CI, and average CI length.}
    \begin{tabular}{cccccc}
    \toprule
    & & Abs. Bias & Std. Error & Coverage (\%) & Avg. CI Length \\
    \midrule
    & $\beta_1$ & 0.2537 & 0.4139 & 87.56 & 2.2668 \\
    & $\beta_2$ & 0.1687 & 0.4038 & 90.19 & 1.3997 \\
    & $\beta_3$ & 0.1936 & 0.3968 & 90.17 & 0.7937 \\
    MLE & $\beta_4$ & 0.1079 & 0.1973 & 89.13 & 1.6577 \\
    & $\beta_5$ & 0.1933 & 0.2210 & 82.84 & 1.3686 \\
    & $\beta_6$ & 0.0805 & 0.2087 & 93.00 & 2.0964\\
    & $\beta_7$ & 0.3087 & 0.8538 & 93.12 & 1.9187\\
    \midrule
    & $\beta_1$ & 0.0162 & 0.3286 & 97.41 & 2.0773 \\
    & $\beta_2$ & 0.0031 & 0.3263 & 96.42 & 1.4088 \\
    & $\beta_3$ & 0.0089 & 0.3198 & 96.16 & 0.7618 \\
    JINI & $\beta_4$ & 0.0025 & 0.1586 & 96.86 & 1.5682 \\
    based on NMLE & $\beta_5$ & 0.0075 & 0.1702 & 96.33 & 1.3245 \\
    & $\beta_6$ & 0.0049 & 0.1568 & 97.82 & 1.9930\\
    & $\beta_7$ & 0.0021 & 0.6733 & 97.81 & 1.7230\\
    \bottomrule
    \end{tabular}
    \label{tab:emulation_subset_res}
\end{table*}

\subsection{Main Results and Contributions}
\label{sec:contributions}

In order to obtain consistent estimators for complex parametric models, we often have to address various analytical and/or numerical obstacles when using classical estimation procedures. In contrast, JINI makes use of an initial estimator that is chosen for its analytical and numerical simplicity (hence typically biased and possibly inconsistent, but readily available), which allows to construct a consistent estimator in a simple manner. In particular, JINI can provide strong bias correction guarantees. Depending on whether the initial estimator is consistent and the form of its bias function, the bias order of JINI varies. For example, we show that JINI has a bias order at least of $\mathcal{O}(n^{-1})$, which is the same as many classical estimators (e.g., MLE) under suitable conditions (see e.g., \citealp{cox1979theoretical}). Under additional smoothness conditions, the bias of JINI is of order $\mathcal{O}(n^{-3})$, or can even be completely eliminated with sufficiently large finite $n$. We further extend our results to settings where $p$ diverges with $n$ but $p<n$. To the best of our knowledge, these results are not known in the literature that considers bias correction for general parametric models with diverging $p$. Our results give a theoretical explanation to the advantageous performance of JINI in settings where $p$ is relatively large compared to $n$, and provide a guideline on how the properties of JINI (in particular its bias order) are affected by the growth of $p$ and $n$. 

Our numerical examples highlight the practical advantages of our approach in complex settings where classical approaches may be difficult to apply. In these examples, JINI shows smaller bias, more accurate coverage, and is simpler to compute compared to classical estimators. Specifically, we consider settings where data exhibit features (e.g., misclassification, rounding) such that classical MLE is relatively difficult to obtain (e.g., requires to use the EM algorithm). We construct JINI based on NMLE (i.e., MLE for a simpler model that neglects these data features), and we observe that JINI has smaller bias and more accurate coverage than MLE. We also consider models for which classical robust estimators are difficult to obtain due to the intractable consistency correction terms, and we propose to construct a robust JINI based on an initial estimator that neglects these correction terms. Our results highlight that the robust JINI is simple to obtain and that it has superior finite sample performance compared to existing robust estimators.  

\subsection{Related Work}
\label{sec:literature}

Different bias correction methods have been proposed for parametric models. For example, bias correction can be achieved by solving an adjusted score function (see e.g., \citealp{Firt:93,KoFi:09,Kosm:14}), which requires analytical computation that varies from model to model. Alternatively, one can define a bias corrected estimator by removing an estimated bias from an initial estimator. Along this idea, many methods have been proposed with different objectives, and JINI is closely related to many of them. For example, the parametric Bootstrap Bias Corrected estimator (BBC) (see e.g., \citealp{efron1994introduction}) and the nonlinear-bias-correcting estimator of \cite{mackinnon1998approximate} both aim to reduce finite sample bias from a consistent initial estimator. Another example related to JINI is the indirect inference method (see e.g., \citealp{gourieroux1993indirect,gallant1996moments,arvanitis2014valid,arvanitis2015class}), which was originally proposed to construct consistent estimators starting from inconsistent ones. \cite{kuk1995asymptotically} independently produced the same idea to correct the asymptotic bias when computing MLE of generalized linear mixed models. They proposed to iterate BBC by updating the plug-in value to its limit, and thus, their approach is known as Iterative Bootstrap (IB). \cite{guerrier2018simulation} studied the finite sample bias correction property of several simulation-based methods, either based on indirect inference or bootstrap. In particular, they showed that one of the indirect inference based estimator enjoys advantageous bias correction property and is equivalent to IB. 

Compared to these existing methods, JINI is shown to enjoy stronger bias correction under weaker conditions. As an example, when considering consistent initial estimators and under the same conditions, JINI can be unbiased with large enough finite $n$ while BBC can only achieve a bias order of $\mathcal{O}(n^{-2})$. \cite{gourieroux1993indirect} assumes that the initial estimator is differentiable with respect to the parameter, which is a strong condition with restricted applicability and is not assumed in our work. Although \cite{guerrier2018simulation} suggests the use of methods based on inconsistent initial estimators, they only derived theoretical results based on consistent initial estimators. Moreover, they require sufficiently smooth bias function in a specific form, which is a relatively stringent condition. In contrast, we provide theoretical results when the initial estimator is inconsistent, and consider more general assumptions on the bias function where the condition of \cite{guerrier2018simulation} can be seen as a special case. Lastly, these works assume fixed parameter dimension $p$, so we fill the gap of diverging $p$ in terms of theory.

\subsection{Organization}
\label{sec:organization}

The rest of the paper is organized as follows. In Section~\ref{sec:JINI}, we formally define JINI. In Section~\ref{sec:results:fixed_p:incons_initial}, we consider inconsistent initial estimators to study the properties of JINI, which include consistency, asymptotic normality and its bias property. In Section~\ref{sec:results:fixed_p:cons_initial}, we further illustrate the bias correction property of JINI when considering consistent initial estimators. In Section~\ref{sec:results:diverge_p}, we extend all theoretical results of JINI to settings where $p$ is allowed to diverge with $n$ and $p<n$. The implementation simplicity and advantageous finite sample performance of JINI are illustrated with simulations in Section~\ref{sec:simu} and a real data analysis on alcohol consumption in Section~\ref{sec:alcohol}. Section~\ref{sec:conclusions} summarizes the article and provides further discussions. Section~\ref{sec:conditions} presents and discusses the assumptions used in this paper. The proofs of all theoretical results as well as some additional discussions are included in supplementary materials.

\section{Just Identified Indirect Inference Estimator (JINI)}
\label{sec:JINI}

Suppose that we observe a random sample of size $n$ generated from a parametric distribution $F_{\bt_0}$, and we aim to estimate $\bt_0\in\bT\subset\real^{p}$. Let $\wh{\bpi}(\bt_0, n)\in\bT$ denote an initial estimator of $\bt_0$ computed on the observed sample, while $\wh{\bpi}(\bt,n)$ denotes the same estimator computed on a generic sample of size $n$ generated from $F_{\bt}$. This initial estimator is typically chosen for its numerical simplicity, so it can be considerably biased or even inconsistent. Moreover, we assume $\wh{\bpi}(\bt, n) \overset{P}{\to} \bpi(\bt)$ for any $\bt\in\bT$. This standard requirement can be guaranteed under regularity conditions, for example, when $\wh{\bpi}(\bt, n)$ is MLE of a possibly misspecified model (see e.g., \citealp{huber1967behavior,white1982maximum}). Naturally, when the initial estimator is consistent, we have $\bpi(\bt)=\bt$, and $\bpi(\bt)\neq\bt$ otherwise. Assuming that $\bpi(\bt,n)\equiv\mathbb{E}_{\bt}\{\wh{\bpi}(\bt, n)\}$ exists, where $\mathbb{E}_{\bt}(\cdot)$ denotes the expectation under $F_{\bt}$, we can write
\bq \label{eqn:decomp_JINI}
    \wh{\bpi}(\bt,n) = \underbrace{\bpi(\bt) + \b(\bt,n)}_{\bpi(\bt,n)} + \v(\bt,n), 
\eq
where $\b(\bt,n)\equiv \bpi(\bt,n)-\bpi(\bt)$ denotes the finite sample bias and $\mathbf{v}(\bt, n) \equiv \widehat{\bpi}(\bt,n) - \bpi(\bt,n)$ is a zero mean random vector. 

In order to estimate $\bt_0$ from $\wh{\bpi}(\bt_0,n)$, we consider the following estimator: 
\bq \label{eqn:def_JINI}
    \wh{\bt} \equiv \argzero_{\bt\in\bT} \wh{\bpi}(\bt_0,n) - \bpi(\bt,n).
\eq
We call $\wh{\bt}$ JINI, as it is closely related to the indirect inference method in the just identified case (i.e., the dimension of $\bpi(\bt,n)$ is equal to $p$).

\begin{Remark}
    From a computational perspective, JINI can be obtained using
    stochastic approximation methods (see e.g., \citealp{robbins1951stochastic,lai1979adaptive,polyak1992acceleration,kuk1995asymptotically}). These
    algorithms are simple to implement and numerically efficient for a
    wide range of models and settings. As an example, we can
    approximate $\bpi(\bt,n)$ by its sample version based on generated
    data, i.e., $\bpi(\bt,n)\approx
    H^{-1}\sum_{h=1}^H\wh\bpi_h(\bt,n)$, where the  approximation can
    be made arbitrarily precise with a sufficiently large $H$. In this
    case, JINI is equivalent to one of the indirect inference based
    estimator presented in \cite{guerrier2018simulation} and can be
    computed using IB.  
\end{Remark}

\section{JINI based on Inconsistent Initial Estimators}
\label{sec:results:fixed_p:incons_initial}

In this section, we study the properties of JINI when considering inconsistent initial estimators (i.e., $\bpi(\bt)\neq\bt$). By the decomposition of the initial estimator in \eqref{eqn:decomp_JINI} and the definition of JINI in \eqref{eqn:def_JINI}, we have $\bpi(\bt_0) + \b(\bt_0,n) + \v(\bt_0,n) = \wh{\bpi}(\bt_0,n) = \bpi(\wh{\bt},n) = \bpi(\wh{\bt}) + \b(\wh{\bt},n)$, and hence we have
\bq \label{eqn:intuition1}
    \bpi(\wh{\bt}) - \bpi(\bt_0) = \b(\bt_0,n) + \v(\bt_0,n) - \b(\wh{\bt},n).    
\eq
Under suitable smoothness conditions on $\bpi(\bt)$ and using Taylor's theorem, we also have 
\bq \label{eqn:intuition2}
    \bpi(\wh{\bt}) - \bpi(\bt_0) = \A(\bt_0)(\wh{\bt}-\bt_0) + \u(\wh{\bt},\bt_0),
\eq
where $\A(\bt_0)\equiv\partial\bpi(\bt_0)/\partial\bt$ and $\u(\wh{\bt},\bt_0)$ is a higher-order term in quadratic form of $\wh{\bt}-\bt_0$. Combining the expressions of $\bpi(\wh{\bt}) - \bpi(\bt_0)$ in \eqref{eqn:intuition1} and \eqref{eqn:intuition2}, we obtain
\bsq
    \A(\bt_0)(\wh{\bt}-\bt_0) + \u(\wh{\bt},\bt_0) = \b(\bt_0,n) + \v(\bt_0,n) - \b(\wh{\bt},n),
\esq
and thus we can write
\be \label{eqn:intuition3}
    \wh{\bt} - \bt_0 &=& \A(\bt_0)^{-1} \left\{\b(\bt_0,n) + \v(\bt_0,n) - \b(\wh{\bt},n)\right\} - \A(\bt_0)^{-1} \u(\wh{\bt},\bt_0) \n\\
    &\approx& \A(\bt_0)^{-1} \left\{\b(\bt_0,n) + \v(\bt_0,n) - \b(\wh{\bt},n)\right\} \n\\
    &=& \A(\bt_0)^{-1} \left\{\wh{\bpi}(\bt_0,n) - \bpi(\bt_0) - \b(\wh{\bt},n)\right\}.
\ee
Under suitable conditions on $\A(\bt_0)$, \eqref{eqn:intuition3} implies the consistency of JINI when $\v(\bt_0,n)\overset{P}{\to} \0$ and $\b(\bt,n)\to \0$ uniformly for $\bt\in\bT$. Moreover, it suggests that the asymptotic normality of JINI is closely related to the asymptotic normality of $\wh{\bpi}(\bt_0,n)-\bpi(\bt_0)$, as formally presented in Theorem~\ref{thm:JINI_consistency_asymp_norm}. We denote $n^{-\alpha}$ as the convergence rate of $\wh{\bpi}(\bt_0,n)$ and $\bSig(\bt_0,n)$ as its variance.

\begin{Theorem}
\label{thm:JINI_consistency_asymp_norm}
    Under Assumptions~\ref{assum:compact}, \ref{assum:v}, \ref{assum:finite_bias}, and \ref{assum:asymp_bias} in Section~\ref{sec:conditions}, we have $\wh{\bt}\overset{P}{\to}\bt_0$. Additionally under Assumption~\ref{assum:initial_asymp_norm} in Section~\ref{sec:conditions}, we have $n^\alpha \left\{\A(\bt_0)^{-1} \bSig(\bt_0,n) \A(\bt_0)^{-1}\right\}^{-1/2} (\wh{\bt}-\bt_0) \overset{D}{\to} \mathcal{N}(\0, \I_p)$, where $\I_p$ is the $p\times p$ identity matrix.
\end{Theorem}

Theorem~\ref{thm:JINI_consistency_asymp_norm} guarantees the consistency of JINI, even if it is based on an inconsistent initial estimator. If $\wh{\bpi}(\bt_0,n)-\bpi(\bt_0)$ is asymptotically normal (Assumption~\ref{assum:initial_asymp_norm}), then JINI is also asymptotically normal at the same convergence rate of $n^{-\alpha}$. For example, when NMLE (i.e., MLE for a misspecified but simpler model) is used as the initial estimator, $\sqrt{n}\{\wh{\bpi}(\bt_0,n)-\bpi(\bt_0)\}$ is asymptotically normal and hence $\alpha=1/2$ (see Section~\ref{sec:conditions} for more details). 

The first equality in \eqref{eqn:intuition3} also gives 
\bsq
    \mathbb{E}(\wh{\bt})-\bt_0 = \A(\bt_0)^{-1} [\b(\bt_0,n) - \mathbb{E}\{\b(\wh{\bt},n)\}] - \A(\bt_0)^{-1} \mathbb{E}\{\u(\wh{\bt},\bt_0)\},
\esq
implying that the bias of JINI is closely related to the order of
$\b(\bt,n)$, which we denote as $\mathcal{O}(n^{-\beta})$, and the
squared error rate $n^{-2\alpha}$ of the initial estimator, as
formally presented in Theorem~\ref{thm:JINI_bias:incons_initial}.

\begin{Theorem}
\label{thm:JINI_bias:incons_initial}
When the initial estimator is inconsistent, under Assumptions~\ref{assum:compact}, \ref{assum:v}, \ref{assum:finite_bias}, and 
\ref{assum:asymp_bias} in Section~\ref{sec:conditions}, we have $\mathbb{E}(\wh{\bt})-\bt_0 = \mathcal{O}\{n^{-\min(2\alpha,\beta)}\}$. 
\end{Theorem}

As an illustration, when the initial estimator is NMLE, we have $\alpha=1/2$ and $\beta=1$ under regularity conditions, hence JINI has a bias order of $\mathcal{O}(n^{-1})$. Since the bias order of MLE is also $\mathcal{O}(n^{-1})$ under suitable conditions (see e.g., \citealp{cox1979theoretical,kosmidis2014bias}), Theorem~\ref{thm:JINI_bias:incons_initial} shows that JINI based on an inconsistent initial estimator can achieve the same bias order as MLE. This is particularly interesting when considering complex models, for which the computation of MLE often entails various analytical and/or numerical obstacles, whereas JINI can be constructed from an inconsistent estimator that is simple to compute and can achieve the same bias order. In practice, we typically observe smaller finite sample bias of JINI than MLE (see e.g., the numerical examples in Sections~\ref{sec:simu} and \ref{sec:alcohol}).

The key message delivered by the results in this section is that, using our approach, we can easily construct a consistent JINI from an inconsistent initial estimator that is simple to compute and readily available. This allows us to circumvent the analytical and computational difficulties typically entailed when computing standard estimators like MLE, while obtaining estimators with advantageous bias correction performance that can lead to accurate inference.

\section{JINI based on Consistent Initial Estimators}
\label{sec:results:fixed_p:cons_initial}

In this section, we present the properties of JINI based on consistent initial estimators. In this case, we have $\bpi(\bt)=\bt$ and hence Assumption~\ref{assum:asymp_bias} in Section~\ref{sec:conditions} is trivially satisfied and $\A(\bt_0)=\I_p$. Then a direct consequence of Theorem~\ref{thm:JINI_consistency_asymp_norm} is that JINI is also consistent and asymptotically normal when using consistent initial estimators, as shown in Corollary~\ref{thm:JINI_consistency_asymp_norm:cons_initial}.

\begin{Corollary}
\label{thm:JINI_consistency_asymp_norm:cons_initial}
    When the initial estimator is consistent, under Assumptions~\ref{assum:compact}, \ref{assum:v} and \ref{assum:finite_bias} in Section~\ref{sec:conditions}, we have $\wh{\bt}\overset{P}{\to}\bt_0$. Additionally under Assumption~\ref{assum:initial_asymp_norm} in Section~\ref{sec:conditions}, we have $n^\alpha \bSig(\bt_0,n)^{-1/2} (\wh{\bt}-\bt_0) \overset{D}{\to} \mathcal{N}(\0, \I_p)$. 
\end{Corollary}

A highlight of Corollary~\ref{thm:JINI_consistency_asymp_norm:cons_initial} is that JINI has the same asymptotic variance as the initial estimator, hence it has no asymptotic efficiency loss when using a consistent initial estimator.

In the rest of this section, we focus on the bias correction property of JINI. Based on \eqref{eqn:intuition1}, the bias of JINI based on a consistent initial estimator is given by 
\bq \label{eqn:intuition4} 
    \mathbb{E}(\wh{\bt})-\bt_0 = \b(\bt_0,n)-\mathbb{E}\{\b(\wh{\bt},n)\}.
    \eq
Since the initial estimator is consistent, $\b(\bt,n)$ represents the bias of the initial estimator and its order, denoted as $\mathcal{O}(n^{-\beta})$, represents the bias order of the initial estimator. So \eqref{eqn:intuition4} suggests that JINI has at least the same bias order as the initial estimator. Moreover, under sufficient smoothness conditions on $\b(\bt,n)$, JINI may achieve better bias orders. As an illustration, consider $\b(\bt,n)$ to be linear, say $\b(\bt,n)=\B(n)\bt$, then we can rewrite \eqref{eqn:intuition4} as $\{\I_p+\B(n)\} \{\mathbb{E}(\wh{\bt})-\bt_0\} = \0$. Since $\b(\bt,n)\to \0$, we have $\{\I_p+\B(n)\}\to\I_p$ and hence $\mathbb{E}(\wh{\bt})-\bt_0 = \0$ when $n$ is sufficiently large. This implies that JINI completely eliminates the linear part of the bias from the initial estimator. This is particularly useful since the leading-order bias of many classical estimators (e.g., MLE) is often linear (see e.g., \citealp{mackinnon1998approximate}), i.e., 
\bq \label{eqn:intuition5}
    \b(\bt,n) = \B(n)\bt + \text{nonlinear bias of higher order}.
\eq
For example, the results of \cite{sur2019modern} suggest that the bias of MLE for a logistic regression is linear under suitable design conditions. \cite{mardia1999bias} provides some examples in which they showed that the first-order bias of MLE is linear. Other examples can be found, for example, in \cite{kendall1954note,marriott1954bias} which consider the estimation of autocorrelations in time series. In these cases, the bias of JINI given in \eqref{eqn:intuition4} becomes
\be \label{eqn:intuition6}
    \mathbb{E}(\wh{\bt}) - \bt_0 &=& \{\I_p+\B(n)\}^{-1} \times \text{higher-order term related to nonlinear bias} \n\\
    &\approx& \text{higher-order term related to nonlinear bias}, 
\ee
and hence the bias of JINI is only related to the nonlinear part of the bias of the initial estimator. In Theorem~\ref{thm:JINI_bias:cons_initial} below, we provide a concrete description on how the bias order of JINI is affected by the form of $\b(\bt,n)$. We denote $\mathcal{O}(n^{-\beta_2})$ as the order of the nonlinear part of $\b(\bt,n)$, and $\mathcal{O}(n^{-\beta_3})$ as the order of the higher-order term of the nonlinear part of $\b(\bt,n)$.

\begin{Theorem}
\label{thm:JINI_bias:cons_initial}
When the initial estimator is consistent, under Assumptions~\ref{assum:compact}, \ref{assum:v}, \ref{assum:finite_bias} in Section~\ref{sec:conditions} and
\begin{enumerate}
    \item under Assumption~\ref{assum:finite_bias:general} in Section~\ref{sec:conditions}, we have $\mathbb{E}(\wh{\bt})-\bt_0 = \mathcal{O}\{n^{-(\alpha+\beta)}\}$.
    \item under Assumption~\ref{assum:finite_bias:comb} in Section~\ref{sec:conditions}, we have $\mathbb{E}(\wh{\bt})-\bt_0 = \mathcal{O}\{n^{-(\alpha+\beta_2)}\}$.
    \item under Assumption~\ref{assum:finite_bias:smooth} in Section~\ref{sec:conditions}, we have $\mathbb{E}(\wh{\bt})-\bt_0 = \mathcal{O}\{n^{-\min(2\alpha+\beta_2,\beta_3)}\}$.
    \item under Assumption~\ref{assum:finite_bias:linear} in Section~\ref{sec:conditions}, we have $\mathbb{E}(\wh{\bt})-\bt_0 = \0$ when $n$ is sufficiently large.
\end{enumerate}
\end{Theorem}

Assumptions~\ref{assum:finite_bias:general} to \ref{assum:finite_bias:linear} make increasingly stronger assumptions on the form of $\b(\bt,n)$ as well as the orders of its linear and nonlinear terms. In short, Assumption~\ref{assum:finite_bias:general} does not require a specific form of $\b(\bt,n)$ except its differentiability. Assumption~\ref{assum:finite_bias:comb} requires the leading-order term of $\b(\bt,n)$ to be linear. Assumption~\ref{assum:finite_bias:smooth} additionally requires that the leading-order term of the nonlinear part of $\b(\bt,n)$ is in quadratic form of $\bt/n$. Assumption~\ref{assum:finite_bias:linear} requires $\b(\bt,n)$ to be linear. 

When using classical estimators (e.g., MLE) as the initial estimators, we have $\alpha=1/2,\beta=1,\beta_2=2,\beta_3=3$ (see Section~\ref{sec:conditions} for more discussions on these values). In this case, the bias orders of JINI are simplified to $\mathcal{O}(n^{-3/2}),\mathcal{O}(n^{-5/2}),\mathcal{O}(n^{-3})$ for the first three results of Theorem~\ref{thm:JINI_bias:cons_initial}, indicating that better bias orders of JINI can be obtained by imposing stronger conditions on $\b(\bt,n)$. These results also show that, even under the weakest Assumption~\ref{assum:finite_bias:general}, JINI outperforms many existing bias correction methods that only guarantee a bias order of $o(n^{-1})$ (see e.g., \citealp{efron1975,Firt:93,Kosm:14}). More generally speaking, the first result of Theorem~\ref{thm:JINI_bias:cons_initial} indicates that JINI can reduce the bias of the initial estimator by at least  $n^{-\alpha}$, where we recall that $\alpha$ denotes the error rate of the initial estimator. Compared to the first result, the second result replaces $\beta$ (bias order of the initial estimator) by $\beta_2$ (the order of the nonlinear bias of the initial estimator). This illustrates that when the initial estimator is consistent, the bias of JINI is determined by the nonlinear bias of the initial estimator, as suggested by \eqref{eqn:intuition6}. When the bias function of the initial estimator is linear, the fourth result shows that JINI is unbiased when $n$ is large enough, validating our observation that the linear bias of the initial estimator can be completely eliminated by JINI. This is in line with the result of \cite{mackinnon1998approximate}, where they showed that their linear-bias-correcting estimator (which is closely related to JINI) is unbiased for a scalar parameter whenever the bias function is linear.

In contrast to JINI which eliminates linear bias of the initial estimator, the bias of BBC is primarily determined by the linear term. To provide an illustration, when the bias of the initial estimator has the form in \eqref{eqn:intuition5}, the bias of BBC $\wt{\bt}$ is given by   
\bse
    \mathbb{E}(\wt{\bt})-\bt_0 &=& \b(\bt_0,n) - \mathbb{E}[\b\{\wh{\bpi}(\bt_0,n),n\}] \\
    &=& -\B(n) \b(\bt_0,n) + \text{higher-order term related to nonlinear bias}.
\ese
So the linear bias from the initial estimator remains in the bias of BBC. In Supplementary Material~\ref{sec:pf:BBC}, we formally study the bias of BBC under our assumption framework, and we find that JINI can achieve better bias correction than BBC.

\section{Extensions to Increasing Dimensional Settings}
\label{sec:results:diverge_p}

In this section, we extend the theoretical results with fixed $p$ presented in Sections~\ref{sec:results:fixed_p:incons_initial} and \ref{sec:results:fixed_p:cons_initial} to increasing dimensional settings, where $p$ diverges with $n$ and $p<n$. We leave the technical assumptions with diverging $p$ in supplementary materials, and focus on the interpretation of the results in this section.

\begin{Theorem}
\label{thm:JINI_consistency:diverge_p}
When $p^{1/2}n^{-\alpha}\to 0$, under the assumptions presented in Supplementary Material~\ref{sec:pf:thm:JINI_consistency:diverge_p} (similar to the ones in Theorem~\ref{thm:JINI_consistency_asymp_norm}), we have $\|\wh{\bt}-\bt_0\|_2=o_p(1)$. 
\end{Theorem}

\begin{Theorem}
\label{thm:JINI_asymp_norm:diverge_p}
Under the assumptions presented in Supplementary Material~\ref{sec:pf:thm:JINI_asymp_norm:diverge_p} (similar to the ones in Theorem~\ref{thm:JINI_consistency_asymp_norm}), when the initial estimator is consistent and $p^{1/2}n^{\alpha-\beta}\to 0$, for any $\s\in\real^p$ such that $\|\s\|_2=1$, we have $n^\alpha\s\trans \bSig(\bt_0,n)^{-1/2} (\wh{\bt}-\bt_0) \overset{D}{\to} \mathcal{N}(0,1)$. When the initial estimator is inconsistent and $\max(p^{1/2}n^{\alpha-\beta}, p^2n^{-\alpha}) \to 0$, we have $n^\alpha\s\trans \left\{\A(\bt_0)^{-1} \bSig(\bt_0,n) \A(\bt_0)^{-1}\right\}^{-1/2} (\wh{\bt}-\bt_0) \overset{D}{\to} \mathcal{N}(0,1)$.
\end{Theorem}

We recall that $n^{-\alpha}$ denotes the convergence rate of the initial estimator, and $\mathcal{O}(n^{-\beta})$ denotes the order of  $\b(\bt,n)$ and is the bias order of the initial estimator when it is consistent. Although the values of $\alpha$ and $\beta$ may vary due to diverging $p$, we use $\alpha=1/2$ and $\beta = 1$ for simplicity to interpret these results. Theorem~\ref{thm:JINI_consistency:diverge_p} guarantees the consistency of JINI when $p/n\to 0$, regardless of inconsistent or consistent initial estimators. As for Theorem~\ref{thm:JINI_asymp_norm:diverge_p}, its first result establishes the asymptotic normality of JINI based on consistent initial estimators when $p/n\to 0$. This requirement on $p/n$ is generally weaker than the one needed to ensure the asymptotic normality of the initial estimator (e.g., MLE). For example, \cite{he2000parameters} studied the asymptotic normality of M-estimators for general parametric models when $p^2\log(n)n^{-1}\to0$. \cite{wang2011gee} established the asymptotic normality of the generalized estimating equations estimator when $p^3n^{-1}\to0$. Nevertheless, when the initial estimator is inconsistent, stronger requirement of $p^4n^{-1}\to 0$ is needed to ensure the asymptotic normality of JINI, as shown in the second result of Theorem~\ref{thm:JINI_asymp_norm:diverge_p}. This is reasonable as more data are needed to compensate the loss of information due to the use of an inconsistent initial estimator. Besides the $p/n$ requirement, the asymptotic normality of JINI based on inconsistent initial estimators also requires the asymptotic normality of $\wh{\bpi}(\bt_0,n)-\bpi(\bt_0)$. As an illustration, in Supplementary Material~\ref{sec:asymo_norm_naive_MLE_logistic_misclas} we consider an example of a logistic regression with misclassification (i.e., the same model used in the real data analysis in Section~\ref{sec:alcohol}), for which we use NMLE (i.e., MLE for a classical logistic regression) as an inconsistent initial estimator of JINI. We show that NMLE is asymptotically normal when $p^2\log(n)n^{-1}\to 0$ and under mild conditions on the design.

\begin{Theorem}
\label{thm:JINI_bias:incons_initial:diverge_p}
When the initial estimator is inconsistent and $p^{1/2}n^{-\alpha}\to 0$, under the assumptions presented in Supplementary Material~\ref{sec:pf:thm:JINI_bias:incons_initial:diverge_p} (similar to the ones in Theorem~\ref{thm:JINI_bias:incons_initial}), for any $\s\in\real^p$ such that $\|\s\|_2=1$, we have $\s\trans\{\mathbb{E}(\wh{\bt})-\bt_0\} = \mathcal{O}(p^2n^{-2\alpha}+p^{1/2}n^{-\beta})$. 
\end{Theorem}

\begin{Theorem}
\label{thm:JINI_bias:cons_initial:diverge_p}
Suppose that the initial estimator is consistent and $p^{1/2}n^{-\alpha}\to 0$. Consider any $\s\in\real^p$ such that $\|\s\|_2=1$.
\begin{enumerate}
    \item Under the assumptions considered in Supplementary Material~\ref{sec:pf:thm:JINI_bias:cons_initial:general:diverge_p} (similar to the ones in the first result of Theorem~\ref{thm:JINI_bias:cons_initial}), we have $\s\trans\{\mathbb{E}(\wh{\bt})-\bt_0\} = \mathcal{O}\{p^{3/2}n^{-(\alpha+\beta)}\}$.
    \item Under the assumptions considered in Supplementary Material~\ref{sec:pf:thm:JINI_bias:cons_initial:comb:diverge_p} (similar to the ones in the second result of Theorem~\ref{thm:JINI_bias:cons_initial}), we have $\s\trans\{\mathbb{E}(\wh{\bt})-\bt_0\} = \mathcal{O}\{p^{3/2}n^{-(\alpha+\beta_2)}\}$.
    \item Under the assumptions considered in Supplementary Material~\ref{sec:pf:thm:JINI_bias:cons_initial:smooth:diverge_p} (similar to the ones in the third result of Theorem~\ref{thm:JINI_bias:cons_initial}), we have $\s\trans\{\mathbb{E}(\wh{\bt})-\bt_0\} = \mathcal{O}\left\{pn^{-(2\alpha+\beta_2)}+p^{1/2}n^{-\beta_3}\right\}$.
    \item Under the assumptions considered in Supplementary Material~\ref{sec:pf:thm:JINI_bias:cons_initial:linear:diverge_p} (similar to the ones in the fourth result of Theorem~\ref{thm:JINI_bias:cons_initial}), we have $\mathbb{E}(\wh{\bt})-\bt_0 = \0$ when $n$ is sufficiently large. 
\end{enumerate}
\end{Theorem}

Compared to Theorems~\ref{thm:JINI_bias:incons_initial} and \ref{thm:JINI_bias:cons_initial} respectively, Theorems~\ref{thm:JINI_bias:incons_initial:diverge_p} and \ref{thm:JINI_bias:cons_initial:diverge_p} show that the bias of JINI may deteriorate due to diverging $p$, whether the initial estimator is consistent or not. Nevertheless, when the initial estimator is consistent, better bias order of JINI can be obtained by imposing stronger conditions on the bias function of the initial estimator. For example, when we use MLE as the initial estimator and hence $\alpha=1/2$ and $\beta_2=2$, the second result of Theorem~\ref{thm:JINI_bias:cons_initial:diverge_p} reduces to $\mathcal{O}(p^{3/2}n^{-5/2})$, or simply $o(n^{-1})$ as $p/n\to 0$. This result is comparable to the bias order of $o(n^{-1})$ achieved by many existing bias correction methods which assume fixed $p$ (see e.g., \citealp{efron1975,Firt:93,Kosm:14}). Finally, under suitable conditions, JINI can still be unbiased with large enough finite $n$, even when $p$ is allowed to diverge.

To summarize, in this section we extend the theoretical results of JINI to settings where $p$ is allowed to diverge with $n$ and $p<n$. To the best of our knowledge, these results are the first in the bias correction literature that consider general parametric models with diverging $p$. Our results provide an explanation for the advantageous finite sample performance of JINI in settings where $p$ is relatively large compared to $n$, as seen in the numerical examples in Sections~\ref{sec:simu} and \ref{sec:alcohol}.

\section{Simulation Studies}
\label{sec:simu}

In this section, we examine the finite sample performance of JINI. The point estimation performance are evaluated by two metrics: finite sample bias and standard error. Then we construct 95\% CI based on asymptotic normality to assess the inference performance, which is evaluated by empirical coverage and average CI length. To be concise, in this section we focus on examples where JINI uses inconsistent initial estimators in order to highlight two key messages: (i) JINI allows to bypass the analytical and/or numerical challenges typically entailed in standard approaches when handling data features (e.g., rounding and misclassification), or in robust estimation. (ii) JINI enjoys advantageous bias correction performance which leads to more accurate coverage. In Supplementary Material~\ref{sec:simu:logistic}, we consider a logistic regression example where JINI uses MLE as a consistent initial estimator, and is compared to MLE and BBC. In this example, JINI shows smaller finite sample bias and more accurate coverage than both MLE and BBC, especially in small sample settings.

\subsection{Beta Regression with Rounded Responses}
\label{sec:simu:beta_rounding}

Rounding of data is ubiquitous. Indeed, continuous random variables are often rounded and treated as discrete data, for example, due to the precision of experimental instruments or the way in which data are recorded and stored (see e.g., \citealp{bai2009statistical}). Such rounding of data is often neglected in the data analysis procedure, but studies have shown that this can lead to biased or even inconsistent estimation, even when the model is correctly specified (see e.g., \citealp{dong2015regression}). In this example, we consider a beta regression (see e.g., \citealp{ferrari2004beta}) to model continuous variables defined on $(0,1)$, such as rates or proportions. In this case, the actual response variable $\wc{Y}_i$ with $i=1,\ldots,n$ has a conditional density given by
\bsq
    f_{\wc{Y}|\x}(y_i, \x_i, \bb, \phi) =
    \frac{\Gamma(\phi)}{\Gamma(\mu_i\phi)\Gamma\{(1-\mu_i)\phi\}}
    y_i^{\mu_i\phi-1} (1-y_i)^{(1-\mu_i)\phi-1}, 
\esq
where $\Gamma(\cdot)$ is the gamma function. We define $\mu_i\equiv\mathbb{E}(\wc{Y}_i|\x_i)=\exp(\x_i\trans\bb)/\{1+\exp(\x_i\trans\bb)\}$, $\phi>0$ is the precision parameter such that $\var(\wc{Y}_i|\x_i)=\mu_i(1-\mu_i)/(1+\phi)$, $\x_i$ is the $(p+1)$-dimensional fixed covariate vector and $\bb$ is the regression coefficient. Instead of observing $\wc{Y}_i$, we observe a rounded $Y_i$ given by $Y_i \equiv \sum_{j=0}^{10} \frac{j}{10} I\left(\frac{j}{10}-0.05 \leq \wc{Y}_i < \frac{j}{10}+0.05\right)$, where $I(\cdot)$ is the indicator function. We assume that the rounding mechanism is known a priori. Moreover, we consider $x_{i,0}=1$ (i.e., intercept) and $x_{i,j}$ as realizations of $\mathcal{N}(0,2n^{-1/2})$ for $j=1,\ldots,p$. The true parameter values are 
\bsq
    \bb_0=(-0.5,\underbrace{1,\ldots,1}_{5},\underbrace{-2,\ldots,-2}_{5},\underbrace{3,\ldots,3}_{5},
    \underbrace{-0.1,0.1,\ldots,-0.1,0.1,-0.1}_{p-15})\trans \quad \text{and} \quad
    \phi_0=10. 
\esq
We consider three settings: (i) $p=50,n=200$, (ii) $p=60,n=500$, and (iii) $p=70,n=1000$, so that we can study the asymptotic behaviors of the estimators. $10,000$ Monte Carlo replications are considered.

In this simulation, we compare the finite sample performance of MLE and JINI. MLE is computed using the EM algorithm. More precisely, the E step at the $k\th$ iteration defines the conditional expectation of the log-likelihood function of $(\bb\trans,\phi)\trans$ with respect to the conditional distribution of $\wc{\Y}\equiv(\wc{Y}_1,\ldots,\wc{Y}_n)\trans$ given $\Y\equiv(Y_1,\ldots,Y_n)\trans$ and the current estimates $\bb^{(k)}$ and $\phi^{(k)}$. The following M step finds $\bb^{(k+1)}$ and $\phi^{(k+1)}$ that maximize the conditional expectation defined in the E step. Naturally, if a more complicated rounding mechanism is used, the conditional expectation of the log-likelihood considered in the E step may have no closed-form expression and approximations may be needed. Alternatively, we implement JINI based on NMLE (i.e., MLE of a classical beta regression). NMLE is inconsistent as it neglects rounding and treats $Y_i$ as the true response variable, but is easily computable using the readily available \texttt{betareg} function in the \texttt{betareg} R package. Since $Y_i$ can take values of $0$ and $1$, we consider a transformation $\{Y_i(n-1)+0.5\}/n$ as the input response variable in the \texttt{betareg} function (see e.g., \citealp{cribari2010beta}). In this case, JINI is easier to obtain than MLE as it avoids the analytical computation of the likelihood function as well as the EM procedure. 

\begin{figure}[!tb]
    \centering
    \includegraphics[width=12cm]{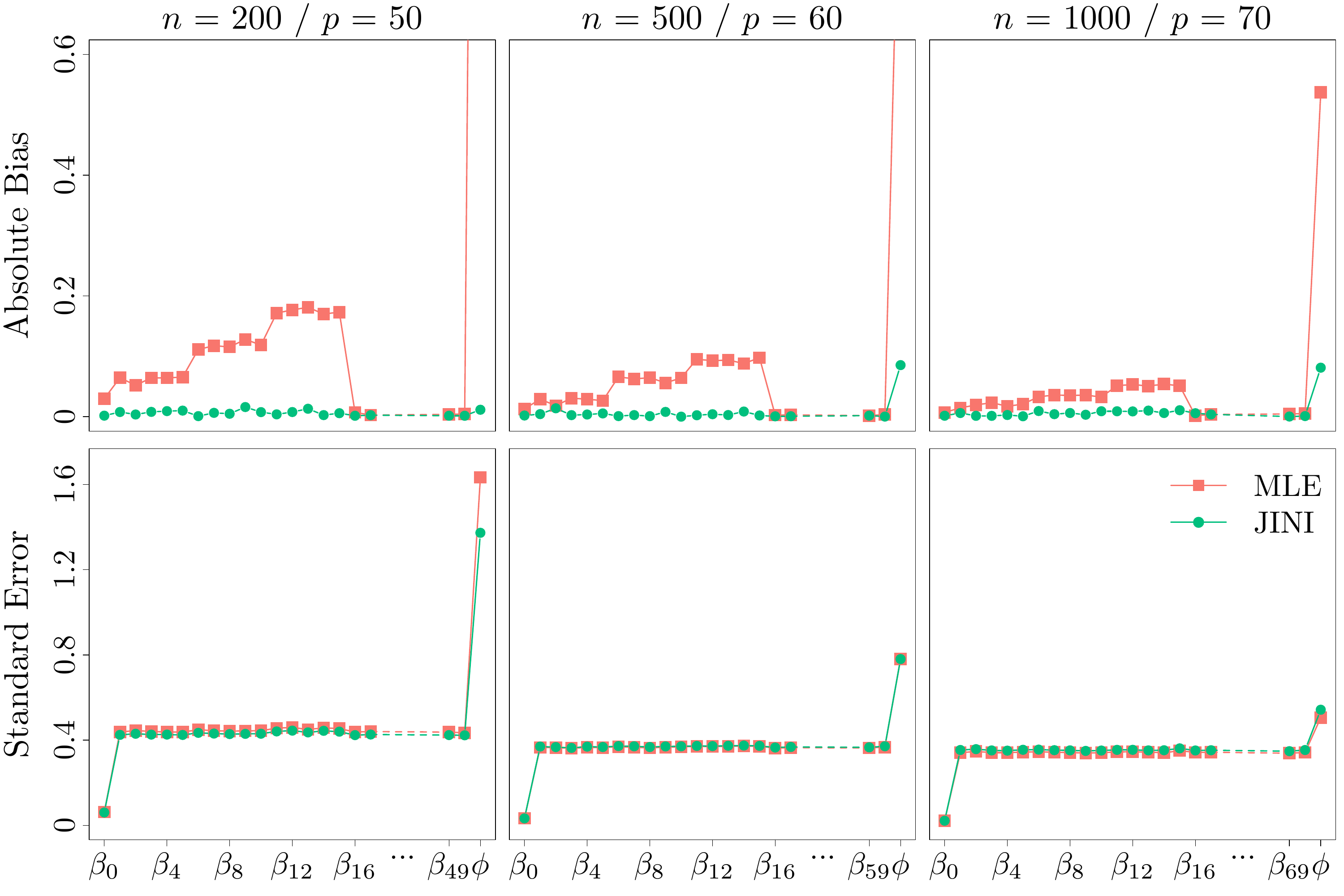}
    \caption{Estimation results for $\bb$ and $\phi$ for the beta regression with rounded responses presented in Section~\ref{sec:simu:beta_rounding}.}  
    \label{fig:beta_point_est_with_rounding}
\end{figure}

The point estimation performance is summarized in Figure~\ref{fig:beta_point_est_with_rounding}, which shows that MLE is severely biased, especially for the parameter $\phi$. In contrast, JINI shows a negligible bias for $\bb$ and a much smaller bias for $\phi$ in all three settings. Moreover, both estimators have comparable standard errors, highlighting that the bias correction of JINI does not come at a price of an increased variance. As $n$ increases, both estimators show decreasing biases and standard errors, suggesting their consistency.

\begin{figure}[!tb]
    \centering
    \includegraphics[width=12cm]{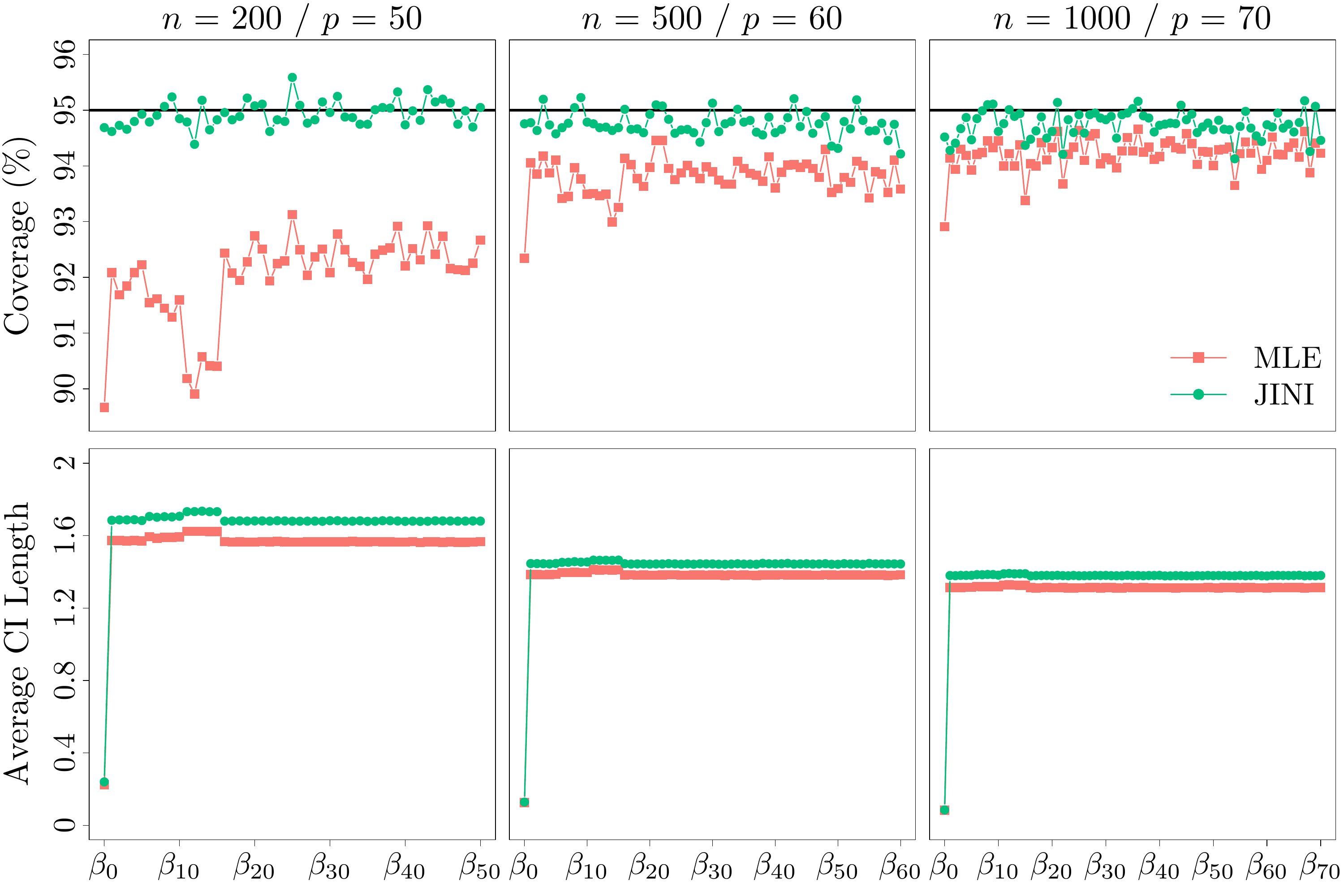}
    \caption{Inference results for $\bb$ for the beta regression with rounded responses presented in Section~\ref{sec:simu:beta_rounding}.}  
    \label{fig:beta_inference_with_rounding}
\end{figure}

We further construct $95\%$ CIs for $\bb$ to compare the coverage accuracy of both estimators. Specifically, we estimate the asymptotic variances of both estimators by parametric bootstrap with $100$ bootstrapped samples, and construct the CIs based on their asymptotic normality. The inference results for $\bb$ are presented in Figure~\ref{fig:beta_inference_with_rounding}. We can see that JINI has a marginally inflated average CI length compared to MLE, but the difference decreases as $n$ increases. However, the narrower CI by MLE is not sufficiently precise, in that its coverage is lower than the nominal level. In contrast,  JINI presents more accurate coverage, due to its negligible bias hence precise variance calculation, especially when $n$ is relatively small.  

\subsection{Robust Logistic Regression}
\label{sec:simu:roblogistic}

Data contamination is frequently encountered in practice. For binary data, it is often associated with misclassification, in the sense that the positive responses can be taken as negative and vice versa. When we have enough prior knowledge about the misclassification mechanism, we can include it into the model, as considered in Section~\ref{sec:alcohol}. When a small amount of data contamination may exist but no prior knowledge is available, the use of a robust estimation approach can be beneficial in order to limit the potential influence of the contamination.  General accounts of robust statistics can be found, for example, in \cite{Hube:81,hampel1986robust}. 

In this section, we consider a logistic regression where the response variable $Y_i$ is generated from a Bernoulli distribution with $\mu_i(\bb)\equiv \pr(Y_i=1|\x_i) = \exp(\x_i\trans\bb)/\{1+\exp(\x_i\trans\bb)\}$, where $\x_i$ is the $(p+1)$-dimensional fixed covariate vector, $\bb$ is the regression coefficient, and $i=1,\ldots,n$. Different robust estimators have been proposed for logistic regression (see e.g., \citealp{Bianco1996,CaRo:01b,MPVF:psycho02,Cize:08}). For example, \cite{CaRo:01b} proposed a robust M-estimator for generalized linear models as the solution to the following estimating equation: 
\bq \label{eqn:cantoni_estimating_equation}
    \sum_{i=1}^n \left[ \psi_c(r_i) w(\x_i) \frac{(\partial/\partial\bb)\mu_i(\bb)}{\sqrt{\mu_i(\bb)\{1-\mu_i(\bb)\}}}-\c_1(\bb)\right]= \0,
\eq
where $\psi_c(\cdot)$ is the Huber function with tuning constant $c$ that is chosen to balance the robustness and asymptotic efficiency loss. $r_i$ is the Pearson residual, $w(\cdot)$ is a weight function on the design, and $\c_1(\bb)$ is the Fisher consistency correction term defined as $\c_1(\bb)\equiv n^{-1}\sum_{i=1}^n \mathbb{E}\{\psi_c(r_i)\} w(\x_i) [\mu_i(\bb)\{1-\mu_i(\bb)\}]^{-1/2} (\partial/\partial\bb)\mu_i(\bb)$. An alternative robust estimator is the Weighted MLE (WMLE) (see e.g., \citealp{FiSm:94,dupuis2002robust}) defined as the solution to the following estimating equation: 
\bq \label{eqn:wmle_estimating_equation}
    \sum_{i=1}^n \left\{w_c(d_i) \s(\bb|y_i,\x_i)-\c_2(\bb)\right\} = \0,
\eq
where $w_c(\cdot)$ is a bounded weight function with tuning constant $c$, $\s(\bb|y_i,\x_i)$ is the score function, $d_i\equiv \|\s(\bb|y_i,\x_i)\|_2$, and $\c_2(\bb)$ is the Fisher consistency correction term defined as $\c_2(\bb)\equiv n^{-1}\sum_{i=1}^n \mathbb{E}\{w_c(d_i) \s(\bb|y_i,\x_i)\}$. 
These classical robust estimators are generally difficult to compute due to the correction terms (see $\c_1(\bb)$ in \eqref{eqn:cantoni_estimating_equation} and $\c_2(\bb)$ in \eqref{eqn:wmle_estimating_equation} as examples) that are often analytically and/or numerically intractable, whereas removing them from the estimating equations will render inconsistent estimators. 

To circumvent these difficulties typically encountered in the classical robust estimation approach, we propose to construct a robust JINI using a Naive WMLE (NWMLE) as the initial estimator. This NWMLE is the solution to $\sum_{i=1}^n w_c(d_i) \s(\bb|y_i,\x_i) = \0$, i.e., neglecting $\c_2(\bb)$ from \eqref{eqn:wmle_estimating_equation}. Although NWMLE is inconsistent, it is significantly simpler to compute, at a computational cost the same as to compute MLE. Moreover, we propose to use the Tukey's biweight function (see \citealp{beaton1974fitting}) given by $w_c(d)=\{1-(d/c)^2\}^2 I(d\leq c)$. Unlike the Huber weight function, the Tukey's biweight function is redescending in the sense that the weights become zero when $d>c$, hence it is not needed to further assign weights on the design. 

We evaluate the performance of our proposed robust JINI at the model (see e.g., \citealp{hampel1986robust}), i.e., when the data are generated from a logistic regression without contamination. We compare it to three benchmark estimators: (i) MLE, which is not robust but is more efficient (asymptotically) than any robust estimator at the model. (ii) The robust M-estimator proposed by \cite{CaRo:01b}, i.e., the solution to \eqref{eqn:cantoni_estimating_equation}. We call it CR for simplicity and use its implementation in the \texttt{glmrob} function in the \texttt{robustbase} R package. (iii) The robust estimator proposed by \cite{Bianco1996}, which we refer to as BY. Compared to CR which is developed for generalized linear models, BY is developed specifically for logistic regression. Its implementation is also available in the \texttt{glmrob} function in R.

In this simulation, we consider $x_{i,0}=1$ (i.e., intercept) and the $n\times p$ covariate matrix $\X= \wt{\X}\bOmega$, where $\wt{x}_{i,j}$ is a realization of $\mathcal{N}(0,4n^{-1/2})$, and $\bOmega$ is the lower triangular Cholesky decomposition matrix of a $p\times p$ Toeplitz matrix whose first row is $(1, 0.8, 0.8^2, \ldots, 0.8^{p-1})$. This choice of covariates is to ensure that $\mu_i(\bb)$ is not trivially equal to either $0$ or $1$, and that the size of the log-odds ratio does not increase with $n$. The true parameter values are $\bb_0 = (0.3, -2, -4, 0, \ldots, 0)\trans$. Although $\bb_0$ is sparse, we note that there is no variable selection conducted in our analysis and a non-sparse choice of $\bb_0$ can lead to similar results. We consider three settings of $p$ and $n$: (i) $p=20,n=200$, (ii) $p=30,n=400$, and (iii) $p=40,n=800$. The tuning constant $c$ is 2.8, 2.6 and 2.2 respectively for robust JINI with Tukey's biweight function, such that it corresponds to approximately 95\% efficiency of MLE. $10,000$ Monte Carlo replications are considered for each setting. 

\begin{figure}[!tb]
    \centering
    \includegraphics[width=12cm]{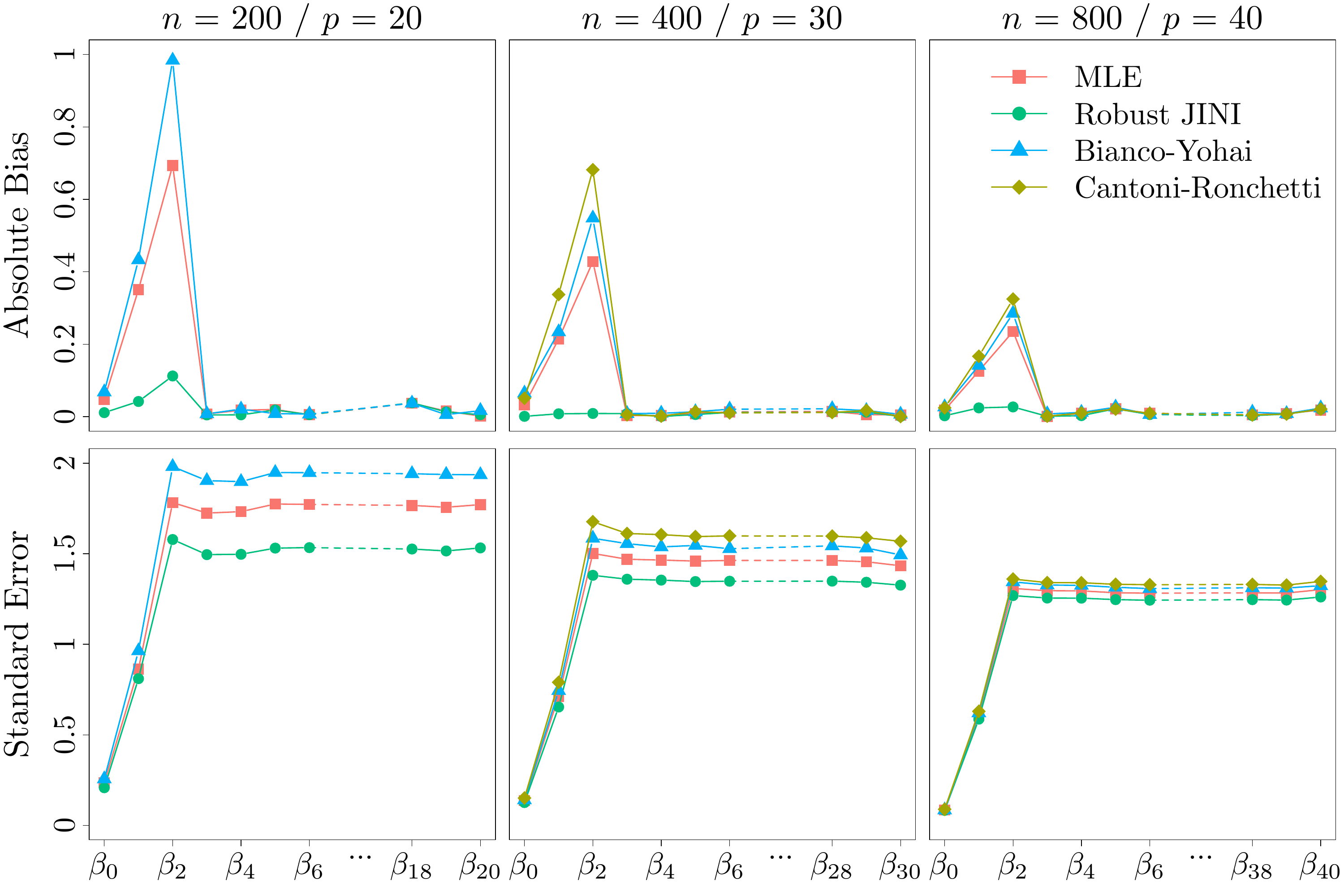}
    \caption{Estimation results for the robust logistic regression presented in Section~\ref{sec:simu:roblogistic}.}  
    \label{fig:roblogistic_point_est_nocont}
\end{figure}

Figure~\ref{fig:roblogistic_point_est_nocont} summarizes the estimation performance of all estimators. We do not include CR in the first setting due to its unstable performance. We can clearly see that robust JINI has much smaller bias and standard error compared to the others, especially in the first setting where $p/n$ is the largest. In particular, robust JINI outperforms MLE, despite the fact that MLE is more (asymptotically) efficient than robust estimators at the model (i.e., with no data contamination). The other robust estimators have larger biases and standard errors than MLE, with BY slightly outperforming CR.

We also compare the inference performance of these estimators by comparing their 95\% CIs, which are constructed based on the asymptotic normality of each estimator. The asymptotic covariance matrix is estimated using plug-in for MLE, using parametric bootstrap with 100 bootstrapped samples for robust JINI, and using the estimates from the \texttt{glmrob} R function for BY and CR. 

\begin{figure}[!tb]
    \centering
    \includegraphics[width=12cm]{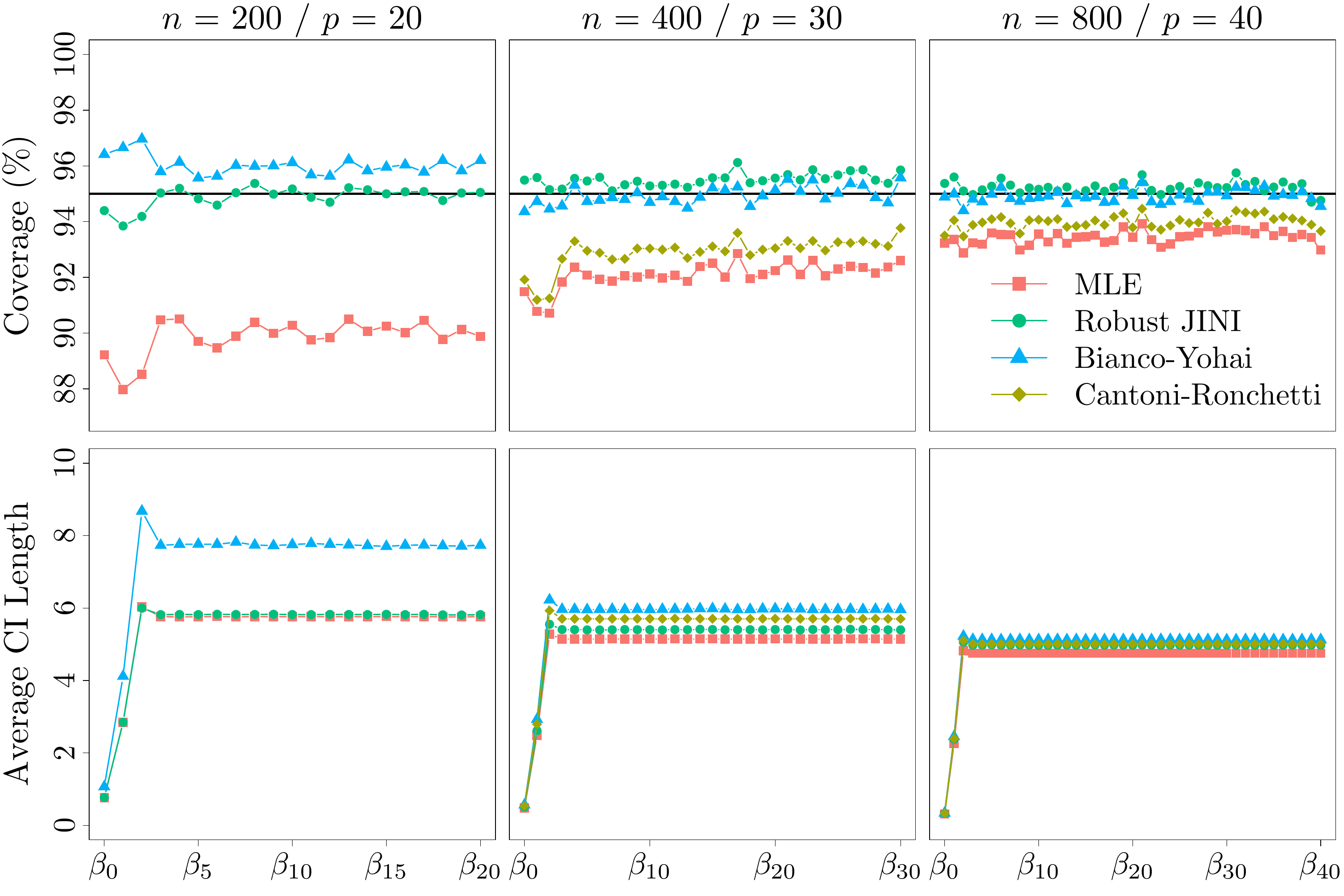}
    \caption{Inference results for the robust logistic regression presented in Section~\ref{sec:simu:roblogistic}.}  
    \label{fig:roblogistic_inference_nocont}
\end{figure}

The inference performance is summarized in Figure~\ref{fig:roblogistic_inference_nocont}, in which we do not include CR in the first setting due to its numerical instability. In all settings, robust JINI has accurate coverage and small average CI length. MLE and CR have comparable average CI lengths, but their CIs have coverages significantly lower than 95\% for all parameters, especially when $p/n$ is large. BY does not perform as well as JINI in the first setting, as it has a heavily inflated average CI length which results in conservative coverage. In the second and third settings, it has similar or slightly worse performance compared to JINI. Overall, JINI is the most reliable robust estimator in this example in that it has small bias, accurate coverage, and small average CI length in all settings, especially when $p/n$ is large.

In order to highlight that our approach allows to construct a robust estimator in a more straightforward manner than the classical approach, in Supplementary Material~\ref{sec:simu:pareto} we consider another simulation on a Pareto regression for which, to the extent of our knowledge, no robust estimator exists. We find that when there is no data contamination, the proposed robust JINI based on NWMLE shows negligible bias and a standard error comparable to MLE. When data are contaminated, the performance of robust JINI remains stable (with negligible bias and similar standard error) and is much better than MLE, suggesting its robustness to data contamination.

\section{Alcohol Consumption Data Analysis}
\label{sec:alcohol}

Alcohol consumption is a significant concern for public health. Its measurements are mostly collected using self-reporting questionnaires (see e.g., \citealp{AUDIT-C:98}), which typically suffer from measurement errors. Indeed, when completing questionnaires, participants often tend to under report their alcohol consumption levels when they consume a large amount of alcohol, whereas those who do not consume much alcohol generally report the truth. This phenomenon is mainly due to the social desirability (see e.g., \citealp{SocialDesire:10}). Consequently, the alcohol consumption data typically display a non-negligible False Negative Rate (FNR), whereas it is often reasonable to think that the False Positive Rate (FPR) is nil or negligible.

FNR and FPR could, in principle, be estimated using MLE (see e.g., \citealp{HAUSMAN1998}). However, a considerably large sample size is necessary to ensure reliable estimates. For example, \cite{LiZh:17} 
found that $n=500$ is already too small to accurately estimate FNR and FPR for a logistic regression with four predictors. Alternative to the likelihood approach, some studies have investigated the potential FNR when using self-reporting questionnaires in the context of alcohol consumption, and suggested a range between 3\% to 10\% (see e.g., \citealp{AlcoholFN:08,AlcoholFN:21}). 

In this section, we consider the data collected during the 2005-2006 school year from two public schools in Portugal. The data were originally collected and first analyzed in \cite{Cortez2008UsingDM} to study the performance of secondary school students in Mathematics and Portuguese language. The data came from two sources: school recordings (e.g., grades, number of school absences) and self-reporting questionnaires (e.g., workday and weekend alcohol consumption, parents' jobs, quality of family relationships). In this analysis, we use the dataset in Mathematics with 395 observations to study students' alcohol consumption levels using a logistic regression. Since both workday and weekend alcohol consumptions take five scales (i.e., very low, low, medium, high, very high), we combine them into binary outcomes to denote students' alcohol consumption levels. The level is low with value $0$ only if the workday alcohol consumption is very low and the weekend alcohol consumption is very low or low. The remaining $44$ attributes, which include binary, numeric and categorical types, are used as regressors. More details on the data description can be found in Table~\ref{tab:alcohol_data_description} in Supplementary Material~\ref{sec:alcohol_additional}. Given that most measurements are self reported, we assume a true FNR of $5\%$, a reasonable value in light of existing studies, and zero FPR. A sensitivity analysis on how the assumed FNR on a range from 3\% to 10\% influences the estimation performance is presented in Figure~\ref{fig:alcohol_sensitivity} in Supplementary Material~\ref{sec:alcohol_additional}, and the same conclusions can be drawn as in this section.

In this study, we consider two estimators: MLE and JINI which uses NMLE (i.e., MLE for a classical logistic regression which neglects FNR) as the initial estimator. We also construct $95\%$ CIs based on asymptotic normality to compare their inference performance. To estimate the asymptotic covariance matrix, we use plug-in for the one of MLE, and parametric bootstrap with $100$ bootstrapped samples for the one of JINI.

Based on this real dataset on alcohol consumption, we first conduct a numerical experiment in Section~\ref{sec:alcohol:simu}. This study shows that JINI outperforms MLE in terms of both point estimation and inference, and allows to better understand the discrepancy of these approaches in the real data analysis presented in Section~\ref{sec:alcohol:real_data}.  

\subsection{Numerical Experiment}
\label{sec:alcohol:simu}

\begin{figure}[!tb]
    \centering
    \includegraphics[width=12cm]{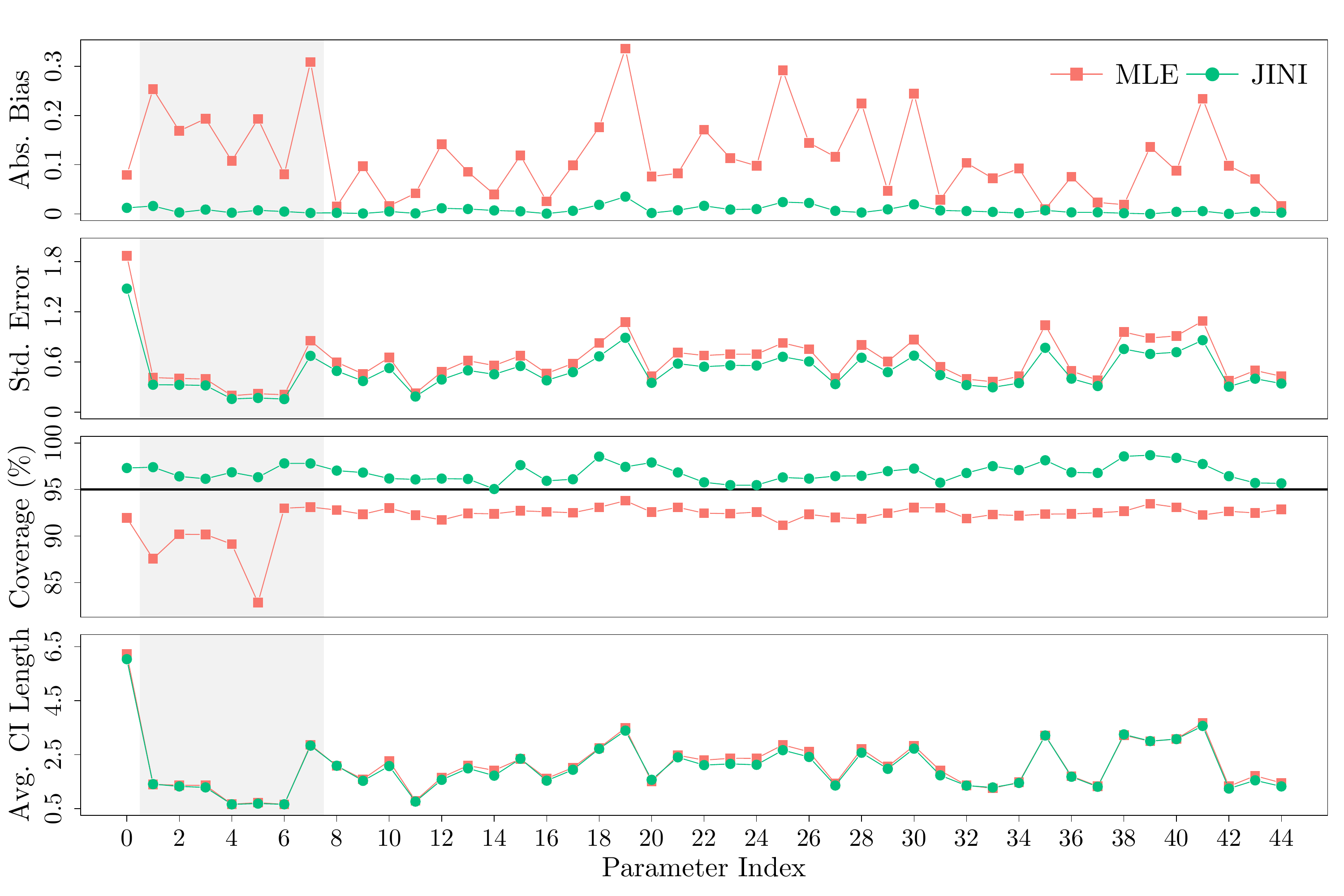}
    \caption{Estimation and inference results for the logistic regression with misclassification presented in Section~\ref{sec:alcohol:simu}. The shaded area covers the  parameters corresponding to covariates that are found significant by either MLE or JINI in the real data analysis presented in Section~\ref{sec:alcohol:real_data}.}  
    \label{fig:alcohol_simu}
\end{figure}

In this section, we conduct a simulation based on the real data on alcohol consumption for a logistic regression with zero FPR and $5\%$ FNR. The covariates include the intercept and the 44 attributes from the real data. 
The true parameter values are set to be the ones estimated by JINI on the real data. We compare the performance of MLE and JINI based on $10,000$ Monte Carlo replications, and the results are summarized in Figure~\ref{fig:alcohol_simu}. 

In Figure~\ref{fig:alcohol_simu}, we first observe that MLE is considerably biased for most parameters, whereas JINI appears nearly unbiased for all parameters. JINI also has smaller standard errors than MLE for all parameters, indicating that its advantageous bias correction performance does not come at an expense of an increased variance. As for inference, the CIs of JINI show more accurate and conservative empirical coverages than the ones of MLE, with comparable average CI lengths. Indeed, for all parameters the coverages of JINI maintain at least $95\%$, whereas MLE always shows less than $95\%$ coverage. In particular, the coverages for $\beta_1$ to $\beta_7$, the parameters corresponding to the significant covariates in the real data, are particularly poor for MLE. This example illustrates that the negligible bias of JINI provides a more reliable basis for accurate inference compared to MLE.

\subsection{Real Data Analysis}
\label{sec:alcohol:real_data}

\begin{figure}[!tb]
    \centering
    \includegraphics[width=12cm]{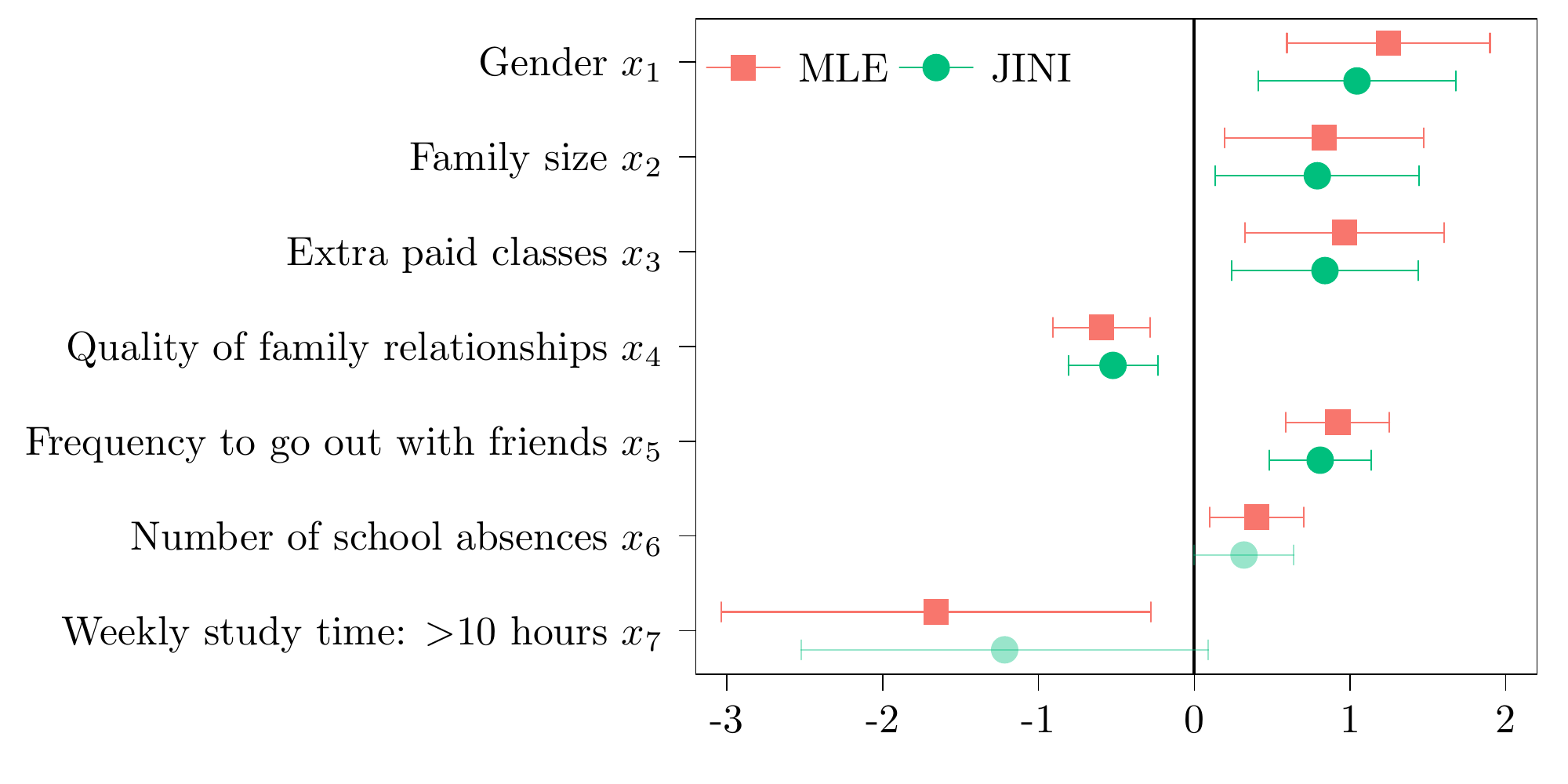}
    \caption{Point estimates and $95\%$ CIs for the real alcohol consumption data analysis presented in Section~\ref{sec:alcohol:real_data}. Only parameters corresponding to covariates that are found significant by either MLE or JINI are included. When the CI does not cover zero, its color is solid. When the CI covers zero, its color is transparent.}  
    \label{fig:alcohol_real_data_significant}
\end{figure}

In this section, we fit the logistic regression with zero FPR and $5\%$ FNR on the real alcohol consumption dataset, and compute the estimates and $95\%$ CIs of MLE and JINI. The complete results are provided in Figure~\ref{fig:alcohol_real_data_all} in Supplementary Material~\ref{sec:alcohol_additional}. In Figure~\ref{fig:alcohol_real_data_significant}, we focus on the parameters which correspond to the covariates found significant by either MLE or JINI, i.e., the ones whose CIs do not cover zero.

In Figure~\ref{fig:alcohol_real_data_significant}, we can see that both MLE and JINI identify $x_1$ to $x_5$ to have significant associations to students' alcohol consumption levels. MLE also detects $x_6$ and $x_7$ as significant, but JINI does not. In the simulation based on this real dataset presented in Section~\ref{sec:alcohol:simu}, we observe that the severe bias of MLE renders unreliable inference. So it is reasonable to conjecture that MLE falsely identifies $x_6$ and $x_7$ to be significant due to its severe biases on $\beta_6$ and $\beta_7$ in this real data analysis. Overall, these results suggest the adequacy of JINI in practical applications, since its advantageous bias correction property allows more reliable inference.

\section{Conclusions}
\label{sec:conclusions}

In this paper, we propose a reliable estimation and inference approach for parametric models based on JINI. We investigate the properties of JINI, including consistency, asymptotic normality and its bias correction property. The main merit of JINI is that it can be constructed in a simple manner, while providing bias correction guarantees that lead to accurate inference. Extended results to settings where $p$ diverges with $n$ and $p<n$ are also included, which (to the best of our knowledge) are the first in the literature that considers bias correction for general parametric models with diverging $p$. Our results provide the theoretical foundation to explain the advantageous finite sample performance of JINI, especially in small sample settings. Our approach is particularly useful, for example, to handle data features (e.g., misclassification and rounding), or to construct robust estimators, as demonstrated in our simulation studies and an alcohol consumption data analysis. 

Various extensions are worth investigating following our work. For
example, a natural methodological extension is to go beyond parametric
models to semi-parametric models or non-parametric models, or to
consider  high dimensional parameters
(i.e., $p\gg n$) in parametric models.

\section{Technical Conditions}
\label{sec:conditions}

In this section, we present and discuss the conditions used to establish the results of JINI in Sections~\ref{sec:results:fixed_p:incons_initial} and \ref{sec:results:fixed_p:cons_initial} when the parameter dimension $p$ is fixed.

\begin{Assumption}
\label{assum:compact}
The parameter space $\bT$ is a compact convex subset of $\real^{p}$ and $\bt_0 \in \Int(\bT)$. 
\end{Assumption}

Assumption~\ref{assum:compact} is a commonly used regularity condition in settings where estimators have no closed-form solutions (see e.g., \citealp{white1982maximum,newey1994large}). 

\begin{Assumption}
\label{assum:v}
There exists some $\alpha>0$ such that $\v(\bt_0,n)=\mathcal{O}_p(n^{-\alpha})$. 
\end{Assumption}

Assumption~\ref{assum:v} is typically satisfied, for example, when $\sqrt{n}\{\wh{\bpi}(\bt_0,n)-\bpi(\bt_0)\}$ is asymptotically normal. For example, \cite{brillinger1983generalized} showed that, under certain conditions on the regressor $\x$, the ordinary least squares estimator is asymptotically normal for a model in which the conditional mean of the response variable is $g(\x\trans\bt_0)$ with an unknown nonlinear function $g(\cdot)$. \cite{brillinger1983generalized} also noted a broad applicability of their results on models such as logistic regression, censored regression (see e.g., \citealp{greene1981asymptotic,nelson1981test}), and Cox proportional hazards model (see e.g., \citealp{CoxDR:72}). \cite{li1989regression} studied that, under appropriate conditions and allowing the true link function to be arbitrary, any MLE is asymptotically normal even if it might be based on a misspecified link function. They also highlighted that distributional violation can be seen as a special kind of link violation, implying a general applicability of their results. Other examples include \cite{huber1967behavior,white1982maximum,czado1992effect}. In these cases where $\wh{\bpi}(\bt_0,n)$ is asymptotically normal, by the decomposition of $\wh{\bpi}(\bt_0,n)$ in \eqref{eqn:decomp_JINI} we directly establish $\v(\bt_0,n)=\mathcal{O}_p(n^{-1/2})$, i.e., Assumption~\ref{assum:v} is satisfied with $\alpha=1/2$.

\begin{Assumption}
\label{assum:finite_bias}
There exists some $\beta > \alpha$, where $\alpha$ is defined in Assumption~\ref{assum:v}, such that 
$\b(\bt,n)=\mathcal{O}(n^{-\beta})$ uniformly for $\bt\in\bT$.
\end{Assumption}

Assumption~\ref{assum:finite_bias} ensures that the rate of the finite sample bias is dominated by the error rate. For example, when the initial estimator is MLE for a possibly misspecified model, we have $\mathbb{E}\{\wh{\bpi}(\bt_0,n)\}-\bpi(\bt_0)=\mathcal{O}(n^{-1})$ under regularity conditions (see e.g., \citealp{cox1968general,cox1979theoretical}), and thus it is reasonable to consider $\beta=1$ in Assumption~\ref{assum:finite_bias}. Moreover, the uniform convergence is a standard regular condition (see e.g., \citealp{newey1994large}) that allows, for example, to guarantee $\b(\wh{\bt},n)\overset{P}{\to}\0.$

\begin{Assumption}
\label{assum:asymp_bias}
The function $\bpi(\bt)$ is injective. Moreover, it is twice continuously differentiable with $\A(\bt_0) \equiv \partial \bpi(\bt_0)/\partial \bt$ to be full rank. 
\end{Assumption}

The injection condition on the non-stochastic limit $\bpi(\bt)$ is a standard regularity condition to ensure identifiability (see e.g., \citealp{gourieroux2000,guerrier2018simulation}). Lower level conditions to ensure identifiability can, for example, be found in \cite{komunjer2012global} and the reference therein. The second part of Assumption~\ref{assum:asymp_bias} essentially requires that $\bpi(\bt)$ is relatively smooth. This can be satisfied, for example, when data exhibit features such as censoring or misclassification (see e.g., \citealp{greene1981asymptotic}). Moreover, the full rank condition on $\A(\bt_0)$ is often considered, for example, in indirect inference (see e.g., \citealp{gourieroux1993indirect}). We highlight that an injective and continuous $\bpi(\bt)$ suffices to ensure the consistency of JINI.

\begin{Assumption}
\label{assum:initial_asymp_norm}
The initial estimator satisfies $n^\alpha \bSig(\bt_0, n)^{-1/2}\left\{\wh{\bpi}(\bt_0,n)-\bpi(\bt_0)\right\} \overset{D}{\to} \mathcal{N}(\0,\I_p)$, where $\alpha$ is defined in Assumption~\ref{assum:v} and $\bSig(\bt_0, n)$ is a $p\times p$ covariance matrix. Moreover, the limit of $\bSig(\bt_0, n)$ as $n\to\infty$ exists and is positive definite. 
\end{Assumption} 

Assumption~\ref{assum:initial_asymp_norm} requires the asymptotic normality of the initial estimator at a convergence rate of $n^{-\alpha}$. As discussed earlier for Assumption~\ref{assum:v}, when we use, for example, MLE based on a possibly misspecified model as the initial estimator (see e.g., \citealp{li1989regression}), we have $\sqrt{n}\{\wh{\bpi}(\bt_0,n)-\bpi(\bt_0)\}$ to be asymptotically normal. This assumption is only needed to establish the asymptotic normality of JINI. 

\begin{Remark}
We can require $n^\alpha \bSig(\bt_0,n)^{-1/2}\left\{\wh{\bpi}(\bt_0,n)-\bpi(\bt_0)\right\}\overset{D}{\to} \Z$ in Assumption~\ref{assum:initial_asymp_norm}, where $\Z$ is a $p$-dimensional random variable whose distribution function does not depend on $\bt$. In this case, JINI satisfies $n^\alpha \left\{\A(\bt_0)^{-1} \bSig(\bt_0,n) \A(\bt_0)^{-1}\right\}^{-1/2} (\wh{\bt}-\bt_0) \overset{D}{\to} \Z$. In other words, the asymptotic distribution of JINI depends on that of the initial estimator. This is of particular interest when the initial estimator converges in distribution to a non-normal random variable. See, for example, \cite{smith1985maximum} for a detailed discussion on nonregular situations where MLE is not asymptotically normal. We prove this generalized result in supplementary materials, but we focus on asymptotic normality in the main article for simplicity of presentation.  
\end{Remark}

The rest of the assumptions are only used when the initial estimator is consistent. In this case, $\b(\bt,n)$ represents the bias of the initial estimator, and hence its order is the bias order of the initial estimator. We define the closed ball centered at $\bt_0$ with radius $\epsilon>0$ as $\boldsymbol{\mB}(\bt_0,\epsilon)\equiv\{\bt\in\bT:\|\bt-\bt_0\|_2\leq\epsilon\}$. 

\setcounter{Assumption}{2}
\renewcommand\theAssumption{\Alph{Assumption}.1}
\begin{Assumption}
\label{assum:finite_bias:general}
The  bias function $\b(\bt,n)$ is differentiable in $\bt\in\boldsymbol{\mB}(\bt_0,\epsilon)$ with some $\epsilon>0$. Moreover, for all $i=1,\ldots,p$, we have $\partial b_i(\bt,n) / \partial \bt = \mathcal{O}(n^{-\beta})$ uniformly for $\bt\in\boldsymbol{\mB}(\bt_0,\epsilon)$, where $\beta$ is defined in Assumption~\ref{assum:finite_bias}.  
\end{Assumption}

Assumption~\ref{assum:finite_bias:general} only requires $\b(\bt,n)$ to be differentiable in a small neighborhood of $\bt_0$, with the order of the derivative at least the same as the bias order of the initial estimator.

\setcounter{Assumption}{2}
\renewcommand\theAssumption{\Alph{Assumption}.2}
\begin{Assumption}
\label{assum:finite_bias:comb}
The bias function $\b(\bt,n)$ can be expressed as 
\bsq
    \b(\bt,n) = \left\{\sum_{j=1}^p b_{ij} \frac{\theta_j}{n^{\beta_1}}\right\}_{i=1,\ldots,p} + \r(\bt,n) + \c(n),
\esq
where $b_{ij}\in\real$, $\beta_1\geq\beta$ with $\beta$ defined in Assumption~\ref{assum:finite_bias}, and $\c(n)\in\real^p$. $\r(\bt,n)\in\real^p$ contains all the nonlinear terms of $\bt$, and is differentiable in $\bt\in\boldsymbol{\mB}(\bt_0,\epsilon)$ with some $\epsilon>0$. For all $i=1,\ldots,p$, we have $\partial r_i(\bt,n)/\partial\bt=\mathcal{O}(n^{-\beta_2})$ uniformly for $\bt\in\boldsymbol{\mB}(\bt_0,\epsilon)$, and $\beta_2\geq\beta$.  
\end{Assumption}

Assumption~\ref{assum:finite_bias:comb} requires that the leading-order term of the bias of the initial estimator is linear. This can be satisfied with $\beta_1=1$ and $\beta_2=2$, for example, when the initial estimator is MLE under regularity conditions (see e.g., \citealp{cox1968general,mardia1999bias,cordeiro2007third}).

\setcounter{Assumption}{2}
\renewcommand\theAssumption{\Alph{Assumption}.3}
\begin{Assumption}
\label{assum:finite_bias:smooth}
The bias function $\b(\bt,n)$ can be expressed as 
\bsq
    \b(\bt,n) = \left\{\sum_{j=1}^p b_{ij} \frac{\theta_j}{n^{\beta_1}} + \sum_{k=1}^p\sum_{l=1}^p r_{ikl}\frac{\theta_k\theta_l}{n^{\beta_2}}\right\}_{i=1,\ldots,p} + \mathcal{O}(n^{-\beta_3}) + \c(n),
\esq
with $b_{ij}\in\real$, $r_{ikl}\in\real$, $\c(n)\in\real^p$ and $\min(\beta_1,\beta_2,\beta_3)\geq\beta$, where $\beta$ is defined in Assumption~\ref{assum:finite_bias}.
\end{Assumption}

Assumption~\ref{assum:finite_bias:smooth} essentially employs a Taylor expansion on the leading-order terms of $\b(\bt,n)$ as a function of $\bt/n$. This can be satisfied, for example, when $\b(\bt,n)$ is a sufficiently smooth function of $\bt/n$. Similar approximations are commonly used when the bias function has no closed-form expression (see e.g., \citealp{tibshirani1993introduction}). When $\beta_1=1$, $\beta_2=2$ and $\beta_3=3$, this assumption corresponds to the condition used in \cite{guerrier2018simulation}.

\setcounter{Assumption}{2}
\renewcommand\theAssumption{\Alph{Assumption}.4}
\begin{Assumption}
\label{assum:finite_bias:linear}
The bias function $\b(\bt,n)$ can be expressed as 
\bsq
    \b(\bt,n) = \left\{\sum_{j=1}^p b_{ij} \frac{\theta_j}{n^{\beta_1}}\right\}_{i=1,\ldots,p} + \c(n),
\esq
with $b_{ij}\in\real$, $\c(n)\in\real^p$ and $\beta_1\geq\beta$, where $\beta$ is defined in Assumption~\ref{assum:finite_bias}.
\end{Assumption}

Assumption~\ref{assum:finite_bias:linear} requires $\b(\bt,n)$ to be linear. A simple example that satisfies this condition is MLE of a uniform distribution. The results of \cite{sur2019modern} suggest that the bias of MLE for a logistic regression is linear under suitable design conditions. In general, a linear bias function may often be a reasonable approximation (see e.g.,
\citealp{mackinnon1998approximate}).




\newpage
\appendix
\centerline{\Large\sc Supplementary Materials}

\section{Proof of Theorem~\ref{thm:JINI_consistency_asymp_norm}}
\label{sec:pf:thm:JINI_consistency_asymp_norm}

The consistency result in Theorem~\ref{thm:JINI_consistency_asymp_norm} is a special case of Theorem~\ref{thm:JINI_consistency:diverge_p} with fixed $p$. Please refer to Section~\ref{sec:pf:thm:JINI_consistency:diverge_p} for the proof of Theorem~\ref{thm:JINI_consistency:diverge_p}.

The asymptotic normality result in Theorem~\ref{thm:JINI_consistency_asymp_norm} is a special case of Theorem~\ref{thm:JINI_asymp_norm:diverge_p} with fixed $p$. Please refer to Section~\ref{sec:pf:thm:JINI_asymp_norm:diverge_p} for the proof of Theorem~\ref{thm:JINI_asymp_norm:diverge_p}.

\section{Proof of Theorem~\ref{thm:JINI_bias:incons_initial}}

Theorem~\ref{thm:JINI_bias:incons_initial} is a special case of Theorem~\ref{thm:JINI_bias:incons_initial:diverge_p} with fixed $p$. Please refer to Section~\ref{sec:pf:thm:JINI_bias:incons_initial:diverge_p} for the proof of Theorem~\ref{thm:JINI_bias:incons_initial:diverge_p}.

\section{Proof of Theorem~\ref{thm:JINI_bias:cons_initial}}

Theorem~\ref{thm:JINI_bias:cons_initial} is a special case of Theorem~\ref{thm:JINI_bias:cons_initial:diverge_p} with fixed $p$. Please refer to Sections~\ref{sec:pf:thm:JINI_bias:cons_initial:general:diverge_p}, \ref{sec:pf:thm:JINI_bias:cons_initial:comb:diverge_p}, \ref{sec:pf:thm:JINI_bias:cons_initial:smooth:diverge_p}, \ref{sec:pf:thm:JINI_bias:cons_initial:linear:diverge_p} respectively for the proof of the first to the fourth results of Theorem~\ref{thm:JINI_bias:cons_initial:diverge_p}.

\section{Assumptions in increasing dimensions}
\label{supp:assumptions:diverge_p}

In this section, we list the assumptions in increasing dimensional
settings, i.e., $p$ is allowed to diverge with $n$ and $p<n$. In
particular, Assumption~X presented in Section~\ref{sec:conditions} of the paper corresponds to Assumption~{S.X} when $p$ is fixed, with X to
be A, B, C, E, C.1, C.2, C.3, C.4. Assumption~D in
Section~\ref{sec:conditions} corresponds to the combination of
Assumptions~{S.D.1} and S.D.2 when $p$ is fixed. 

\begin{Assumption}
\label{assum:compact:diverge_p}
For any $p$, the parameter space $\bT$ is a compact convex subset of $\real^{p}$ and $\bt_0 \in \Int(\bT)$. 
\end{Assumption}

\begin{Assumption}
\label{assum:v:diverge_p}
There exists some $\alpha>0$ such that $\|\v(\bt_0,n)\|_2=
\mathcal{O}_p(p^{1/2}n^{-\alpha})$ and $p^{1/2}n^{-\alpha}=o(1)$.  
\end{Assumption}

\begin{Assumption}
\label{assum:finite_bias:diverge_p}
There exists some $\beta > \alpha$, where $\alpha$ is defined in Assumption~\ref{assum:v:diverge_p}, such that the finite sample bias function $\b(\bt,n)$ satisfies $\sup_{\bt\in\bT}\|\b(\bt,n)\|_2= \mathcal{O}(p^{1/2}n^{-\beta})$. 
\end{Assumption}

\setcounter{Assumption}{3}
\renewcommand\theAssumption{S.\Alph{Assumption}.1}
\begin{Assumption}
\label{assum:asymp_bias1:diverge_p}
The function $\bpi(\bt)$ is injective and continuous.
\end{Assumption}

\setcounter{Assumption}{3}
\renewcommand\theAssumption{S.\Alph{Assumption}.2}
\begin{Assumption}
\label{assum:asymp_bias2:diverge_p}
The function $\bpi(\bt)$ is twice continuously
differentiable. Moreover, it satisfies the following:  
\begin{enumerate}
    \item Let $\A(\bt_0)\equiv \partial \bpi(\bt_0)/\partial
      \bt$. There exist some $C>0$ and $c>0$ such that $c \leq
      \sigma_{\min}\{\A(\bt_0)\} \leq \sigma_{\max}\{\A(\bt_0)\}
      \leq C$, where $\sigma(\cdot)$ denotes the singular value. 

    \item There exists some $\epsilon>0$ and $C>0$ such that
      $\max_{i=1,\ldots,p}\sup_{\bt\in\boldsymbol{\mB}(\bt_0,\epsilon)}\|\partial^2
      \pi_i(\bt)/\partial\bt\partial\bt\trans\|_2 \leq
      C$. 
\end{enumerate} 
\end{Assumption}

\setcounter{Assumption}{4}
\renewcommand\theAssumption{S.\Alph{Assumption}}
\begin{Assumption}
\label{assum:initial_asymp_norm:diverge_p}
For any $\s\in\real^p$ such that $\|\s\|_2=1$, we have 
\bsq
    n^{\alpha}\s\trans\bSig(\bt_0,
    n)^{-1/2}\left\{\wh{\bpi}(\bt_0,n)-\bpi(\bt_0)\right\}
    \overset{D}{\to} Z, 
\esq
where $\alpha$ is defined in Assumption~\ref{assum:v:diverge_p}, and
$Z$ is a random variable whose distribution function does not depend
on $\s$ and $\bt$. 
Moreover, $\bSig(\bt_0, n)$ is a $p\times p$ positive definite covariance matrix, which is such that the limit of $\s\trans\bSig(\bt_0, n)\s$ exists as $n\to\infty$ for any $\s\in\real^p$ with $\|\s\|_2=1$, and $c\leq \lim_{n\to\infty}\s\trans\bSig(\bt_0, n)\s\leq C$ where $C,c$ are finite positive constants.
\end{Assumption}

\setcounter{Assumption}{2}
\renewcommand\theAssumption{S.\Alph{Assumption}.1}
\begin{Assumption}
\label{assum:finite_bias:general:diverge_p}
There exists some $\epsilon>0$ such that the finite sample bias function
  $\b(\bt,n)$ is differentiable in
  $\bt\in\boldsymbol{\mB}(\bt_0,\epsilon)$ and $\max_{i=1,\ldots,p}
  \sup_{\bt\in\boldsymbol{\mB}(\bt_0,\epsilon)} \|\partial b_i(\bt,n)
  / \partial \bt\|_2 = \mathcal{O}(p^{1/2}n^{-\beta})$, where $\beta$
  is defined in Assumption~\ref{assum:finite_bias:diverge_p}.
\end{Assumption}

\setcounter{Assumption}{2}
\renewcommand\theAssumption{S.\Alph{Assumption}.2}
\begin{Assumption}
\label{assum:finite_bias:comb:diverge_p}
The finite sample bias function $\b(\bt,n)$ can be expressed
as $\b(\bt,n)=\B(n)\bt+\c(n)+\r(\bt,n)$, with
$\B(n)\in\real^{p\times p}$, $\c(n)\in\real^p$ and
$\r(\bt,n)\in\real^p$ which contains all the nonlinear terms of
$\bt$. Moreover, the following conditions are satisfied:
\begin{enumerate}
    \item There exists some $\beta_1\geq\beta$, where $\beta$ is
      defined in Assumption~\ref{assum:finite_bias:diverge_p}, such
      that $\|\B(n)\|_2 = \mathcal{O}(n^{-\beta_1})$.
    \item There exists some $\epsilon>0$ such that $\r(\bt,n)$ is differentiable in
      $\bt\in\boldsymbol{\mB}(\bt_0,\epsilon)$. Moreover, there exists
      some $\beta_2\geq\beta$ such
      that $\max_{i=1,\ldots,p}\sup_{\bt\in\boldsymbol{\mB}(\bt_0,\epsilon)}\|\partial
      r_i(\bt,n)/\partial\bt\|_2 = \mathcal{O}(p^{1/2}n^{-\beta_2})$.
\end{enumerate}
\end{Assumption}

\setcounter{Assumption}{2}
\renewcommand\theAssumption{S.\Alph{Assumption}.3}
\begin{Assumption}
\label{assum:finite_bias:smooth:diverge_p}
The finite sample bias function $\b(\bt,n)$ can be expressed as
$\b(\bt,n)=\B(n)\bt+\c(n)+\r(\bt,n)$, where $\B(n)\in\real^{p\times
  p}$, $\c(n)\in\real^p$ and 
$\r(\bt,n)\in\real^p$ and the following conditions are satisfied: 
\begin{enumerate}
    \item There exists some $\beta_1\geq\beta$, where $\beta$ is defined in Assumption~\ref{assum:finite_bias:diverge_p}, such that $\|\B(n)\|_2 = \mathcal{O}(n^{-\beta_1})$.
    \item $\r(\bt,n) = n^{-\beta_2}\left(\bt\trans\R_i
        \bt\right)_{i=1,\ldots,p} + \e(\bt,n)$, where $\R_i \in
      \real^{p\times p}$ is symmetric, $\e(\bt,n)\in\real^p$ which does not contain linear and quadratic terms of $\bt$, and
      $\beta_2\geq\beta$. 
    \item Let $\U(\bt_0)$ be a $p\times p$ matrix with the $i\th$ row
      to be $(\R_i\bt_0)\trans$. 
There exists some $C>0$ such that $\|\U(\bt_0)\|_2 \leq C$. 
    \item For any $\s\in\real^p$ such that $\|\s\|_2=1$, there exists
      some $C>0$ such that $|\lambda|_{\max}(\sum_{i=1}^p s_i\R_i)
      \leq C$, where $|\lambda|_{\max}(\cdot)$ denotes
        the maximum absolute eigenvalue.
    \item There exist some $\beta_3\geq \beta$ and
      some $\epsilon>0$ such that
      $\sup_{\bt\in\boldsymbol{\mB}(\bt_0,\epsilon)}\|\e(\bt,n)\|_2 =
      \mathcal{O}(p^{1/2}n^{-\beta_3})$. 
\end{enumerate}
\end{Assumption}

\setcounter{Assumption}{2}
\renewcommand\theAssumption{S.\Alph{Assumption}.4}
\begin{Assumption}
\label{assum:finite_bias:linear:diverge_p}
The finite sample bias function $\b(\bt,n)$ can be expressed as $\b(\bt,n)=\B(n)\bt+\c(n)$, where $\B(n)\in\real^{p\times p}$ and $\c(n)\in\real^p$. Moreover, there exists some $\beta_1\geq \beta$, where $\beta$ is defined in Assumption~\ref{assum:finite_bias:diverge_p}, such that $\|\B(n)\|_2=\mathcal{O}(n^{-\beta_1})$.
\end{Assumption}

\section{Complete statement and proof of Theorem~\ref{thm:JINI_consistency:diverge_p}}
\label{sec:pf:thm:JINI_consistency:diverge_p}

Under Assumptions~\ref{assum:compact:diverge_p}, \ref{assum:v:diverge_p}, \ref{assum:finite_bias:diverge_p}, and \ref{assum:asymp_bias1:diverge_p}, JINI is consistent with $\|\wh{\bt}-\bt_0\|_2=o_p(1)$.

\begin{proof}
We recall that $\bpi(\bt, n)=\bpi(\bt)+\b(\bt, n)$. We define the functions $Q(\bt)$, $Q(\bt,n)$ and $\wh{Q}(\bt,n)$ as follows:
\bse
    && Q(\bt) \equiv \left\|\bpi(\bt_0) - \bpi(\bt)\right\|_2, \\
    && Q(\bt, n) \equiv \left\|\bpi(\bt_0, n) - \bpi(\bt, n)\right\|_2, \\
    && \wh{Q}(\bt, n) \equiv \left\|\wh{\bpi}(\bt_0, n) -
      \bpi(\bt, n)\right\|_2. 
\ese
Then this proof is directly obtained by verifying the conditions of
Theorem 2.1 of \cite{newey1994large} on the functions $Q(\bt)$ and
$\wh{Q}(\bt,n)$. Reformulating the requirements of this theorem to our
setting, we want to show that (i) $\bT$ is compact, (ii) $Q(\bt)$ is
continuous, (iii) $Q(\bt)$ is uniquely minimized at $\bt_0$, and (iv)
$\wh{Q}(\bt,n)$ converges uniformly in probability to $Q(\bt)$. 

Assumption~\ref{assum:compact:diverge_p} ensures the compactness of $\bT$ and
Assumption~\ref{assum:asymp_bias1:diverge_p} ensures that $Q(\bt)$ is continuous
since $\bpi(\bt)$ is continuous. Moreover, $\bpi(\bt)$ is an injective
function of $\bt$ by Assumption~\ref{assum:asymp_bias1:diverge_p}, so $Q(\bt)$
achieves the minimum of zero if and only if $\bpi(\bt)=\bpi(\bt_0)$,
i.e., if and only if $\bt=\bt_0$. So requirements (i), (ii) and (iii)
are satisfied and what remains to be shown is that $\wh{Q}(\bt,n)$
converges uniformly in probability to $Q(\bt)$.

Using the above definitions, we have
\be \label{eqn:pf_consistency_1}
    \underset{\bt \in \bT}{\sup} \left|\wh{Q}(\bt, n) - Q(\bt)\right|
    &\leq& \underset{\bt \in \bT}{\sup} \left\{\left|\wh{Q}(\bt, n) -
        Q(\bt, n)\right| + \left|Q(\bt, n) - Q(\bt)\right| \right\}
    \n\\ 
    &\leq& \underset{\bt \in \bT}{\sup}\left|\wh{Q}(\bt, n) - Q(\bt,
      n)\right| + \underset{\bt \in \bT}{\sup}\left|Q(\bt, n) -
      Q(\bt)\right|. 
\ee
Considering the first term on the right hand side of
\eqref{eqn:pf_consistency_1}, we have 
\be \label{eqn:pf_consistency_2}
    \underset{\bt \in \bT}{\sup}\left|\wh{Q}(\bt, n) - Q(\bt,
      n)\right| &\leq& \underset{\bt \in
      \bT}{\sup}\left\|\wh{\bpi}(\bt_0, n) - \bpi(\bt, n) - \bpi(\bt_0, n) + \bpi(\bt, n)\right\|_2
    \n\\ 
    &=& \left\|\wh{\bpi}(\bt_0, n) - \bpi(\bt_0, n)\right\|_2 \n\\
    &=& \left\|\v(\bt_0,n)\right\|_2
    \n\\ 
    &=& \mathcal{O}_p(p^{1/2}n^{-\alpha}) \n\\
    &=& o_p(1),
\ee 
where the inequality is by reverse triangle inequality and the
last two equalities are by Assumption~\ref{assum:v:diverge_p}. Similarly, the second
term on the right hand side of \eqref{eqn:pf_consistency_1} can be
computed as 
\be \label{eqn:pf_consistency_3}
    \underset{\bt \in \bT}{\sup}\left|Q(\bt, n) - Q(\bt)\right| &\leq&
    \underset{\bt \in \bT}{\sup}\left\| \bpi(\bt_0, n) - \bpi(\bt, n)
      - \bpi(\bt_0) + \bpi(\bt) \right\|_2 \n\\ 
    &\leq& \left\| \bpi(\bt_0, n) - \bpi(\bt_0) \right\|_2 +
    \underset{\bt \in \bT}{\sup}\left\| \bpi(\bt, n) -  \bpi(\bt)
    \right\|_2 \n\\ 
    &=& \left\| \mathbf{b}(\bt_0,n) \right\|_2 + \underset{\bt \in
      \bT}{\sup}\left\| \mathbf{b}(\bt,n) \right\|_2 \n\\ 
    &=& \mathcal{O}(p^{1/2}n^{-\beta})\n\\
    &=& o(1),
\ee
where the last two equalities are by
Assumption~\ref{assum:finite_bias:diverge_p}. Combining the results in
\eqref{eqn:pf_consistency_2} and \eqref{eqn:pf_consistency_3} into
\eqref{eqn:pf_consistency_1}, we obtain
\bsq
    \underset{\bt \in \bT}{\sup} \left|\wh{Q}(\bt, n) - Q(\bt)\right|
    = o_p(1)+o(1) = o_p(1), 
\esq
which completes the proof.
\end{proof}

\section{Complete statement and proof of Theorem~\ref{thm:JINI_bias:incons_initial:diverge_p}}
\label{sec:pf:thm:JINI_bias:incons_initial:diverge_p}

When the initial estimator is inconsistent, under Assumptions~\ref{assum:compact:diverge_p}, \ref{assum:v:diverge_p},
\ref{assum:finite_bias:diverge_p}, \ref{assum:asymp_bias1:diverge_p} and \ref{assum:asymp_bias2:diverge_p}, for any $\s\in\real^p$ such that $\|\s\|_2=1$, we have 
\bsq
    \s\trans \left\{\mathbb{E}(\wh{\bt})-\bt_0\right\} =
    \mathcal{O}\left(p^2n^{-2\alpha} + p^{1/2}n^{-\beta}\right).
\esq

\begin{proof}
By definition in \eqref{eqn:def_JINI}, we have
\bsq
    \wh{\bpi}(\bt_0,n) = \bpi(\bt_0) + \b(\bt_0,n) + \v(\bt_0,n) = \bpi(\wh{\bt},n) = \bpi(\wh{\bt}) + \b(\wh{\bt},n).
\esq
By rearranging the terms, we obtain
\bq \label{eqn:diff_pi}
    \bpi(\wh{\bt}) - \bpi(\bt_0) = \b(\bt_0,n) - \b(\wh{\bt},n) + \v(\bt_0,n).
\eq 
Let $\bDel\equiv \wh{\bt}-\bt_0$. Since $\bpi(\bt)$ is twice continuously differentiable by Assumption~\ref{assum:asymp_bias2:diverge_p}, by Taylor's theorem we have $\bpi(\wh{\bt}) - \bpi(\bt_0) =
\A(\bt_0)\bDel+\u$, where we denote $\u \equiv (u_1,\ldots, u_p)\trans$ with $u_i\equiv \sum_{j=1}^p\sum_{k=1}^p g_{ijk}(\wh{\bt},\bt_0)\Delta_j\Delta_k$ and
\bsq
    g_{ijk}(\wh{\bt},\bt_0) \equiv \int_0^1 (1-t) \frac{\partial^2 \pi_i\{\bt_0+t(\wh{\bt}-\bt_0)\}}{\partial\theta_j\partial\theta_k} dt.
\esq
Moreover, $\u$ is such that $\|\u\|_2/\|\bDel\|_2 \to 0$. So we can rewrite \eqref{eqn:diff_pi} as 
\bsq
    \A(\bt_0)\bDel+\u = \b(\bt_0,n) - \b(\wh{\bt},n) + \v(\bt_0,n).
\esq
Since $\A(\bt_0)$ is nonsingular by Assumption~\ref{assum:asymp_bias2:diverge_p}, we further have
\bq \label{eqn:express_delta}
    \bDel + \A(\bt_0)^{-1}\u = - \A(\bt_0)^{-1}\left\{\b(\wh{\bt},n) - \b(\bt_0,n)\right\} + \A(\bt_0)^{-1}\v(\bt_0,n).
\eq
So we obtain
\be \label{eqn:order_l2_norm_delta}
    \left\| \bDel + \A(\bt_0)^{-1}\u \right\|_2 &\leq& \left\|\A(\bt_0)^{-1}\left\{\b(\wh{\bt},n) - \b(\bt_0,n)\right\}\right\|_2 + \left\|\A(\bt_0)^{-1}\v(\bt_0,n)\right\|_2 \n\\
    &\leq& \sigma_{\min}^{-1} \left\{\A(\bt_0)\right\} \left\{\left\|\b(\wh{\bt},n) - \b(\bt_0,n)\right\|_2 + \left\|\v(\bt_0,n)\right\|_2\right\} \n\\
    &\leq& \sigma_{\min}^{-1} \left\{\A(\bt_0)\right\} \left\{2\sup_{\bt\in\bT}\|\b(\bt,n)\|_2 + \|\v(\bt_0,n)\|_2\right\} \n\\
    &=& \mathcal{O}(p^{1/2}n^{-\beta}) + \mathcal{O}_p(p^{1/2}n^{-\alpha}) \n\\
    &=& \mathcal{O}_p(p^{1/2}n^{-\alpha}),    
\ee
where the equalities use Assumptions~\ref{assum:v:diverge_p},
\ref{assum:finite_bias:diverge_p} and \ref{assum:asymp_bias2:diverge_p}. Moreover, by
Assumption~\ref{assum:asymp_bias2:diverge_p} we have
\bsq
    c\|\u\|_2 \leq \sigma_{\max}^{-1} \{\A(\bt_0)\} \|\u\|_2 \leq \left\|\A(\bt_0)^{-1}\u\right\|_2 \leq \sigma_{\min}^{-1} \{\A(\bt_0)\} \|\u\|_2 \leq C\|\u\|_2,
\esq
where $C,c$ are finite positive constants. Together with $\|\u\|_2/\|\bDel\|_2 \to 0$, we can write $\left\|\A(\bt_0)^{-1}\u\right\|_2 = c_n\|\bDel\|_2$ with $c_n\to 0$. So we can further express \eqref{eqn:order_l2_norm_delta} as
\bse
  \mathcal{O}_p(p^{1/2}n^{-\alpha}) = \left\| \bDel + \A(\bt_0)^{-1}\u \right\|_2 &\geq& \left|\|\bDel\|_2 - \left\|\A(\bt_0)^{-1}\u\right\|_2\right| \\
    &=& \left|\|\bDel\|_2 - c_n\|\bDel\|_2 \right| \\
    &=& |1-c_n| \|\bDel\|_2 \\
    &\geq& \|\bDel\|_2/2,
\ese
for sufficiently large $n$, where the first inequality is by reverse triangle inequality. Thus we obtain $\|\bDel\|_2= \mathcal{O}_p(p^{1/2}n^{-\alpha})$.

Next we study the order of $\s\trans\mathbb{E}(\bDel)$ for any $\s\in\real^p$ such that $\|\s\|_2=1$. By \eqref{eqn:express_delta} we have
\be \label{eqn:order_bias}
    \left|\s\trans\mathbb{E}(\bDel)\right| &=& \left|\s\trans\A(\bt_0)^{-1}\mathbb{E}(\u) + \s\trans\A(\bt_0)^{-1} \mathbb{E}\left\{\b(\wh{\bt},n) - \b(\bt_0,n)\right\}\right| \n\\
    &\leq& \left|\s\trans\A(\bt_0)^{-1}\mathbb{E}(\u)\right| + \left|\s\trans\A(\bt_0)^{-1} \mathbb{E}\left\{\b(\wh{\bt},n) - \b(\bt_0,n)\right\}\right|.
\ee
We aim to evaluate the two terms on the right hand side
of \eqref{eqn:order_bias}.

For the first term, let $\c\trans\equiv \s\trans\A(\bt_0)^{-1}/
\|\s\trans\A(\bt_0)^{-1}\|_2$ such that $\|\c\|_2=1$, then we can
write 
\bq \label{eqn:order_bias_part1}
    \left|\s\trans\A(\bt_0)^{-1}\mathbb{E}(\u)\right| =
    \left\|\s\trans\A(\bt_0)^{-1}\right\|_2
    \left|\c\trans\mathbb{E}(\u)\right| \leq \sigma_{\min}^{-1}
    \left\{\A(\bt_0)\right\} \mathbb{E}\left(|\c\trans\u|\right). 
\eq
We note that
\bsq
    \c\trans\u = \sum_{k=1}^p\sum_{j=1}^p \left\{\sum_{i=1}^p c_ig_{ijk}(\wh{\bt},\bt_0)\right\}\Delta_j\Delta_k = \sum_{k=1}^p (\g_k\trans\bDel)\Delta_k = \z\trans\bDel,
\esq
where $\g_k\equiv (g_{1k},\ldots,g_{pk})\trans$ with $g_{jk}\equiv \sum_{i=1}^p c_ig_{ijk}(\wh{\bt},\bt_0)$ for $j=1,\ldots,p$, and $\z\equiv (z_1,\ldots,z_p)\trans$ with $z_k\equiv \g_k\trans\bDel$ for $k=1,\ldots,p$. We also note that
\bse
    \sum_{k=1}^p \|\g_k\|_2^2 &=& \sum_{k=1}^p \sum_{j=1}^p \left\{\sum_{i=1}^p c_ig_{ijk}(\wh{\bt},\bt_0)\right\}^2 \\
    &\leq& p \sum_{k=1}^p \sum_{j=1}^p \sum_{i=1}^p c_i^2
    g_{ijk}(\wh{\bt},\bt_0)^2 \\
    &\leq& p\max_{i=1,\ldots,p}\sum_{k=1}^p \sum_{j=1}^p g_{ijk}(\wh{\bt},\bt_0)^2 \\
    &=& p\max_{i=1,\ldots,p}\sum_{k=1}^p \sum_{j=1}^p \left[\int_0^1 (1-t) \frac{\partial^2 \pi_i\{\bt_0+t(\wh{\bt}-\bt_0)\}}{\partial\theta_j\partial\theta_k} dt\right]^2 \\
    &\leq& p\max_{i=1,\ldots,p} \sum_{k=1}^p \sum_{j=1}^p \int_0^1 (1-t)^2 \left[\frac{\partial^2 \pi_i\{\bt_0+t(\wh{\bt}-\bt_0)\}}{\partial\theta_j\partial\theta_k}\right]^2 dt.
\ese
Since $\wh{\bt}$ is consistent with $\|\wh{\bt}-\bt_0\|_2=o_p(1)$ by
Section~\ref{sec:pf:thm:JINI_consistency:diverge_p}, we have $\{\bt_0+t(\wh{\bt}-\bt_0)\}$
with $t\in(0,1)$ to be in a small neighborhood
$\boldsymbol{\mB}(\bt_0,\epsilon)$ of $\bt_0$ with probability
approaching one. So we further have
\bse
    \sum_{k=1}^p \|\g_k\|_2^2 &\leq& \frac{p}{3} \max_{i=1,\ldots,p} \sup_{\bt\in\boldsymbol{\mB}(\bt_0,\epsilon)} \sum_{k=1}^p \sum_{j=1}^p \left\{\frac{\partial^2 \pi_i(\bt)}{\partial\theta_j\partial\theta_k}\right\}^2 \\
    &=& \frac{p}{3} \max_{i=1,\ldots,p} \sup_{\bt\in\boldsymbol{\mB}(\bt_0,\epsilon)} \left\|\frac{\partial^2 \pi_i(\bt)}{\partial\bt\partial\bt\trans}\right\|_F^2 \\
    &\leq& \frac{p^2}{3} \max_{i=1,\ldots,p} \sup_{\bt\in\boldsymbol{\mB}(\bt_0,\epsilon)} \left\|\frac{\partial^2 \pi_i(\bt)}{\partial\bt\partial\bt\trans}\right\|_2^2\\
    &=& \mathcal{O}(p^2),
\ese
where the last equality is by
Assumption~\ref{assum:asymp_bias2:diverge_p}. So under
Assumption~\ref{assum:compact:diverge_p} we have
\bq \label{eqn:linear_order_u}
    \mathbb{E}(|\c\trans\u|) = \mathbb{E}(|\z\trans\bDel|) \leq
    \mathbb{E}(\|\z\|_2\|\bDel\|_2) \leq
    \mathbb{E}\left\{\|\bDel\|_2^2 \left(\sum_{k=1}^p
        \|\g_k\|_2^2\right)^{1/2}\right\} =
    \mathcal{O}(p^2n^{-2\alpha}). 
\eq 
Together with
  \eqref{eqn:order_bias_part1}, we obtain 
\bq \label{eqn:order_bias_part1_final}
    \left|\s\trans\A(\bt_0)^{-1}\mathbb{E}(\u)\right| \leq
    \sigma_{\min}^{-1} \left\{\A(\bt_0)\right\}
    \mathbb{E}\left(|\c\trans\u|\right)= \mathcal{O}(p^2n^{-2\alpha}),
\eq
where the last equality uses Assumption~\ref{assum:asymp_bias2:diverge_p}.

For the second term on the right hand side of \eqref{eqn:order_bias}, we have
\be \label{eqn:order_bias_part2_final}
    && \left|\s\trans\A(\bt_0)^{-1} \mathbb{E}\left\{\b(\wh{\bt},n) - \b(\bt_0,n)\right\}\right| \n\\
    &\leq& \left|\s\trans\A(\bt_0)^{-1}\mathbb{E}\left\{\b(\wh{\bt},n)\right\}\right| + \left|\s\trans\A(\bt_0)^{-1}\b(\bt_0,n)\right| \n\\
    &\leq& \left\|\s\trans\A(\bt_0)^{-1}\right\|_2 \mathbb{E}\left\{\left\|\b(\wh{\bt},n)\right\|_2\right\} + \left\|\s\trans\A(\bt_0)^{-1}\right\|_2 \left\|\b(\bt_0,n)\right\|_2 \n\\
    &\leq& 2 \sigma_{\min}^{-1}\{\A(\bt_0)\} \sup_{\bt\in\bT}\left\|\b(\bt,n)\right\|_2 \n\\
    &=& \mathcal{O}(p^{1/2}n^{-\beta}),
\ee 
where the last equality is by Assumptions~\ref{assum:finite_bias:diverge_p} and \ref{assum:asymp_bias2:diverge_p}. Plugging the results of \eqref{eqn:order_bias_part1_final} and \eqref{eqn:order_bias_part2_final} into \eqref{eqn:order_bias}, we obtain
\bsq
    \left|\s\trans\mathbb{E}(\bDel)\right| =\mathcal{O}(p^2n^{-2\alpha}) + \mathcal{O}(p^{1/2}n^{-\beta}) = \mathcal{O}(p^2n^{-2\alpha} + p^{1/2}n^{-\beta}).
\esq
\end{proof}

\section{Complete statement and proof of Theorem~\ref{thm:JINI_asymp_norm:diverge_p}}
\label{sec:pf:thm:JINI_asymp_norm:diverge_p}

\subsection{With inconsistent initial estimators}
\label{sec:pf:thm:JINI_asymp_norm:diverge_p:inconsistent_initial}

Suppose $\max(p^{1/2}n^{\alpha-\beta}, p^2n^{-\alpha})
  \to 0$, where $\alpha$ and $\beta$ are given in
  Assumptions~\ref{assum:v:diverge_p} and
  \ref{assum:finite_bias:diverge_p} respectively.

When the initial estimator is inconsistent, under
Assumptions~\ref{assum:compact:diverge_p}, \ref{assum:v:diverge_p},
\ref{assum:finite_bias:diverge_p}, \ref{assum:asymp_bias1:diverge_p},
\ref{assum:asymp_bias2:diverge_p} and
\ref{assum:initial_asymp_norm:diverge_p}, for any $\s\in\real^p$ such
that $\|\s\|_2=1$ we have 
\bsq
    n^{\alpha} \s\trans \left\{\A(\bt_0)^{-1} \bSig(\bt_0,n)
      \A(\bt_0)^{-1}\right\}^{-1/2} \left(\wh{\bt}-\bt_0\right)
    \overset{D}{\to} Z, 
\esq
where $\A(\bt_0)$ is given in
Assumption~\ref{assum:asymp_bias2:diverge_p}, and $Z$ is the random
variable given in Assumption~\ref{assum:initial_asymp_norm:diverge_p}.  

\begin{proof}
When the initial estimator is inconsistent, i.e., $\bpi(\bt)\neq \bt$,
by definition in \eqref{eqn:def_JINI} we have
\bsq
    \wh{\bpi}(\bt_0,n) = \bpi(\bt_0)+\b(\bt_0,n)+\v(\bt_0,n) = \bpi(\wh{\bt},n)=\bpi(\wh{\bt})+\b(\wh{\bt},n).
\esq 
By rearranging the terms, we obtain
\bsq
    \bpi(\wh{\bt})-\bpi(\bt_0) = \b(\bt_0,n) - \b(\wh{\bt},n) + \v(\bt_0,n).
\esq 
Let $\bDel\equiv \wh{\bt}-\bt_0$. Since $\bpi(\bt)$ is twice continuously differentiable by Assumption~\ref{assum:asymp_bias2:diverge_p}, by Taylor's theorem we have $\bpi(\wh{\bt}) - \bpi(\bt_0) =
\A(\bt_0)\bDel+\u$, where we denote $\u \equiv (u_1,\ldots, u_p)\trans$ with $u_i\equiv \sum_{j=1}^p\sum_{k=1}^p g_{ijk}(\wh{\bt},\bt_0)\Delta_j\Delta_k$ and
\bsq
    g_{ijk}(\wh{\bt},\bt_0) \equiv \int_0^1 (1-t) \frac{\partial^2 \pi_i\{\bt_0+t(\wh{\bt}-\bt_0)\}}{\partial\theta_j\partial\theta_k} dt.
\esq 
So we can further write
\bsq
    \A(\bt_0)\bDel + \u = \b(\bt_0,n) - \b(\wh{\bt},n) + \v(\bt_0,n).
\esq
Moreover, since $\A(\bt_0)$ is nonsingular by Assumption~\ref{assum:asymp_bias2:diverge_p}, we further have 
\bsq
    \bDel = -\A(\bt_0)^{-1}\u + \A(\bt_0)^{-1}\left\{\b(\bt_0,n)+\v(\bt_0,n)\right\} - \A(\bt_0)^{-1} \b(\wh{\bt},n),
\esq
and hence, for any $\s\in\real^p$ such that $\|\s\|_2=1$ we have 
\be \label{eqn:JINI_asymp_norm_decomp}
    n^{\alpha} \s\trans\bDel &=& n^{\alpha} \s\trans\A(\bt_0)^{-1}\left\{\b(\bt_0,n) + \v(\bt_0,n)\right\} \n\\
    && - n^{\alpha} \s\trans\A(\bt_0)^{-1}\b(\wh{\bt},n) - n^{\alpha} \s\trans\A(\bt_0)^{-1}\u,
\ee 
where $\alpha$ is defined in Assumption~\ref{assum:v:diverge_p}. We
aim to evaluate the three terms on the right hand side of 
\eqref{eqn:JINI_asymp_norm_decomp}.

For the first term on the right hand side of
\eqref{eqn:JINI_asymp_norm_decomp}, by
Assumption~\ref{assum:initial_asymp_norm:diverge_p} we know that for
any $\s\in\real^p$ such that $\|\s\|_2=1$ we have 
\bsq
    n^{\alpha} \s\trans\bSig(\bt_0, n)^{-1/2}\left\{\b(\bt_0,n) + \v(\bt_0,n)\right\} = n^{\alpha} \s\trans\bSig(\bt_0, n)^{-1/2}\left\{\wh{\bpi}(\bt_0,n)-\bpi(\bt_0)\right\} \overset{D}{\to} Z,
\esq
where $Z$ is the random variable given in Assumption~\ref{assum:initial_asymp_norm:diverge_p}. So we have
\be \label{eqn:JINI_asymp_norm_decomp_part1}
    && n^{\alpha} \s\trans\A(\bt_0)^{-1}\left\{\b(\bt_0,n) + \v(\bt_0,n)\right\} \n\\
    &=& n^{\alpha}
    \s\trans\A(\bt_0)^{-1}\bSig(\bt_0,n)^{1/2}\bSig(\bt_0,
    n)^{-1/2}\left\{\b(\bt_0,n) + \v(\bt_0,n)\right\} \n\\  
    &=& \left\|\s\trans\A(\bt_0)^{-1}\bSig(\bt_0, n)^{1/2}\right\|_2
    n^{\alpha} \frac{\s\trans\A(\bt_0)^{-1}\bSig(\bt_0,
      n)^{1/2}}{\left\|\s\trans\A(\bt_0)^{-1}\bSig(\bt_0,
        n)^{1/2}\right\|_2} \n\\
    &&\bSig(\bt_0, n)^{-1/2} \left\{\b(\bt_0,n)
      + \v(\bt_0,n)\right\} \n\\ 
    &\overset{D}{\to}& \sigma(\bt_0)Z,
\ee
where we define $\sigma^2(\bt_0)\equiv \lim_{n\to\infty}
\left\|\s\trans\A(\bt_0)^{-1}\bSig(\bt_0, n)^{1/2}\right\|_2^2$ and
$\sigma(\bt_0)\equiv\sqrt{\sigma^2(\bt_0)}$. We can show that
$\sigma^2(\bt_0)$ exists and is strictly positive. Indeed, we can
write
\bse
    \sigma^2(\bt_0) &=& \lim_{n\to\infty} \left\|\s\trans\A(\bt_0)^{-1}\bSig(\bt_0, n)^{1/2}\right\|_2^2 \\
    &=& \lim_{n\to\infty} \s\trans\A(\bt_0)^{-1}\bSig(\bt_0, n)\A(\bt_0)^{-1}\s \\
    &=& \|\s\trans\A(\bt_0)^{-1}\|_2^2 \lim_{n\to\infty} \left\{\frac{\s\trans\A(\bt_0)^{-1}}{\|\s\trans\A(\bt_0)^{-1}\|_2} \bSig(\bt_0, n) \frac{\A(\bt_0)^{-1}\s}{\|\s\trans\A(\bt_0)^{-1}\|_2}\right\}. 
\ese
Under Assumption~\ref{assum:initial_asymp_norm:diverge_p}, we know $\|\s\trans\A(\bt_0)^{-1}\|_2^{-2}\lim_{n\to\infty} \s\trans\A(\bt_0)^{-1}\bSig(\bt_0, n)\A(\bt_0)^{-1}\s$ exists and 
\bsq
    c_1 \leq \lim_{n\to\infty} \left\{\frac{\s\trans\A(\bt_0)^{-1}}{\|\s\trans\A(\bt_0)^{-1}\|_2} \bSig(\bt_0, n) \frac{\A(\bt_0)^{-1}\s}{\|\s\trans\A(\bt_0)^{-1}\|_2}\right\} \leq C_1,
\esq
where $C_1,c_1$ are finite positive constants. Moreover, under Assumption~\ref{assum:asymp_bias2:diverge_p} we have 
\bq \label{eqn:order_eigen_A_inv}
    c_2 \leq \sigma_{\max}^{-1}\{\A(\bt_0)\} \leq \|\s\trans\A(\bt_0)^{-1}\|_2 \leq \sigma_{\min}^{-1}\{\A(\bt_0)\}\leq C_2,
\eq
where $C_2,c_2$ are finite positive constants. So we show that $\sigma^2(\bt_0)$ exists and is strictly positive.

For the second term on the right hand side of
\eqref{eqn:JINI_asymp_norm_decomp}, since $\|\bDel\|_2=o_p(1)$ by
Section~\ref{sec:pf:thm:JINI_consistency:diverge_p}, $\wh{\bt}$ is in
a small neighborhood $\boldsymbol{\mB}(\bt_0,\epsilon)$ of $\bt_0$
with probability approaching one. So we have 
\bq \label{eqn:JINI_asymp_norm_decomp_part2}
    \left|n^{\alpha}\s\trans\A(\bt_0)^{-1}\b(\wh{\bt},n)\right| \leq n^{\alpha} \|\s\trans\A(\bt_0)^{-1}\|_2 \|\b(\wh{\bt},n)\|_2 = \mathcal{O}_p(p^{1/2}n^{\alpha-\beta}) = o_p(1),
\eq
where the first equality uses \eqref{eqn:order_eigen_A_inv} and
Assumption~\ref{assum:finite_bias:diverge_p}, and the second equality
is because $p^{1/2}n^{\alpha-\beta} \to 0$. 

Lastly, we evaluate the third term on the right hand side of
\eqref{eqn:JINI_asymp_norm_decomp}, i.e., $n^{\alpha}
\s\trans\A(\bt_0)^{-1}\u$. Using \eqref{eqn:linear_order_u} in
Section~\ref{sec:pf:thm:JINI_bias:incons_initial:diverge_p} and
Markov's inequality, we have $|\s\trans\A(\bt_0)^{-1}\u|
\|\s\trans\A(\bt_0)^{-1}\|_2^{-1} =
\mathcal{O}_p(p^2n^{-2\alpha})$. So we have 
\bq \label{eqn:JINI_asymp_norm_decomp_part3}
    \left|n^{\alpha} \s\trans\A(\bt_0)^{-1}\u\right| = n^{\alpha} \left\|\s\trans\A(\bt_0)^{-1}\right\|_2 \left|\frac{\s\trans\A(\bt_0)^{-1}}{\left\|\s\trans\A(\bt_0)^{-1}\right\|_2}\u\right| = \mathcal{O}_p\left(p^2n^{-\alpha}\right) = o_p(1),
\eq
where the second equality uses \eqref{eqn:order_eigen_A_inv} and the last equality uses $p^2n^{-\alpha} \to 0$. 

Therefore, putting the results of \eqref{eqn:JINI_asymp_norm_decomp_part1}, \eqref{eqn:JINI_asymp_norm_decomp_part2} and \eqref{eqn:JINI_asymp_norm_decomp_part3} into \eqref{eqn:JINI_asymp_norm_decomp}, for any $\s\in\real^p$ such that $\|\s\|_2=1$ we have
\bsq
    n^\alpha \s\trans\bDel \overset{D}{\to} \sigma(\bt_0)Z.
\esq
Since $\sigma^2(\bt_0) = \lim_{n\to\infty}\s\trans\A(\bt_0)^{-1}\bSig(\bt_0, n)\A(\bt_0)^{-1}\s$, we can equivalently write 
\bsq
    n^\alpha \s\trans \left\{\A(\bt_0)^{-1}\bSig(\bt_0, n)\A(\bt_0)^{-1}\right\}^{-1/2} \left(\wh{\bt}-\bt_0\right) \overset{D}{\to} Z.
\esq

\end{proof}

\subsection{With consistent initial estimators}
\label{sec:pf:thm:JINI_asymp_norm:diverge_p:consistent_initial}

Suppose $p^{1/2}n^{\alpha-\beta} \to 0$, where $\alpha$ and $\beta$ are given in Assumptions~\ref{assum:v:diverge_p} and \ref{assum:finite_bias:diverge_p} respectively. When the initial estimator is consistent, under
Assumptions~\ref{assum:compact:diverge_p}, \ref{assum:v:diverge_p},
\ref{assum:finite_bias:diverge_p}, and
\ref{assum:initial_asymp_norm:diverge_p}, for any $\s\in\real^p$ such that $\|\s\|_2=1$ we have 
\bsq
    n^{\alpha} \s\trans \bSig(\bt_0,n)^{-1/2} \left(\wh{\bt}-\bt_0\right)
    \overset{D}{\to} Z, 
\esq
where $Z$ is the random
variable given in Assumption~\ref{assum:initial_asymp_norm:diverge_p}.  

\begin{proof}
When the initial estimator is consistent, i.e., $\bpi(\bt)=\bt$, by definition in \eqref{eqn:def_JINI} we have
\bsq
    \wh{\bpi}(\bt_0,n) = \bt_0+\b(\bt_0,n)+\v(\bt_0,n) = \bpi(\wh{\bt},n)=\wh{\bt}+\b(\wh{\bt},n).
\esq 
By rearranging the terms, we obtain
\bsq
    \wh{\bt}-\bt_0 = \b(\bt_0,n) - \b(\wh{\bt},n) + \v(\bt_0,n).
\esq
So for any $\s\in\real^p$ such that $\|\s\|_2=1$ we have
\bq \label{eqn:JINI_asymp_norm:decomp:consistent_initial}
    n^{\alpha}\s\trans\left(\wh{\bt}-\bt_0\right) = n^{\alpha}\s\trans\left\{\b(\bt_0,n) + \v(\bt_0,n)\right\} - n^{\alpha}\s\trans\b(\wh{\bt},n),
\eq
where $\alpha$ is defined in Assumption~\ref{assum:v:diverge_p}. In
the rest of the proof, we evaluate the two terms on the right hand
side of \eqref{eqn:JINI_asymp_norm:decomp:consistent_initial}.

For the first term on the right hand side of \eqref{eqn:JINI_asymp_norm:decomp:consistent_initial}, by Assumption~\ref{assum:initial_asymp_norm:diverge_p} we know that for any $\s\in\real^p$ such that $\|\s\|_2=1$ we have
\bsq
    n^\alpha \s\trans \bSig(\bt_0,n)^{-1/2} \left\{\b(\bt_0,n) + \v(\bt_0,n)\right\} = n^\alpha \s\trans \bSig(\bt_0,n)^{-1/2} \left\{\wh{\bpi}(\bt_0,n)-\bt_0\right\} \overset{D}{\to} Z,
\esq
where $Z$ is the random variable given in Assumption~\ref{assum:initial_asymp_norm:diverge_p}. So we have  
\be \label{eqn:JINI_asymp_norm:decomp:consistent_initial:part1}
    && n^\alpha \s\trans \left\{\b(\bt_0,n) + \v(\bt_0,n)\right\} \n\\
    &=& n^\alpha \s\trans \bSig(\bt_0,n)^{1/2} \bSig(\bt_0,n)^{-1/2} \left\{\b(\bt_0,n) + \v(\bt_0,n)\right\} \n\\
    &=& \|\s\trans\bSig(\bt_0,n)^{1/2}\|_2 n^\alpha \frac{\s\trans \bSig(\bt_0,n)^{1/2}}{\|\s\trans \bSig(\bt_0,n)^{1/2}\|_2} \bSig(\bt_0,n)^{-1/2} \left\{\b(\bt_0,n) + \v(\bt_0,n)\right\} \n\\
    &\overset{D}{\to}& \sigma(\bt_0)Z,
\ee
where we define $\sigma^2(\bt_0) \equiv
\lim_{n\to\infty}\s\trans\bSig(\bt_0,n)\s$ and $\sigma(\bt_0)\equiv
\sqrt{\sigma^2(\bt_0)}$. By
Assumption~\ref{assum:initial_asymp_norm:diverge_p}, $\sigma^2(\bt_0)$
exists and is strictly positive.  

For the second term on the right hand side of \eqref{eqn:JINI_asymp_norm:decomp:consistent_initial}, we have
\bq \label{eqn:JINI_asymp_norm:decomp:consistent_initial:part2}
    \left|n^\alpha \s\trans\b(\wh{\bt},n)\right| \leq n^{\alpha} \|\b(\wh{\bt},n)\|_2 = \mathcal{O}(p^{1/2}n^{\alpha-\beta}) = o_p(1),
\eq
where the first equality uses
Assumption~\ref{assum:finite_bias:diverge_p} and the second equality uses $p^{1/2}n^{\alpha-\beta}\to 0$. 

Putting the results of \eqref{eqn:JINI_asymp_norm:decomp:consistent_initial:part1} and \eqref{eqn:JINI_asymp_norm:decomp:consistent_initial:part2} into \eqref{eqn:JINI_asymp_norm:decomp:consistent_initial}, for any $\s\in\real^p$ such that $\|\s\|_2=1$ we have
\bsq
    n^{\alpha} \s\trans\left(\wh{\bt}-\bt_0\right) \overset{D}{\to} \sigma(\bt_0)Z.
\esq
Since $\sigma^2(\bt_0)=\lim_{n\to\infty}\s\trans\bSig(\bt_0,n)\s$, we can equivalently write
\bsq
    n^\alpha \s\trans \bSig(\bt_0,n)^{-1/2} \left(\wh{\bt}-\bt_0\right) \overset{D}{\to} Z. 
\esq
\end{proof}

\section{Complete statement and proof of Theorem~\ref{thm:JINI_bias:cons_initial:diverge_p} (First result)}
\label{sec:pf:thm:JINI_bias:cons_initial:general:diverge_p}

When the initial estimator is consistent, under
Assumptions~\ref{assum:compact:diverge_p}, \ref{assum:v:diverge_p},
\ref{assum:finite_bias:diverge_p} and
\ref{assum:finite_bias:general:diverge_p}, for any $\s\in\real^p$ such
that $\|\s\|_2=1$ we have
\bsq
    \s\trans \left\{\mathbb{E}(\wh{\bt})-\bt_0\right\} =
    \mathcal{O}\left\{p^{3/2}n^{-(\alpha+\beta)}\right\}. 
\esq

\begin{proof}
When the initial estimator is consistent, we have $\bpi(\bt)=\bt$. Then by definition in \eqref{eqn:def_JINI}, we have 
\bsq
    \wh{\bpi}(\bt_0, n) = \bt_0 + \b(\bt_0, n) + \v(\bt_0,n) = \bpi(\wh{\bt},n) = \wh{\bt} + \b(\wh{\bt},n). 
\esq
By rearranging the terms, we have
\bsq
    \wh{\bt}-\bt_0 = \b(\bt_0,n) - \b(\wh{\bt},n) + \v(\bt_0,n),
\esq
and hence, under Assumptions~\ref{assum:v:diverge_p} and \ref{assum:finite_bias:diverge_p} we obtain 
\be \label{eqn:pf_finbias_1}
    \left\|\wh{\bt}-\bt_0\right\|_2 &\leq&
    2\sup_{\bt\in\bT}\left\|\b(\bt,n)\right\|_2 +
    \left\|\v(\bt_0,n)\right\|_2 \n\\
    &=& \mathcal{O}(p^{1/2}n^{-\beta}) +
    \mathcal{O}_p(p^{1/2}n^{-\alpha}) \n\\
    &=&
    \mathcal{O}_p(p^{1/2}n^{-\alpha}). 
\ee
Moreover, since $\wh{\bt}$ is consistent with
$\|\wh{\bt}-\bt_0\|_2=o_p(1)$ by
Section~\ref{sec:pf:thm:JINI_consistency:diverge_p}, for any
$\s\in\real^p$ such that $\|\s\|_2=1$, with probability approaching
one we have
\bsq 
    \s\trans \left\{\mathbb{E}(\wh{\bt}) - \bt_0\right\} = -
    \s\trans \mathbb{E} \left\{\b(\wh{\bt},n) - \b(\bt_0, n)\right\}
    = -\s\trans \mathbb{E}\left\{\B(\bt_1^*,\ldots, \bt_p^*, n)
      (\wh{\bt} - \bt_0)\right\},
\esq
where the last equality is by mean value theorem, $\B(\bt_1^*,\ldots, \bt_p^*, n)$ is a $p\times p$ matrix with
the $i\th$ row to be $\partial b_i(\bt_i^*,n)/\partial\bt\trans$,
$\bt_i^* \in \bT$ lies between $\wh{\bt}$ and $\bt_0$ for
$i=1,\ldots,p$, and hence $\bt_i^*$ lies in a small
neighborhood $\boldsymbol{\mB}(\bt_0,\epsilon)$ of $\bt_0$ for
$i=1,\ldots,p$. So by Assumption~\ref{assum:finite_bias:general:diverge_p} we
have
\be \label{eqn:pf_finbias_2}
    \left\|\s\trans \B(\bt_1^*,\ldots,\bt_p^*,n)\right\|_2 &\leq& \left\|\B(\bt_1^*,\ldots,\bt_p^*,n)\right\|_F \n\\
    &\leq& p^{1/2} \max_{i=1,\ldots,p} \left\|\frac{\partial b_i(\bt_i^*,n)}{\partial \bt\trans}\right\|_2 \n\\
    &\leq&  p^{1/2} \max_{i=1,\ldots,p} \sup_{\bt\in\boldsymbol{\mB}(\bt_0,\epsilon)}\left\|\frac{\partial b_i(\bt,n)}{\partial \bt\trans}\right\|_2 \n\\
    &=& \mathcal{O}(pn^{-\beta}), 
\ee
and hence we obtain
\bse
    \left|\s\trans \left\{\mathbb{E}(\wh{\bt}) - \bt_0\right\}\right| &\leq& \mathbb{E}\left\{\left|\s\trans\B(\bt_1^*,\ldots,\bt_p^*,n)(\wh{\bt}-\bt_0)\right|\right\} \n\\
    &\leq& \mathbb{E}\left\{\left\|\s\trans\B(\bt_1^*,\ldots,\bt_p^*,n)\right\|_2 \left\|\wh{\bt}-\bt_0\right\|_2\right\} \n\\
    &=& \mathcal{O}\left\{p^{3/2}n^{-(\alpha+\beta)}\right\},
\ese
where the last equality uses \eqref{eqn:pf_finbias_1}, \eqref{eqn:pf_finbias_2} and Assumption~\ref{assum:compact:diverge_p}.
\end{proof}

\section{Complete statement and proof of Theorem~\ref{thm:JINI_bias:cons_initial:diverge_p} (Second result)}
\label{sec:pf:thm:JINI_bias:cons_initial:comb:diverge_p}

When the initial estimator is consistent, under Assumptions~\ref{assum:compact:diverge_p}, \ref{assum:v:diverge_p}, \ref{assum:finite_bias:diverge_p} and \ref{assum:finite_bias:comb:diverge_p}, for any $\s\in\real^p$ such that $\|\s\|_2=1$ we have 
\bsq
    \s\trans \left\{\mathbb{E}(\wh{\bt})-\bt_0\right\} = \mathcal{O}\left\{p^{3/2}n^{-(\alpha+\beta_2)}\right\}.
\esq

\begin{proof}
When the initial estimator is consistent, we have $\bpi(\bt)=\bt$. Then by definition in \eqref{eqn:def_JINI} we have
\bsq
    \wh{\bpi}(\bt_0, n) = \bt_0 + \b(\bt_0, n) + \v(\bt_0,n) = \bpi(\wh{\bt},n) = \wh{\bt} + \b(\wh{\bt},n). 
\esq
By rearranging the terms, we have
\bsq
    \wh{\bt}-\bt_0 = \b(\bt_0,n) - \b(\wh{\bt},n) + \v(\bt_0,n),
\esq
and hence, under Assumptions~\ref{assum:v:diverge_p} and \ref{assum:finite_bias:diverge_p} we obtain 
\be \label{eqn:pf_combbias_1}
    \left\|\wh{\bt}-\bt_0\right\|_2 &\leq&
    2\sup_{\bt\in\bT}\left\|\b(\bt,n)\right\|_2 +
    \left\|\v(\bt_0,n)\right\|_2 \n\\
    &=& \mathcal{O}(p^{1/2}n^{-\beta}) +
    \mathcal{O}_p(p^{1/2}n^{-\alpha}) \n\\
    &=&
    \mathcal{O}_p(p^{1/2}n^{-\alpha}). 
\ee
Moreover, since $\b(\bt,n)=\B(n)\bt+\c(n)+\r(\bt,n)$ by
Assumption~\ref{assum:finite_bias:comb:diverge_p}, we also obtain  
\bsq 
    \wh{\bt}-\bt_0 = -\B(n)(\wh{\bt}-\bt_0) + \r(\bt_0,n)-\r(\wh{\bt},n)+\v(\bt_0,n),
\esq
and thus,
\bq \label{eqn:pf_combbias_2}
    \{\I+\B(n)\}(\wh{\bt}-\bt_0) =  \r(\bt_0,n)-\r(\wh{\bt},n)+\v(\bt_0,n).
\eq 
Under Assumption~\ref{assum:finite_bias:comb:diverge_p}, we have $\|\B(n)\|_2 = \mathcal{O}(n^{-\beta_1}) = o(1)$, so $\{\I+\B(n)\}$ is nonsingular for sufficiently large $n$, $\{\I+\B(n)\}^{-1}=\sum_{k=0}^\infty \{-\B(n)\}^k$, and 
\bq \label{eqn:pf_combbias_2.2}
    \left\|\{\I+\B(n)\}^{-1}\right\|_2 = \left\| \I+\sum_{k=1}^\infty \{-\B(n)\}^k\right\|_2 \leq \|\I\|_2 + \sum_{k=1}^\infty \|\B(n)\|_2^k = \mathcal{O}(1).
\eq
Since $\wh{\bt}$ is consistent with $\|\wh{\bt}-\bt_0\|_2=o_p(1)$ by Section~\ref{sec:pf:thm:JINI_consistency:diverge_p}, we can rewrite \eqref{eqn:pf_combbias_2}, with probability approaching one, as
\be \label{eqn:pf_combbias_2.1}
    \wh{\bt}-\bt_0 &=& \{\I+\B(n)\}^{-1}\left\{\r(\bt_0,n)-\r(\wh{\bt},n)\right\} + \{\I+\B(n)\}^{-1}\v(\bt_0,n) \n\\
    &=& -\{\I+\B(n)\}^{-1} \R(\bt_1^*,\ldots,\bt_p^*,n)(\wh{\bt}-\bt_0) + \{\I+\B(n)\}^{-1}\v(\bt_0,n),
\ee
where the second equality is by mean value theorem,
$\R(\bt_1^*,\ldots,\bt_p^*,n)$ is a $p\times p$ matrix with the 
$i\th$ row to be $\partial r_i(\bt_i^*,n)/\partial \bt\trans$,
$\bt_i^*\in\bT$ lies between $\wh{\bt}$ and $\bt_0$ for
$i=1,\ldots,p$, and hence $\bt_i^*$ lies in a small
neighborhood $\boldsymbol{\mB}(\bt_0,\epsilon)$ of $\bt_0$ for
$i=1,\ldots,p$. Taking expectations on both sides of
\eqref{eqn:pf_combbias_2.1}, we have
\bsq
    \mathbb{E}(\wh{\bt}) - \bt_0 = - \mathbb{E}\left[\{\I+\B(n)\}^{-1} \R(\bt_1^*,\ldots,\bt_p^*,n)(\wh{\bt}-\bt_0)\right].
\esq
So for any $\s\in\real^p$ such that $\|\s\|_2=1$, we have
\bsq
    \s\trans\left\{\mathbb{E}(\wh{\bt}) - \bt_0\right\} = -\mathbb{E}\left[\s\trans\{\I+\B(n)\}^{-1} \R(\bt_1^*,\ldots,\bt_p^*,n)(\wh{\bt}-\bt_0)\right],
\esq
and hence we have
\be \label{eqn:pf_combbias_2.3}
    \left|\s\trans\left\{\mathbb{E}(\wh{\bt}) - \bt_0\right\}\right| &\leq& \mathbb{E}\left[ \left| \s\trans\{\I+\B(n)\}^{-1} \R(\bt_1^*,\ldots,\bt_p^*,n)(\wh{\bt}-\bt_0)\right| \right] \n\\
    &=& \left\|\s\trans\{\I+\B(n)\}^{-1}\right\|_2 \mathbb{E}\left\{\left|\a\trans \R(\bt_1^*,\ldots,\bt_p^*,n)(\wh{\bt}-\bt_0)\right|\right\} \n\\
    &\leq& \left\|\{\I+\B(n)\}^{-1}\right\|_2 \mathbb{E}\left\{\left|\a\trans \R(\bt_1^*,\ldots,\bt_p^*,n)(\wh{\bt}-\bt_0)\right|\right\} \n\\
    &=& \mathcal{O}(1)\;  \mathbb{E}\left\{\left|\a\trans \R(\bt_1^*,\ldots,\bt_p^*,n)(\wh{\bt}-\bt_0)\right|\right\},
\ee
where we denote $\a\trans \equiv \s\trans\{\I+\B(n)\}^{-1}/ \|\s\trans\{\I+\B(n)\}^{-1}\|_2$ and the last equality uses \eqref{eqn:pf_combbias_2.2}. Moreover, we note that 
\bse
    \left\|\a\trans\R(\bt_1^*,\ldots,\bt_p^*,n) \right\|_2^2 &=& \sum_{j=1}^p \left\{\sum_{i=1}^p a_i R_{ij}(\bt_1^*,\ldots,\bt_p^*,n) \right\}^2 \\
    &\leq& p\sum_{i=1}^p \sum_{j=1}^p a_i^2 R_{ij}(\bt_1^*,\ldots,\bt_p^*,n)^2 \\
    &\leq& p \max_{i=1,\ldots,p} \left\|\frac{\partial r_i(\bt_i^*,n)}{\partial \bt}\right\|_2^2 \\
    &\leq&  p \max_{i=1,\ldots,p} \sup_{\bt\in\boldsymbol{\mB}(\bt_0,\epsilon)}\left\|\frac{\partial r_i(\bt,n)}{\partial \bt}\right\|_2^2 \\
    &=& \mathcal{O}(p^2n^{-2\beta_2}),
\ese
where the last equality is by
Assumption~\ref{assum:finite_bias:comb:diverge_p}. Together with \eqref{eqn:pf_combbias_1} and under Assumption~\ref{assum:compact:diverge_p}, we have
\bse
    \mathbb{E}\left\{\left|\a\trans \R(\bt_1^*,\ldots,\bt_p^*,n)(\wh{\bt}-\bt_0)\right|\right\} &\leq& \mathbb{E}\left\{\left\|\a\trans \R(\bt_1^*,\ldots,\bt_p^*,n)\right\|_2 \left\|\wh{\bt}-\bt_0\right\|_2\right\} \\
    &=& \mathcal{O}\left\{p^{3/2}n^{-(\alpha+\beta_2)}\right\}.
\ese
Plugging this result into \eqref{eqn:pf_combbias_2.3}, we obtain
\bsq
    \left|\s\trans\left\{\mathbb{E}(\wh{\bt})-\bt_0\right\}\right| = \mathcal{O}\left\{p^{3/2}n^{-(\alpha+\beta_2)}\right\},
\esq
which completes the proof.
\end{proof}

\section{Complete statement and proof of Theorem~\ref{thm:JINI_bias:cons_initial:diverge_p} (Third result)}
\label{sec:pf:thm:JINI_bias:cons_initial:smooth:diverge_p}

When the initial estimator is consistent, under
Assumptions~\ref{assum:compact:diverge_p}, \ref{assum:v:diverge_p},
\ref{assum:finite_bias:diverge_p} and
\ref{assum:finite_bias:smooth:diverge_p}, for any $\s\in\real^p$ such
that $\|\s\|_2=1$ we have  
\bsq
    \s\trans \left\{\mathbb{E}(\wh{\bt})-\bt_0\right\} =
    \mathcal{O}\left\{pn^{-(2\alpha+\beta_2)} +
      p^{1/2}n^{-\beta_3}\right\}. 
\esq

\begin{proof}
When the initial estimator is consistent, we have $\bpi(\bt)=\bt$. Then by definition in \eqref{eqn:def_JINI} we have
\bsq
    \wh{\bpi}(\bt_0, n) = \bt_0 + \b(\bt_0, n) + \v(\bt_0,n) = \bpi(\wh{\bt},n) = \wh{\bt} + \b(\wh{\bt},n). 
\esq
By rearranging the terms, we have
\bsq
    \wh{\bt}-\bt_0 = \b(\bt_0,n) - \b(\wh{\bt},n) + \v(\bt_0,n),
\esq
and hence, under Assumptions~\ref{assum:v:diverge_p} and \ref{assum:finite_bias:diverge_p} we obtain 
\be \label{eqn:pf_cor_combbias_1}
    \left\|\wh{\bt}-\bt_0\right\|_2 &\leq&
    2\sup_{\bt\in\bT}\left\|\b(\bt,n)\right\|_2 +
    \left\|\v(\bt_0,n)\right\|_2 \n\\
    &=& \mathcal{O}(p^{1/2}n^{-\beta}) +
    \mathcal{O}_p(p^{1/2}n^{-\alpha}) \n\\
    &=&
    \mathcal{O}_p(p^{1/2}n^{-\alpha}). 
\ee 
By Assumption~\ref{assum:finite_bias:smooth:diverge_p}, we can write 
\bsq
    \b(\bt,n)=\B(n)\bt+\c(n)+\r(\bt,n) = \B(n)\bt+\c(n)+\u(\bt,n)+\e(\bt,n),
\esq
where we define $\u(\bt,n)\equiv
\{u_i(\bt,n)\}_{i=1,\ldots,p}$ with $u_i(\bt,n)\equiv
n^{-\beta_2}\bt\trans\R_i \bt$. 
Moreover, by Taylor's theorem, we have
\bsq
    \u(\wh{\bt},n)-\u(\bt_0,n) = 2n^{-\beta_2} \U(\bt_0)(\wh{\bt}-\bt_0) + \w(\wh{\bt},n),
\esq
where $\U(\bt_0)$ is a $p\times p$ matrix with the $i\th$ row to be $(\R_i\bt_0)\trans$ and $\w(\wh{\bt},n) \equiv \{w_i(\wh{\bt},n)\}_{i=1,\ldots,p}$ with 
\bsq
    w_i(\wh{\bt},n) \equiv n^{-\beta_2} (\wh{\bt}-\bt_0)\trans \R_i (\wh{\bt}-\bt_0).
\esq 
Then we can obtain
\bse
    \wh{\bt}-\bt_0 &=& \b(\bt_0,n) - \b(\wh{\bt},n) + \v(\bt_0,n) \\
    &=& -\B(n) (\wh{\bt}-\bt_0) + \u(\bt_0,n) - \u(\wh{\bt},n) + \e(\bt_0,n) - \e(\wh{\bt},n) + \v(\bt_0,n) \\
    &=& -\B(n) (\wh{\bt}-\bt_0) -2n^{-\beta_2} \U(\bt_0) (\wh{\bt}-\bt_0) - \w(\wh{\bt},n) + \e(\bt_0,n) - \e(\wh{\bt},n) + \v(\bt_0,n),
\ese 
and thus,
\bq \label{eqn:pf_cor_combbias_2}
    \left\{\I+\B(n)+2n^{-\beta_2} \U(\bt_0)\right\}(\wh{\bt}-\bt_0) = - \w(\wh{\bt},n) + \e(\bt_0,n) - \e(\wh{\bt},n) + \v(\bt_0,n).
\eq
For simplicity, we denote $\D\equiv\B(n)+2n^{-\beta_2} \U(\bt_0)$ and we have
\bsq
    \|\D\|_2 \leq \|\B(n)\|_2 + 2n^{-\beta_2} \|\U(\bt_0)\|_2 = \mathcal{O}\left\{n^{-\min(\beta_1,\beta_2)}\right\} = o(1), 
\esq
where the equalities uses Assumption~\ref{assum:finite_bias:smooth:diverge_p}. So $(\I+\D)$ is
  nonsingular for large enough $n$, $(\I+\D)^{-1} =
  \sum_{k=0}^\infty (-\D)^k$ and  
\bq \label{eqn:pf_cor_combbias_3}
    \|(\I+\D)^{-1}\|_2 = \left\|\I + \sum_{k=1}^\infty (-\D)^k \right\|_2 \leq \|\I\|_2 + \sum_{k=1}^\infty \|\D\|_2^k = \mathcal{O}(1).
\eq 
So we can rewrite \eqref{eqn:pf_cor_combbias_2} as \bsq
    \wh{\bt}-\bt_0 = -(\I+\D)^{-1}\w(\wh{\bt},n) + (\I+\D)^{-1} \left\{\e(\bt_0,n)-\e(\wh{\bt},n)\right\} + (\I+\D)^{-1}\v(\bt_0,n). 
\esq 
Taking expectations on both sides, we have 
\bsq
    \mathbb{E}(\wh{\bt})-\bt_0 = -(\I+\D)^{-1}\mathbb{E}\left\{\w(\wh{\bt},n)\right\} +(\I+\D)^{-1} \mathbb{E}\left\{\e(\bt_0,n)-\e(\wh{\bt},n)\right\}.
\esq
So for any $\s\in\real^p$ such that $\|\s\|_2=1$, we have
\bsq 
    \s\trans \left\{\mathbb{E}(\wh{\bt})-\bt_0\right\} = -\s\trans(\I+\D)^{-1}\mathbb{E}\left\{\w(\wh{\bt},n)\right\} +\s\trans(\I+\D)^{-1} \mathbb{E}\left\{\e(\bt_0,n)-\e(\wh{\bt},n)\right\},
\esq 
and hence we have
\be \label{eqn:pf_cor_combbias_4}
    \left|\s\trans \left\{\mathbb{E}(\wh{\bt})-\bt_0\right\}\right| &\leq& \left|\s\trans(\I+\D)^{-1}\mathbb{E}\left\{\w(\wh{\bt},n)\right\}\right| \n\\
    && + \left|\s\trans(\I+\D)^{-1} \mathbb{E}\left\{\e(\bt_0,n)-\e(\wh{\bt},n)\right\}\right|. 
\ee
Below we aim to evaluate the two terms on the right hand side of \eqref{eqn:pf_cor_combbias_4}. For the first term, we first note that for any $\s\in\real^p$ such that $\|\s\|_2=1$, we have 
\bse
    \mathbb{E}\left\{\left|\s\trans\w(\wh{\bt},n)\right|\right\} &=& n^{-\beta_2}\mathbb{E}\left\{\left|(\wh{\bt}-\bt_0)\trans \left(\sum_{i=1}^p s_i\R_i\right) (\wh{\bt}-\bt_0)\right|\right\} \\
    &\leq& n^{-\beta_2} \mathbb{E} \left\{\|\wh{\bt}-\bt_0\|_2^2 \; |\lambda|_{\max} \left(\sum_{i=1}^p s_i\R_i\right)\right\} \\
    &=& \mathcal{O}\left\{pn^{-(2\alpha+\beta_2)}\right\},
\ese
where the last equality uses \eqref{eqn:pf_cor_combbias_1}, Assumptions~\ref{assum:compact:diverge_p} and \ref{assum:finite_bias:smooth:diverge_p}. So we can obtain 
\be \label{eqn:pf_cor_combbias_5}
    \left|\s\trans(\I+\D)^{-1}\mathbb{E}\left\{\w(\wh{\bt},n)\right\}\right| &=& \left|\left\|\s\trans(\I+\D)^{-1}\right\|_2 \frac{\s\trans(\I+\D)^{-1}}{\left\|\s\trans(\I+\D)^{-1}\right\|_2} \mathbb{E}\left\{\w(\wh{\bt},n)\right\} \right| \n\\
    &\leq& \left\|\s\trans(\I+\D)^{-1}\right\|_2 \mathbb{E}\left\{\left|\frac{\s\trans(\I+\D)^{-1}}{\left\|\s\trans(\I+\D)^{-1}\right\|_2} \w(\wh{\bt},n)\right|\right\} \n\\
    &\leq& \|(\I+\D)^{-1}\|_2 \; \mathcal{O}\left\{pn^{-(2\alpha+\beta_2)}\right\}\n\\ 
    &=& \mathcal{O}\left\{pn^{-(2\alpha+\beta_2)}\right\},
\ee
where the last equality uses \eqref{eqn:pf_cor_combbias_3}. Next we
evaluate the second term on the right hand side of
\eqref{eqn:pf_cor_combbias_4}. We first note that $\wh{\bt}$ is
consistent with $\|\wh{\bt}-\bt_0\|_2=o_p(1)$ by
Section~\ref{sec:pf:thm:JINI_consistency:diverge_p}, so $\wh{\bt}$ lies in a small
neighborhood $\boldsymbol{\mB}(\bt_0,\epsilon)$ of $\bt_0$ with
probability approaching one. So we have
\bse
    \left|\s\trans(\I+\D)^{-1} \left\{\e(\bt_0,n)-\e(\wh{\bt},n)\right\}\right| &\leq& \left\|\s\trans(\I+\D)^{-1}\right\|_2 \left\|\e(\bt_0,n)-\e(\wh{\bt},n)\right\|_2 \n\\
    &\leq& 2\left\|(\I+\D)^{-1}\right\|_2 \sup_{\bt\in\boldsymbol{\mB}(\bt_0,\epsilon)}\|\e(\bt,n)\|_2 \n\\
    &=& \mathcal{O}(p^{1/2}n^{-\beta_3}),
\ese
where the last equality uses \eqref{eqn:pf_cor_combbias_3} and
Assumption~\ref{assum:finite_bias:smooth:diverge_p}, leading to
\be \label{eqn:pf_cor_combbias_6}
    \left|\s\trans(\I+\D)^{-1}
      \mathbb{E}\left\{\e(\bt_0,n)-\e(\wh{\bt},n)\right\}\right| &\leq& \mathbb{E}\left[\left|\s\trans(\I+\D)^{-1} \left\{\e(\bt_0,n)-\e(\wh{\bt},n)\right\}\right|\right] \n\\
      &=& \mathcal{O}(p^{1/2}n^{-\beta_3}).
\ee
Combining the results of \eqref{eqn:pf_cor_combbias_5} and
\eqref{eqn:pf_cor_combbias_6} into \eqref{eqn:pf_cor_combbias_4}, we
obtain 
\bsq
    \left|\s\trans \left\{\mathbb{E}(\wh{\bt})-\bt_0\right\}\right| =
    \mathcal{O}\left\{pn^{-(2\alpha+\beta_2)}\right\} +
    \mathcal{O}\left(p^{1/2}n^{-\beta_3}\right) =
    \mathcal{O}\left\{pn^{-(2\alpha+\beta_2)} +
      p^{1/2}n^{-\beta_3}\right\}, 
\esq
which completes the proof.
\end{proof}

\section{Complete statement and proof of Theorem~\ref{thm:JINI_bias:cons_initial:diverge_p} (Fourth result)}
\label{sec:pf:thm:JINI_bias:cons_initial:linear:diverge_p}

When the initial estimator is consistent, under Assumptions~\ref{assum:compact:diverge_p}, \ref{assum:v:diverge_p}, \ref{assum:finite_bias:diverge_p} and \ref{assum:finite_bias:linear:diverge_p}, we have $\mathbb{E}(\wh{\bt})-\bt_0 = \0$ when $n$ is sufficiently large.

\begin{proof}
When the initial estimator is consistent, we have $\bpi(\bt)=\bt$. Then by definition in \eqref{eqn:def_JINI}, we have 
\bsq
    \wh{\bpi}(\bt_0, n) = \bt_0 + \b(\bt_0, n) + \v(\bt_0,n) = \bpi(\wh{\bt},n) = \wh{\bt} + \b(\wh{\bt},n). 
\esq 
By rearranging the terms, we have
\bsq
    \wh{\bt}-\bt_0 = \b(\bt_0,n) - \b(\wh{\bt},n) + \v(\bt_0,n) = -\B(n) (\wh{\bt}-\bt_0) + \v(\bt_0,n),
\esq
where the last equality is because $\b(\bt,n) = \B(n)\bt+\c(n)$ by Assumption~\ref{assum:finite_bias:linear:diverge_p}, and hence we have
\bsq
    \left\{\I+\B(n)\right\}(\wh{\bt}-\bt_0) = \v(\bt_0,n).
\esq 
Under Assumption~\ref{assum:finite_bias:linear:diverge_p}, we have
$\|\B(n)\|_2 = \mathcal{O}(n^{-\beta_1}) = o(1)$, so $\{\I+\B(n)\}$ is
nonsingular for sufficiently large $n$ and we can further write  
\bsq
    \wh{\bt}-\bt_0 = \left\{\I+\B(n)\right\}^{-1}\v(\bt_0,n).
\esq
Taking expectations on both sides, we directly have
$\mathbb{E}(\wh{\bt}) - \bt_0 = \0$ when $n$ is sufficiently large.
\end{proof}

\section{Bias correction results of BBC}
\label{sec:pf:BBC}

Using our notations, BBC is defined as
\be \label{def:BBC}
    \wt{\bt} &\equiv& \wh{\bpi}(\bt_0,n) + \left[\wh{\bpi}(\bt_0,n) - \frac{1}{H}\sum_{h=1}^H \wh{\bpi}_h\left\{\wh{\bpi}(\bt_0,n),n\right\}\right] \n\\
    &=& 2\wh{\bpi}(\bt_0,n) - \frac{1}{H}\sum_{h=1}^H \wh{\bpi}_h\left\{\wh{\bpi}(\bt_0,n),n\right\},
\ee 
where $H\in\mathbb{N}$ is the number of simulated samples that is sufficiently large, and $\wh{\bpi}_h\left\{\wh{\bpi}(\bt_0,n),n\right\}$ denotes the value of the initial estimator computed on the $h\th$ independent simulated sample of size $n$ under $F_{\wh{\bpi}(\bt_0,n)}$. 

\subsection{First result}
When the initial estimator is consistent, under Assumptions~\ref{assum:compact:diverge_p}, \ref{assum:v:diverge_p}, \ref{assum:finite_bias:diverge_p} and \ref{assum:finite_bias:general:diverge_p}, for any $\s\in\real^p$ such that $\|\s\|_2=1$ we have 
\bsq
    \s\trans \left\{\mathbb{E}(\wt{\bt})-\bt_0\right\} =
    \mathcal{O}\left\{p^{3/2}n^{-(\alpha+\beta)}\right\}. 
\esq

\begin{proof}
By the definition in \eqref{def:BBC}, BBC satisfies
\bse
    \wt{\bt} &=& 2\wh{\bpi}(\bt_0,n) - \frac{1}{H}\sum_{h=1}^H\wh{\bpi}_h\left\{\wh{\bpi}(\bt_0,n),n\right\} \\
    &=& \wh{\bpi}(\bt_0,n)+\bt_0 + \b(\bt_0,n)+\v(\bt_0,n) \\
    && - \left[\wh{\bpi}(\bt_0,n) + \b\left\{\wh{\bpi}(\bt_0,n),n\right\} + \frac{1}{H}\sum_{h=1}^H \v_h\left\{\wh{\bpi}(\bt_0,n),n\right\}\right] \\
    &=& \bt_0+\b(\bt_0,n)-\b\left\{\wh{\bpi}(\bt_0,n),n\right\}+\v(\bt_0,n)-\frac{1}{H}\sum_{h=1}^H \v_h\left\{\wh{\bpi}(\bt_0,n),n\right\},
\ese 
where $\v_h\left\{\wh{\bpi}(\bt_0,n),n\right\}$ corresponds to the zero mean noise of the $h\th$ simulated sample under $F_{\wh{\bpi}(\bt_0,n)}$. By rearranging terms and taking expectations on both sides, we obtain
\bq \label{eqn:pf_BBC_1}
    \mathbb{E}(\wt{\bt})-\bt_0 = \b(\bt_0,n)-\mathbb{E}\left[\b\left\{\wh{\bpi}(\bt_0,n),n\right\}\right].
\eq    

Since $\wh{\bpi}(\bt_0,n)$ is consistent, for any $\s\in\real^p$ such
that $\|\s\|_2=1$, with probability approaching one we have
\be \label{eqn:pf_BBC_2}
    \left|\s\trans \left\{\mathbb{E}(\wt{\bt})-\bt_0\right\}\right| &=& \left|\s\trans \left(\b(\bt_0,n)-\mathbb{E}\left[\b\left\{\wh{\bpi}(\bt_0,n),n\right\}\right]\right)\right| \n\\
    &=& \left|\mathbb{E}\left[\s\trans \B(\bt_1^*,\ldots,\bt_p^*,n) \left\{\wh{\bpi}(\bt_0,n)-\bt_0\right\}\right] \right| \n\\
    &\leq& \mathbb{E}\left\{\left\|\s\trans \B(\bt_1^*,\ldots,\bt_p^*,n)\right\|_2 \left\|\wh{\bpi}(\bt_0,n)-\bt_0\right\|_2\right\},
\ee
where the second equality is by the mean value theorem,
$\B(\bt_1^*,\ldots,\bt_p^*,n)$ is a $p\times p$ matrix with the $i\th$
row to be $\partial b_i(\bt_i^*,n)/\partial\bt\trans$, $\bt_i^*\in\bT$
lies between $\wh{\bpi}(\bt_0,n)$ and $\bt_0$ for $i=1,\ldots,p$, and
hence $\bt_i^*$ lies in a small 
neighborhood $\boldsymbol{\mB}(\bt_0,\epsilon)$ of $\bt_0$ for
$i=1,\ldots,p$. Note that 
\be \label{eqn:pf_BBC_3}
    \left\|\s\trans \B(\bt_1^*,\ldots,\bt_p^*,n)\right\|_2 &\leq& \left\|\B(\bt_1^*,\ldots,\bt_p^*,n)\right\|_F \n\\
    &\leq& p^{1/2} \max_{i=1,\ldots,p} \left\|\frac{\partial b_i(\bt_i^*,n)}{\partial \bt\trans}\right\|_2 \n\\
    &\leq& p^{1/2} \max_{i=1,\ldots,p}\sup_{\bt\in\boldsymbol{\mB}(\bt_0,\epsilon)} \left\|\frac{\partial b_i(\bt,n)}{\partial \bt\trans}\right\|_2 \n\\
    &=& \mathcal{O}(pn^{-\beta}),
\ee 
where the last equality is by Assumption~\ref{assum:finite_bias:general:diverge_p}. Moreover, we recall that $\wh{\bpi}(\bt_0,n)=\bt_0+\b(\bt_0,n)+\v(\bt_0,n)$, so we have 
\be \label{eqn:pf_BBC_4}
    \left\|\wh{\bpi}(\bt_0,n) - \bt_0\right\|_2 &\leq& \|\b(\bt_0,n)\|_2
    + \|\v(\bt_0,n)\|_2 \n\\
    &=& \mathcal{O}(p^{1/2}n^{-\beta}) +
    \mathcal{O}_p(p^{1/2}n^{-\alpha}) \n\\
    &=&
    \mathcal{O}_p(p^{1/2}n^{-\alpha}), 
\ee
where the equalities use Assumptions~\ref{assum:v:diverge_p} and \ref{assum:finite_bias:diverge_p}. Therefore, plugging in the results of \eqref{eqn:pf_BBC_3} and \eqref{eqn:pf_BBC_4} into \eqref{eqn:pf_BBC_2}, under Assumption~\ref{assum:compact:diverge_p} we obtain
that for any $\s\in\real^p$ such that $\|\s\|_2=1$, we have
\bsq
    \s\trans \left\{\mathbb{E}(\wt{\bt})-\bt_0\right\} = \mathcal{O}\left\{p^{3/2}n^{-(\alpha+\beta)}\right\}.
\esq
\end{proof}

\subsection{Second result}
When the initial estimator is consistent, under Assumptions~\ref{assum:compact:diverge_p}, \ref{assum:v:diverge_p}, \ref{assum:finite_bias:diverge_p} and \ref{assum:finite_bias:comb:diverge_p}, for any $\s\in\real^p$ such that $\|\s\|_2=1$ we have 
\bsq
    \s\trans \left\{\mathbb{E}(\wt{\bt})-\bt_0\right\} = \mathcal{O}\left\{p^{1/2}n^{-(\beta+\beta_1)} + p^{3/2}n^{-(\alpha+\beta_2)}\right\}.
\esq

\begin{proof}

By the definition in \eqref{def:BBC}, BBC satisfies
\bse
    \wt{\bt} &=& 2\wh{\bpi}(\bt_0,n) - \frac{1}{H}\sum_{h=1}^H\wh{\bpi}_h\left\{\wh{\bpi}(\bt_0,n),n\right\} \\
    &=& \wh{\bpi}(\bt_0,n)+\bt_0 + \b(\bt_0,n)+\v(\bt_0,n) \\
    && - \left[\wh{\bpi}(\bt_0,n) + \b\left\{\wh{\bpi}(\bt_0,n),n\right\} + \frac{1}{H}\sum_{h=1}^H \v_h\left\{\wh{\bpi}(\bt_0,n),n\right\}\right] \\
    &=& \bt_0+\b(\bt_0,n)-\b\left\{\wh{\bpi}(\bt_0,n),n\right\}+\v(\bt_0,n)-\frac{1}{H}\sum_{h=1}^H \v_h\left\{\wh{\bpi}(\bt_0,n),n\right\}.
\ese 
By rearranging terms and taking expectations on both sides, we obtain
\bsq
    \mathbb{E}(\wt{\bt})-\bt_0 = \b(\bt_0,n)-\mathbb{E}\left[\b\left\{\wh{\bpi}(\bt_0,n),n\right\}\right].
\esq 
Since $\b(\bt,n)=\B(n)\bt+\c(n)+\r(\bt,n)$ by Assumption~\ref{assum:finite_bias:comb:diverge_p}, we can further write
\bse
    \mathbb{E}(\wt{\bt})-\bt_0 &=& \B(n)\bt_0+\c(n)+\r(\bt_0,n) - \B(n)\mathbb{E}\{\wh{\bpi}(\bt_0,n)\} - \c(n) - \mathbb{E}[\r\{\wh{\bpi}(\bt_0,n),n\}] \\
    &=& -\B(n)\b(\bt_0,n) + \r(\bt_0,n) - \mathbb{E}[\r\{\wh{\bpi}(\bt_0,n),n\}].
\ese 
So for any $\s\in\real^p$ such that $\|\s\|_2=1$ we have
\be \label{eqn:pf_BBC_comb_1}
    \left|\s\trans\left\{\mathbb{E}(\wt{\bt})-\bt_0\right\}\right| &\leq& \left|\s\trans\B(n)\b(\bt_0,n)\right| + \mathbb{E}\left(\left|\s\trans\left[\r(\bt_0,n) - \r\{\wh{\bpi}(\bt_0,n),n\}\right]\right|\right) \n\\
    &\leq& \|\B(n)\|_2 \|\b(\bt_0,n)\|_2 + \mathbb{E}\left(\left|\s\trans\left[\r(\bt_0,n) - \r\{\wh{\bpi}(\bt_0,n),n\}\right]\right|\right) \n\\
    &=& \mathcal{O}\left\{p^{1/2}n^{-(\beta+\beta_1)}\right\} + \mathbb{E}\left(\left|\s\trans\left[\r(\bt_0,n) - \r\{\wh{\bpi}(\bt_0,n),n\}\right]\right|\right), 
\ee
where the last equality uses
Assumptions~\ref{assum:finite_bias:diverge_p} and
\ref{assum:finite_bias:comb:diverge_p}. Moreover, since
$\wh{\bpi}(\bt_0,n)$ is consistent, with probability approaching one
we have 
\bse
    \mathbb{E}\left(\left|\s\trans\left[\r(\bt_0,n) - \r\{\wh{\bpi}(\bt_0,n),n\}\right]\right|\right) &=& \mathbb{E}\left[\left|\s\trans\R(\bt_1^*,\ldots,\bt_p^*,n) \{\wh{\bpi}(\bt_0,n)-\bt_0\}\right|\right] \\
    &\leq& \mathbb{E} \left\{\left\|\s\trans\R(\bt_1^*,\ldots,\bt_p^*,n)\right\|_2 \left\|\wh{\bpi}(\bt_0,n)-\bt_0\right\|_2\right\},
\ese
where the first equality is by mean value theorem, $\R(\bt_1^*,\ldots,\bt_p^*,n)$ is a $p\times p$ matrix with the $i\th$ row to be $\partial r_i(\bt_i^*,n)/\partial \bt\trans$, $\bt_i^* \in\bT$ lies between $\wh{\bpi}(\bt_0,n)$ and $\bt_0$ for $i=1,\ldots,p$, and hence $\bt_i^*$ lies in a small
neighborhood $\boldsymbol{\mB}(\bt_0,\epsilon)$ of $\bt_0$ for
$i=1,\ldots,p$. We also note that by Assumption~\ref{assum:finite_bias:comb:diverge_p} we have
\bse
    \left\|\s\trans\R(\bt_1^*,\ldots,\bt_p^*,n) \right\|_2^2 &=& \sum_{j=1}^p \left\{\sum_{i=1}^p s_i R_{ij}(\bt_1^*,\ldots,\bt_p^*,n) \right\}^2 \\
    &\leq& p\sum_{i=1}^p \sum_{j=1}^p s_i^2 R_{ij}(\bt_1^*,\ldots,\bt_p^*,n)^2 \\
    &\leq& p \max_{i=1,\ldots,p} \left\|\frac{\partial r_i(\bt_i^*,n)}{\partial \bt}\right\|_2^2 \\
    &\leq& p \max_{i=1,\ldots,p} \sup_{\bt\in\boldsymbol{\mB}(\bt_0,\epsilon)}\left\|\frac{\partial r_i(\bt,n)}{\partial \bt}\right\|_2^2 \\
    &=& \mathcal{O}(p^2n^{-2\beta_2}),
\ese
and by Assumptions~\ref{assum:v:diverge_p} and \ref{assum:finite_bias:diverge_p} we have 
\bsq
    \left\|\wh{\bpi}(\bt_0,n)-\bt_0\right\|_2 \leq \left\|\b(\bt_0,n)\right\|_2 + \left\|\v(\bt_0,n)\right\|_2 = \mathcal{O}_p(p^{1/2}n^{-\alpha}).
\esq 
So under Assumption~\ref{assum:compact:diverge_p} we have 
\be \label{eqn:pf_BBC_comb_2}
    \mathbb{E}\left(\left|\s\trans\left[\r(\bt_0,n) - \r\{\wh{\bpi}(\bt_0,n),n\}\right]\right|\right) &\leq& \mathbb{E} \left\{\left\|\s\trans\R(\bt_1^*,\ldots,\bt_p^*,n)\right\|_2 \left\|\wh{\bpi}(\bt_0,n)-\bt_0\right\|_2\right\} \n\\
    &=& \mathcal{O}\left\{p^{3/2}n^{-(\alpha+\beta_2)}\right\}.
\ee
Plugging the result of \eqref{eqn:pf_BBC_comb_2} into \eqref{eqn:pf_BBC_comb_1}, we obtain 
\bsq
    \s\trans \left\{\mathbb{E}(\wt{\bt})-\bt_0\right\} = \mathcal{O}\left\{p^{1/2}n^{-(\beta+\beta_1)} + p^{3/2}n^{-(\alpha+\beta_2)}\right\}.
\esq
\end{proof}

\subsection{Third result}

When the initial estimator is consistent, under Assumptions~\ref{assum:compact:diverge_p}, \ref{assum:v:diverge_p}, \ref{assum:finite_bias:diverge_p} and \ref{assum:finite_bias:linear:diverge_p}, for any $\s\in\real^p$ such that $\|\s\|_2=1$ we have 
\bsq
    \s\trans \left\{\mathbb{E}(\wt{\bt})-\bt_0\right\} =
    \mathcal{O}\left\{p^{1/2}n^{-(\beta+\beta_1)}\right\}. 
\esq

\begin{proof}

By the definition in \eqref{def:BBC}, BBC satisfies
\bse
    \wt{\bt} &=& 2\wh{\bpi}(\bt_0,n) - \frac{1}{H}\sum_{h=1}^H\wh{\bpi}_h\left\{\wh{\bpi}(\bt_0,n),n\right\} \\
    &=& \wh{\bpi}(\bt_0,n)+\bt_0 + \b(\bt_0,n)+\v(\bt_0,n) \\
    && - \left[\wh{\bpi}(\bt_0,n) + \b\left\{\wh{\bpi}(\bt_0,n),n\right\} + \frac{1}{H}\sum_{h=1}^H \v_h\left\{\wh{\bpi}(\bt_0,n),n\right\}\right] \\
    &=& \bt_0+\b(\bt_0,n)-\b\left\{\wh{\bpi}(\bt_0,n),n\right\}+\v(\bt_0,n)-\frac{1}{H}\sum_{h=1}^H \v_h\left\{\wh{\bpi}(\bt_0,n),n\right\}.
\ese 
By rearranging terms and taking expectations on both sides, we obtain
\bsq 
    \mathbb{E}(\wt{\bt})-\bt_0 = \b(\bt_0,n)-\mathbb{E}\left[\b\left\{\wh{\bpi}(\bt_0,n),n\right\}\right].
\esq  
Since $\b(\bt,n)=\B(n)\bt+\c(n)$ by Assumption~\ref{assum:finite_bias:linear:diverge_p}, we can further write
\bsq
    \mathbb{E}(\wt{\bt})-\bt_0 = \B(n)\bt_0 + \c(n) - \B(n)\mathbb{E}\left\{\wh{\bpi}(\bt_0,n)\right\} - \c(n) = -\B(n)\b(\bt_0,n).
\esq
So for any $\s\in\real^p$ such that $\|\s\|_2=1$ we have 
\bsq
    \left|\s\trans\left\{\mathbb{E}(\wt{\bt})-\bt_0\right\}\right|
    \leq \|\B(n)\|_2 \|\b(\bt_0,n)\|_2 =
    \mathcal{O}\left\{p^{1/2}n^{-(\beta+\beta_1)}\right\}, 
\esq
where the last equality is by Assumptions~\ref{assum:finite_bias:diverge_p} and \ref{assum:finite_bias:linear:diverge_p}.    
\end{proof}

\section{Asymptotic normality of NMLE for a logistic regression with misclassification}
\label{sec:asymo_norm_naive_MLE_logistic_misclas}

Consider a binary random variable $Y$ with $\pr(Y=1) =
\exp(\x\trans\bb_0)/\{1+\exp(\x\trans\bb_0)\}$ and $\pr(Y=0) =
1/\{1+\exp(\x\trans\bb_0)\}$, where $\x$ is a $p$-dimensional fixed
covariate vector, $\bb_0\in\bT$ is the vector of the true regression coefficients, and $\bT$ is a compact convex subset of $\real^p$. Given the fixed covariate $\x_i$ with $i=1,\ldots,n$, instead of observing $Y_i$, we observe a random sample $Z_i$ which is a misclassified version of $Y_i$. For simplicity, we assume that the false positive rate is zero and that the false negative rate is $\alpha \in [0,1)$, i.e.,
\bsq
    \pr(Z=1|Y=0) = 0 \quad \text{and} \quad \pr(Z=0|Y=1) = \alpha. 
\esq
Therefore, the true model for the random variable $Z$ is given by 
\bse
    \pr(Z=0) &=& \pr(Z=0|Y=0)\pr(Y=0) + \pr(Z=0|Y=1)\pr(Y=1) \\
    &=& \frac{1+\alpha\exp(\x\trans\bb_0)}{1+\exp(\x\trans\bb_0)},
\ese
and 
\bse
    \mu(\bb_0) \equiv \pr(Z=1) &=& \pr(Z=1|Y=0)\pr(Y=0) + \pr(Z=1|Y=1)\pr(Y=1) \\
    &=& \frac{(1-\alpha)\exp(\x\trans\bb_0)}{1+\exp(\x\trans\bb_0)}.
\ese

We consider a postulated model for $Z$, which is the classical logistic regression without misclassification. In other words, we assume that the model for $Z$ is 
\bsq
    \mu^*(\bb)\equiv \pr^*(Z=1) = \frac{\exp(\x\trans\bb)}{1+\exp(\x\trans\bb)} \quad \text{and} \quad \pr^*(Z=0) = \frac{1}{1+\exp(\x\trans\bb)}. 
\esq
In this case, the score function corresponding to this postulated model is $\s(\bb) = \sum_{i=1}^n \left\{z_i - \mu_i^*(\bb)\right\}\x_i$. So the corresponding MLE is 
\bsq
    \wh{\bpi} \equiv \argzero_{\bb\in\bT} \s(\bb) = \argzero_{\bb\in\bT} \sum_{i=1}^n \left\{z_i - \mu_i^*(\bb)\right\}\x_i. 
\esq
Since the postulated model does not take into account the misclassification, we call the resulting $\wh{\bpi}$ Naive MLE (NMLE) and it is inconsistent to the true parameter $\bb_0$. We also define
\bsq
    \bpi_0 \equiv \argzero_{\bb\in\bT} \sum_{i=1}^n \left\{\mathbb{E}(Z_i) - \mu_i^*(\bb)\right\}\x_i = \argzero_{\bb\in\bT} \sum_{i=1}^n \left\{\mu_i(\bb_0)- \mu_i^*(\bb)\right\}\x_i.       
\esq 

To show the asymptotic normality of $\sqrt{n}(\wh{\bpi}-\bpi_0)$, we consider the following assumptions:
\begin{itemize}
    \item[(L1)] The parameter dimension $p$ satisfies $p^2\log(n)n^{-1} \to 0$.
    \item[(L2)] $\sum_{i=1}^n\|\x_i\|_2^2 = \mathcal{O}(np)$ and $\sup_{\s\in\real^p:\|\s\|_2=1} \sum_{i=1}^n (\s\trans\x_i)^4 = \mathcal{O}(n)$.
    \item[(L3)] $\liminf_{n\to\infty} \lambda_{\min}(\D_n)\geq C$, where $\D_n\equiv n^{-1}\sum_{i=1}^n \exp(\x_i\trans\bpi_0)/\{1+\exp(\x_i\trans\bpi_0)\}^2 \x_i\x_i\trans$, $C$ is a finite positive constant, and $\lambda_{\min}(\cdot)$ refers to the minimum eigenvalue. 
    \item[(L4)] For any $\s\in\real^p$ such that $\|\s\|_2=1$, we have $\liminf_{n\to\infty} n^{-1}\sum_{i=1}^n (\s\trans\x_i)^2\geq C$, where $C$ is a finite positive constant.  
\end{itemize}

To start, we verify conditions (C0)-(C5) of \cite{he2000parameters} so
that we can use their Theorem~2.2. We define $\bpsi(Z_i,\x_i,\bb) \equiv \{Z_i-\mu_i^*(\bb)\}\x_i$ and
\bse
    \boldeta_i(\bb_1,\bb_2) &\equiv& \bpsi(Z_i,\x_i,\bb_1) - \bpsi(Z_i,\x_i,\bb_2) - \mathbb{E}_{\bb_1}\{\bpsi(Z_i,\x_i,\bb_1)\} + \mathbb{E}_{\bb_2}\{\bpsi(Z_i,\x_i,\bb_2)\} \\
    &=& \{\mu_i(\bb_2)-\mu_i(\bb_1)\}\x_i.
\ese

Condition (C0) is trivially satisfied by the definition of $\wh{\bpi}$. 

To verify condition (C1), we first note that by assumption (L2) we have
\bse
    \max_{i=1,\ldots,n}\sup_{\bb_1:\|\bb_1-\bb_2\|_2\leq d} \{\mu_i^*(\bb_2)-\mu_i^*(\bb_1)\}^2 \leq c \max_{i=1,\ldots,n}\sup_{\bb_1:\|\bb_1-\bb_2\|_2\leq d}\|\x_i\|_2^2 \|\bb_1-\bb_2\|_2^2 \leq cnpd^2,
\ese
with large enough $n$, where $c$ is a finite positive constant and $0<d\leq 1$. So we have
\bse
    && \max_{i=1,\ldots,n} \mathbb{E}\left\{\sup_{\bb_1:\|\bb_1-\bb_2\|_2\leq d} \|\boldeta_i(\bb_1,\bb_2)\|_2^2\right\} \\
    &\leq& (1-\alpha)^2 \max_{i=1,\ldots,n}\sup_{\bb_1:\|\bb_1-\bb_2\|_2\leq d} \{\mu_i^*(\bb_2)-\mu_i^*(\bb_1)\}^2  \sum_{i=1}^n \|\x_i\|_2^2 \\
    &\leq& n^3d^2,
\ese
with large enough $n$, where the last inequality is by assumption (L1). So condition (C1) is satisfied with $C=3$ and $r=2$. 

To verify condition (C2), we first note that
\bse
    \left\|\sum_{i=1}^n\bpsi(Z_i,\x_i,\bpi_0)\right\|_2^2 &=& \sum_{j=1}^p \left[\sum_{i=1}^n \{Z_i-\mu_i(\bb_0)\}x_{ij} + \sum_{i=1}^n \{\mu_i(\bb_0)-\mu_i^*(\bpi_0)\}x_{ij} \right]^2 \\
    &=& \sum_{j=1}^p \left[\sum_{i=1}^n \{Z_i-\mu_i(\bb_0)\}x_{ij} \right]^2 + \sum_{j=1}^p \left[ \sum_{i=1}^n \{\mu_i(\bb_0)-\mu_i^*(\bpi_0)\}x_{ij} \right]^2 \\
    && +2\sum_{j=1}^p \left[\sum_{i=1}^n \{Z_i-\mu_i(\bb_0)\}x_{ij} \right] \left[ \sum_{i=1}^n \{\mu_i(\bb_0)-\mu_i^*(\bpi_0)\}x_{ij} \right] \\
    &=& \sum_{j=1}^p \left[\sum_{i=1}^n \{Z_i-\mu_i(\bb_0)\}x_{ij} \right]^2,
\ese
where the last equality uses the definition of $\bpi_0$, i.e., $\sum_{i=1}^n \{\mu_i(\bb_0)-\mu_i^*(\bpi_0)\}\x_i=\0$. Moreover, by assumption (L2) we have
\bsq
    \sum_{j=1}^p\mathbb{E} \left(\left[\sum_{i=1}^n \{Z_i-\mu_i(\bb_0)\}x_{ij} \right]^2\right) = \sum_{i=1}^n \sum_{j=1}^p x_{ij}^2 \var(Z_i) = \mathcal{O}(np).
\esq
So we have $\sum_{j=1}^p [\sum_{i=1}^n \{Z_i-\mu_i(\bb_0)\}x_{ij} ]^2=\mathcal{O}_p(np)$ by Markov's inequality, and hence $\|\sum_{i=1}^n \bpsi(Z_i,\x_i,\bpi_0)\|_2=\mathcal{O}_p(n^{1/2}p^{1/2})$, which verifies condition (C2). 

To verify condition (C3), for any $\s\in\real^p$ such that $\|\s\|_2=1$ and any $\bb\in\bT$, by Taylor's theorem we have
\bse
    && \s\trans\sum_{i=1}^n\left\{\bpsi(Z_i,\x_i,\bpi_0) - \bpsi(Z_i,\x_i,\bb)\right\} \\
    &=& \sum_{i=1}^n b'(\x_i\trans\bpi_0) \s\trans\x_i \x_i\trans(\bb-\bpi_0) + \frac{1}{2}\sum_{i=1}^n (\bb-\bpi_0)\trans b''(\x_i\trans\bb') \s\trans\x_i\x_i\x_i\trans(\bb-\bpi_0) \\
    &=& n\s\trans\D_n(\bb-\bpi_0) + \frac{1}{2}\sum_{i=1}^n (\bb-\bpi_0)\trans b''(\x_i\trans\bb') \s\trans\x_i\x_i\x_i\trans(\bb-\bpi_0),
\ese
where $b'(\cdot),b''(\cdot)$ denote the first and second derivatives of the expit function respectively and are both bounded, and $\bb'\in\bT$ lies between $\bb$ and $\bpi_0$. So for any $B>0$, we have
\bse
    && \sup_{\s\in\real^p:\|\s\|_2=1} \sup_{\bb:\|\bb-\bpi_0\|_2\leq B(p/n)^{1/2}}  \left|\s\trans\sum_{i=1}^n \mathbb{E}\left\{\bpsi(Z_i,\x_i,\bpi_0) - \bpsi(Z_i,\x_i,\bb)\right\} - n\s\trans\D_n(\bb-\bpi_0)\right| \\
    &\leq& C \sup_{\s\in\real^p:\|\s\|_2=1}\sup_{\bb:\|\bb-\bpi_0\|_2\leq B(p/n)^{1/2}} (\bb-\bpi_0)\trans \sum_{i=1}^n |\s\trans\x_i| \x_i\x_i\trans (\bb-\bpi_0) \\
    &\leq& C \sup_{\bb:\|\bb-\bpi_0\|_2\leq B(p/n)^{1/2}}\|\bb-\bpi_0\|_2^2 \sup_{\s\in\real^p:\|\s\|_2=1}\sup_{\a\in\real^p:\|\a\|_2=1} \sum_{i=1}^n |\a\trans\x_i|^2 |\s\trans\x_i| \\
    &=& \mathcal{O}(p)\\
    &=& o(n^{1/2}),
\ese
where $C$ is a finite positive constant. The first equality uses Cauchy-Schwarz inequality and assumption (L2). The last equality uses assumption (L1). So condition (C3) is verified. 

Lastly, to verify conditions (C4) and (C5), for any $B>0$ we have
\bse
    && \sup_{\s\in\real^p:\|\s\|_2=1} \sup_{\bb_1:\|\bb_1-\bb_2\|_2\leq B(p/n)^{1/2}} \sum_{i=1}^n \left\{\s\trans\boldeta_i(\bb_1,\bb_2)\right\}^2 \\
    &=& \sup_{\s\in\real^p:\|\s\|_2=1} \sup_{\bb_1:\|\bb_1-\bb_2\|_2\leq B(p/n)^{1/2}} \sum_{i=1}^n \{\mu_i(\bb_2)-\mu_i(\bb_1)\}^2 (\s\trans\x_i)^2 \\
    &\leq& C \sup_{\s\in\real^p:\|\s\|_2=1} \sup_{\bb_1:\|\bb_1-\bb_2\|_2\leq B(p/n)^{1/2}} \sum_{i=1}^n \{\x_i\trans(\bb_2-\bb_1)\}^2 (\s\trans\x_i)^2 \\ 
    &\leq& C \sup_{\s\in\real^p:\|\s\|_2=1} \sup_{\a\in\real^p:\|\a\|_2=1} \sup_{\bb_1:\|\bb_1-\bb_2\|_2\leq B(p/n)^{1/2}}  \sum_{i=1}^n (\a\trans\x_i)^2 (\s\trans\x_i)^2 \|\bb_2-\bb_1\|_2^2 \\ 
    &=& \mathcal{O}(p),
\ese
where $C$ is a finite positive constant and the last equality is by assumption (L2). So conditions (C4) and (C5) hold with $A(n,m)=p$. 

Therefore, by Theorem~2.2 of \cite{he2000parameters}, we have 
\bsq
    \wh{\bpi}-\bpi_0 = -n^{-1}\sum_{i=1}^n \D_n^{-1}\{Z_i-\mu_i^*(\bpi_0)\}\x_i + \r_n,
\esq
with $\|\r_n\|_2=o_p(n^{-1/2})$. Since $|\s\trans\r_n| \leq \|\r_n\|_2=o_p(n^{-1/2})$ for any $\s\in\real^p$ with $\|\s\|_2=1$, we can further obtain
\bsq
    \sqrt{n}\s\trans (\wh{\bpi}-\bpi_0) = -n^{-1/2} \sum_{i=1}^n \{Z_i-\mu_i^*(\bpi_0)\} \s\trans\D_n^{-1}\x_i + o_p(1). 
\esq
We remain to study the asymptotic behavior of $-n^{-1/2} \sum_{i=1}^n \{Z_i-\mu_i^*(\bpi_0)\} \s\trans\D_n^{-1}\x_i$. Note that
\bsq
    \sigma_n^2 \equiv \var\left[-n^{-1/2} \sum_{i=1}^n \{Z_i-\mu_i(\bb_0)\} \s\trans\D_n^{-1}\x_i\right] = n^{-1}\s\trans\D_n^{-1} \left\{\sum_{i=1}^n \var(Z_i)\x_i\x_i\trans\right\}\D_n^{-1}\s,
\esq
which is bounded above and below with large enough $n$ by assumptions
(L2) to (L4).

Let $\sigma_n \equiv \sqrt{\sigma_n^2}$, we define $U_i \equiv
\sigma_n^{-1}n^{-1/2} \{Z_i-\mu_i(\bb_0)\} \s\trans\D_n^{-1}\x_i$ such
that $\mathbb{E}(U_i)=0$
and $\sum_{i=1}^n \var(U_i)=1$. Then by assumptions (L2) and (L3) we have
\bsq
    \sum_{i=1}^n \mathbb{E}(U_i^4) = \sum_{i=1}^n \frac{(\s\trans\D_n^{-1}\x_i)^4}{n^2\sigma_n^4} \mathbb{E}\left[\{Z_i-\mu_i(\bb_0)\}^4\right] =o(1).
\esq
So the Lyapunov condition is verified and $\sum_{i=1}^n U_i \overset{D}{\to}\mathcal{N}(0,1)$. Note that by the definition of $\bpi_0$, we have $\sum_{i=1}^n \{\mu_i(\bb_0)-\mu_i^*(\bpi_0)\}\x_i = \0$. So we have
\bse
    && \sqrt{n}\s\trans (\wh{\bpi}-\bpi_0) = -n^{-1/2} \sum_{i=1}^n \{Z_i-\mu_i^*(\bpi_0)\} \s\trans\D_n^{-1}\x_i + o_p(1) \\
    &=& -n^{-1/2} \sum_{i=1}^n \{Z_i-\mu_i(\bb_0)\} \s\trans\D_n^{-1}\x_i - n^{-1/2} \sum_{i=1}^n \{\mu_i(\bb_0)-\mu_i^*(\bpi_0)\} \s\trans\D_n^{-1}\x_i + o_p(1) \\
    &=& -n^{-1/2} \sum_{i=1}^n \{Z_i-\mu_i(\bb_0)\} \s\trans\D_n^{-1}\x_i + o_p(1) \\
    &=& -\sigma_n \sum_{i=1}^n U_i+o_p(1) \\
    &\overset{D}{\to}& \mathcal{N}(0, \sigma^2),
\ese
with $\sigma^2\equiv \lim_{n\to\infty}\sigma_n^2$. In other words, NMLE $\wh{\bpi}$ is asymptotically normal with respect to a shifted target $\bpi_0$ instead of $\bb_0$.

\section{Additional simulation: Logistic regression}
\label{sec:simu:logistic}

In this section, we consider a logistic regression with the same
simulation settings as the ones used in
Section~\ref{sec:simu:roblogistic} of the paper, except that for
each setting we consider five additional parameters with true values
to be zero such that each setting is more challenging with a larger
$p/n$ ratio compared to Section~\ref{sec:simu:roblogistic}. Despite
the sparsity of the true parameter values, there is no variable
selection conducted in our analysis. In this example, we consider
three estimators: (i) MLE, (ii) JINI using MLE as the
initial estimator, and (iii) BBC defined in \eqref{def:BBC} which
uses MLE as the initial estimator and $100$ simulated samples. We
also construct $95\%$ CIs based on their
asymptotic normality, where the asymptotic covariance matrices are
estimated by plugging in the point estimates to the asymptotic
covariance matrices which have closed-form expressions.

\begin{figure}
    \centering
    \includegraphics[width=12cm]{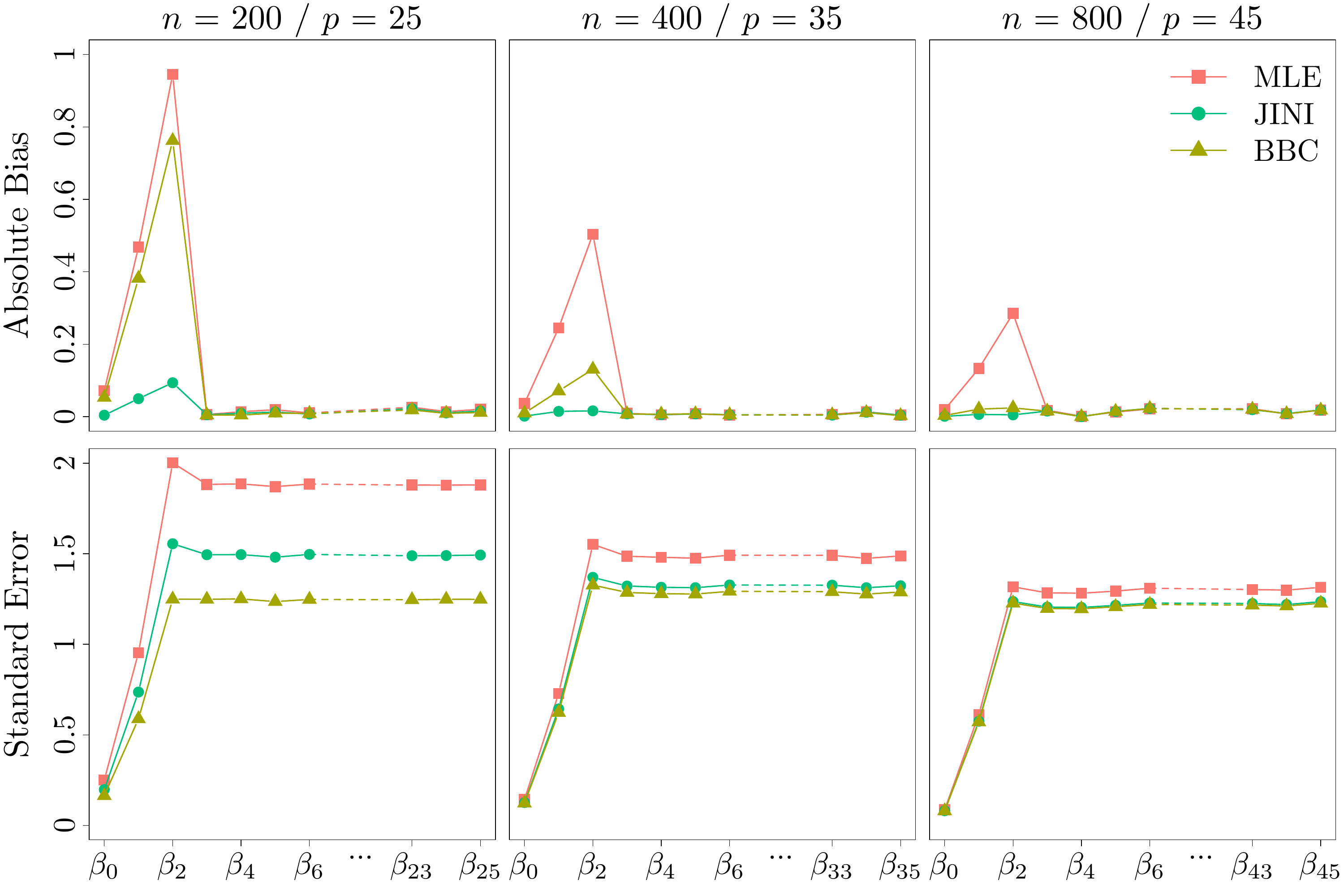}
    \caption{Estimation results for the logistic regression presented in Section~\ref{sec:simu:logistic}.}  
    \label{fig:logistic_point_est}
\end{figure}

The estimation performance of all estimators are presented in
Figure~\ref{fig:logistic_point_est}. We can see that JINI has
a considerably smaller bias than the others in all settings, especially
in the first setting with the largest $p/n$ ratio. BBC also has
a smaller bias than MLE in all settings, but the bias reduction is not as much as
JINI. Moreover, BBC has the smallest standard error and MLE
has the largest. The standard error of JINI is marginally inflated compared to BBC, but the difference quickly shrinks to zero as $p$ and $n$ increase.

\begin{figure}
    \centering
    \includegraphics[width=12cm]{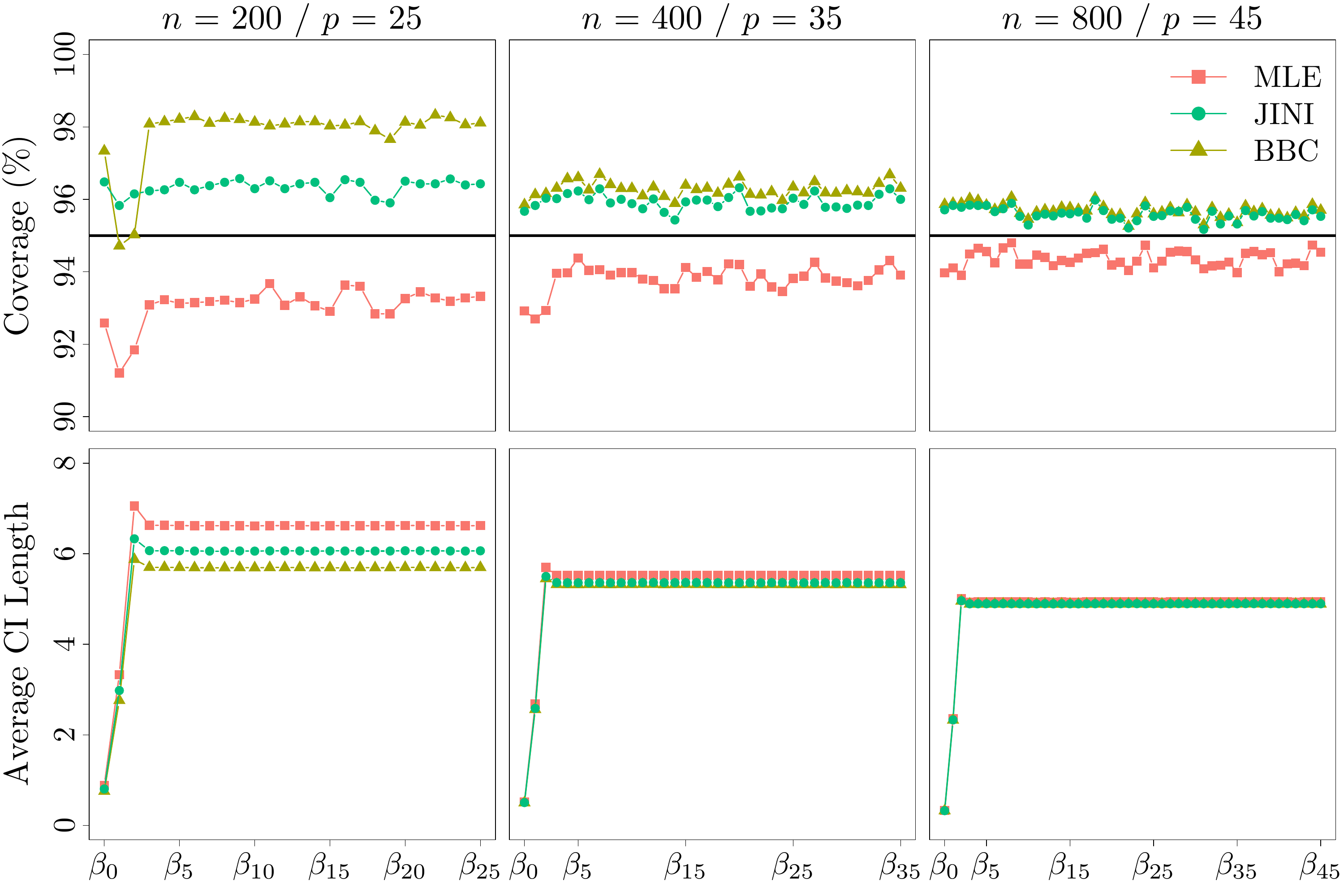}
    \caption{Inference results for the logistic regression presented
      in Section~\ref{sec:simu:logistic}.}
    \label{fig:logistic_inference}
\end{figure}

The inference results of all estimators are presented in Figure~\ref{fig:logistic_inference}. First of all, MLE has either larger or similar average CI length compared to JINI and BBC, while its coverage severely falls below the nominal level, especially in the first setting with large $p/n$. This is caused by the significant bias of MLE. Moreover, both JINI and BBC produce conservative CIs with similar average lengths. However, the CIs of JINI have coverage closer to the nominal level than BBC, especially in the first setting, which reflects the superiority of JINI.

\section{Additional simulation: Robust Pareto regression}
\label{sec:simu:pareto}

Pareto regression is an important model in domains where extreme
events are observed and risk factors influencing these extreme events
are essential to determine. This is the case, for example, in finance,
insurance and natural sciences. Although a generalized Pareto
distribution (see \citealp{Pick:75}) is more commonly used in practice (see
e.g., \citealp{DaSm:90,BeGo:03,HaGrKn:18}), in this section we focus
on a two parameter Pareto distribution (i.e., with only scale and
shape parameters) for simplicity.

  In the (two parameter) Pareto regression,
the response variable $Y_i$ with $i=1,\ldots,n$ follows a Pareto
distribution with support $[\gamma,\infty)$, where $\gamma>0$ is the
scale parameter. The conditional distribution of $Y_i$ given a fixed
covariate vector $\x_i$ is $f_{Y|\x}(y_i,\x_i,\bb,\gamma) = \alpha_i
\gamma^{\alpha_i} y_i^{-(\alpha_i+1)}$, where $\alpha_i \equiv
\exp(\x_i\trans\bb)$ is the shape parameter, also known as the tail
index. To the best of our knowledge, there is currently no available
robust estimator for the Pareto regression. In this section, we aim to use JINI to construct a robust estimator in a simple manner.

Specifically, we propose to construct a robust JINI using a Naive WMLE (NWMLE)
as the initial estimator, similarly to the approach presented in
Section~\ref{sec:simu:roblogistic}. In particular, NWMLE is
the solution to the estimating equation $\sum_{i=1}^n w_c(d_i)
\s(\bb,\gamma|y_i,\x_i)=\0$, where $\s(\bb,\gamma|y_i,\x_i)$ is the
score function and $d_i \equiv
\|\s(\bb,\gamma|y_i,\x_i)\|_2$. Moreover, $w_c(d)$ is the Tukey's
biweight function (see \citealp{beaton1974fitting}) given by
$w_c(d)=\{1-(d/c)^2\}^2I(d\leq c)$, where $c$ is the tuning constant
chosen to balance the robustness and asymptotic efficiency loss. Since
the estimating equation neglects the consistency correction term, the
resulting estimator (i.e., NWMLE) is inconsistent but is
simple to compute. Given that there is no alternative robust
estimator, in this simulation we consider MLE as a benchmark
estimator, which is more efficient (asymptotically) than any robust
estimator when there is no data contamination.

To investigate the finite sample performance of the proposed robust
JINI compared to MLE, we consider a simulation setting with
$n=150$ and $p=10$. We let $x_{i,0}=1$ be the intercept and $x_{i,j}$
be realizations of $\mathcal{N}(0,n^{-1/2})$ with $j=1,\ldots,p$. The
true parameter values are $\bb_0 = (1.5, -1, -1, 1, 1, -0.1, 0.1,
-0.1, 0.1, -0.1, 0.1)\trans$ and $\gamma_0 = 5$. The covariates and
parameter values are chosen such that $\alpha_i>2$ for all
$i=1,\ldots,n$, and hence the mean and variance of $Y_i$ exist. For
robust JINI, we choose the tuning constant $c=10$ for the Tukey's
biweight function such that it corresponds to approximately 95\%
efficiency of MLE. In order to evaluate the robustness of robust JINI, we consider data contamination in the following form:
Given the same covariates, 10\% of the observed responses are
generated from another Pareto distribution based on the same $\bb_0$
and a different scale parameter of 50. We consider $10,000$ Monte Carlo
replications.  

\begin{figure}
    \centering
    \includegraphics[width=12cm]{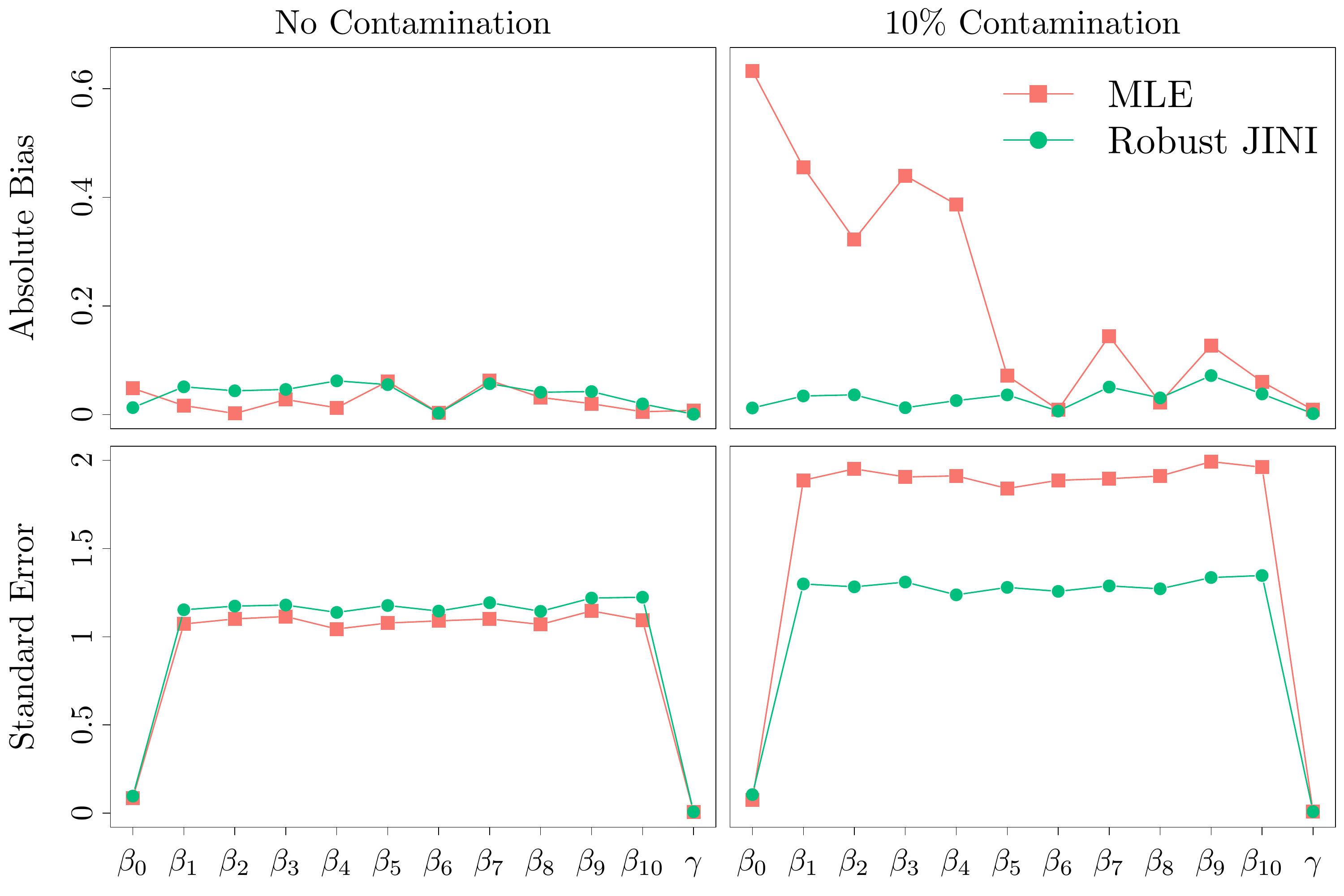}
    \caption{Estimation results for the robust Pareto regression presented in Section~\ref{sec:simu:pareto}.}  
    \label{fig:pareto_point_est}
\end{figure}

The estimation performance of robust JINI and MLE is
presented in Figure~\ref{fig:pareto_point_est}, considering both
uncontaminated and contaminated data. When the data are not
contaminated, both MLE and robust JINI show comparable and
almost negligible biases, with comparable standard errors. When 10\%
of the data are contaminated, MLE shows very different and
unstable performance, with heavily inflated bias and standard
error. This is expected because, as a non-robust estimator, MLE is
very sensitive to the presence of contaminated data. On the other
hand, robust JINI does not appear to be much affected by
the presence of contaminated data, with stable bias and standard error
that are comparable to (or slightly large than) the ones of
robust JINI when there is no data contamination. This is a desired
finite sample property for a robust estimator, i.e., the ability to
maintain stable performance when data present slight model deviation.

The simulation studies presented in this section and
Section~\ref{sec:simu:roblogistic} highlight the following main message: Our approach avoids the analytical and computational difficulties typically encountered in the classical robust estimation approach, such as the intractable computation on the consistency correction term. It allows to construct a robust estimator in a significantly simpler manner, while providing advantageous finite sample performance even when $p$ is relatively large compared to $n$.

\section{Additional materials for the alcohol consumption data analysis in Section~\ref{sec:alcohol}}
\label{sec:alcohol_additional}

\begin{table}
\centering
\caption{Alcohol Consumption Data Description (adapted from Table~1 in \cite{Cortez2008UsingDM} to our analysis)}
\begin{tabular}{lll}
\toprule
Notation & Attribute & Description \\
\midrule
$Y$ & Alcohol consumption level & Binary, $1$ for high and $0$ for low. \\
$x_1$ & Gender & Binary, $1$ for male and $0$ for female. \\
$x_2$ & Family size & Binary, $1$ for $\leq 3$ and $0$ for $>3$. \\ 
$x_3$ & Extra paid classes & Binary, $1$ for yes and $0$ for no. \\
$x_4$ & Quality of family relationships & Numeric from $1$ to $5$. \\
$x_5$ & Frequency to go out with friends & Numeric from $1$ to $5$. \\
$x_6$ & Number of school absences & Numeric from $0$ to $75$. \\
$x_{7}$-$x_{9}$ & Weekly study time & $>10$ hours, $5$-$10$ hours, $2$-$5$ hours, \\
&& $<2$ hours (as baseline).\\
$x_{10}$ & Student's school & Binary, $1$ for \textit{Mousinho da Silveira} and \\
&& $0$ for \textit{Gabriel Pereira}. \\
$x_{11}$ & Age & Numeric from $15$ to $22$. \\
$x_{12}$ & Student's address & Binary, $1$ for urban and $0$ for rural. \\
$x_{13}$ & Parent's cohabitation status & Binary, $1$ for together and $0$ for living apart. \\
$x_{14}$-$x_{17}$ & Mother's job & At home, health care, civil services, teacher,\\
&& others (as baseline).\\
$x_{18}$-$x_{21}$ & Father's job & At home, health care, civil services, teacher,\\
&& others (as baseline).\\
$x_{22}$-$x_{24}$ & School choice reason & Course preference, close to home, reputation,\\ 
&& others (as baseline). \\
$x_{25}$-$x_{26}$ & Student's guardian & Father, mother, others (as baseline). \\
$x_{27}$-$x_{28}$ & Travel time & $<15$ mins (as baseline), $15$-$30$ mins, $>30$ mins. \\
$x_{29}$-$x_{30}$ & Number of past class failures & $0$ (as baseline), $1$, $\geq 2$. \\
$x_{31}$ & Extra school support & Binary, $1$ for yes and $0$ for no. \\
$x_{32}$ & Family educational support & Binary, $1$ for yes and $0$ for no. \\
$x_{33}$ & Extracurricular activities & Binary, $1$ for yes and $0$ for no. \\
$x_{34}$ & Attended nursery school & Binary, $1$ for yes and $0$ for no. \\
$x_{35}$ & Consider higher education & Binary, $1$ for yes and $0$ for no. \\
$x_{36}$ & Internet access at home & Binary, $1$ for yes and $0$ for no. \\
$x_{37}$ & In a romantic relationship & Binary, $1$ for yes and $0$ for no. \\
$x_{38}$-$x_{41}$ & Free time after school & Very low (as baseline), low, medium, high, \\
& & very high. \\
$x_{42}$ & First period grade & Numeric from $0$ to $20$. \\
$x_{43}$ & Second period grade & Numeric from $0$ to $20$. \\
$x_{44}$ & Final grade & Numeric from $0$ to $20$. \\
\bottomrule
\end{tabular}
\label{tab:alcohol_data_description}
\end{table}

\begin{figure}[h!]
    \centering
    \includegraphics[width=12cm]{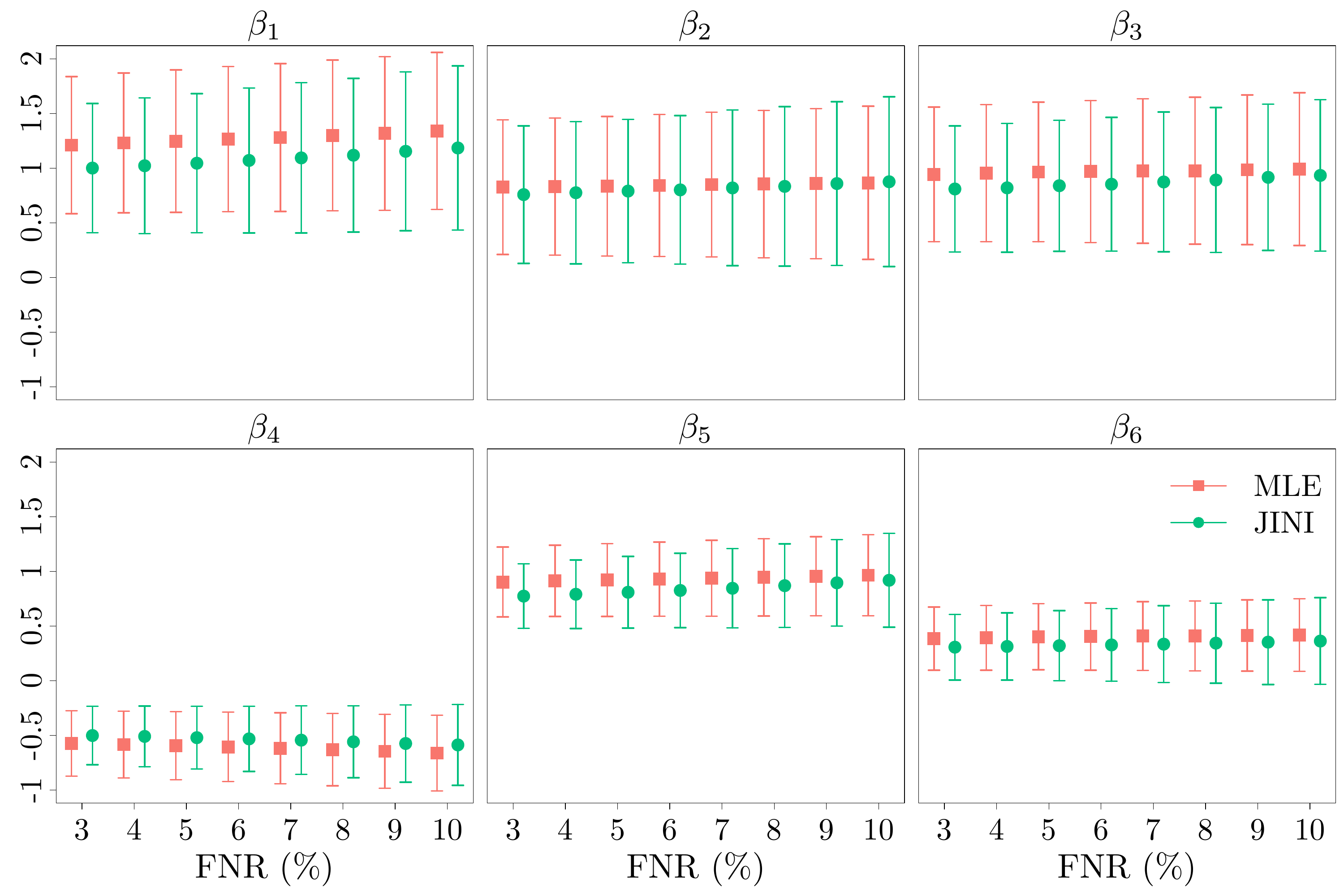}
    \caption{Sensitivity analysis to evaluate how the assumed False
      Negative Rate (FNR) for the misclassified logistic regression
      considered in Section~\ref{sec:alcohol} influences the
      performance of the estimators. The dots correspond to the point
      estimates and the bars correspond to the 95\% CIs.}
    \label{fig:alcohol_sensitivity}
\end{figure}

\begin{figure}[h!]
    \centering
\includegraphics[width=\linewidth]{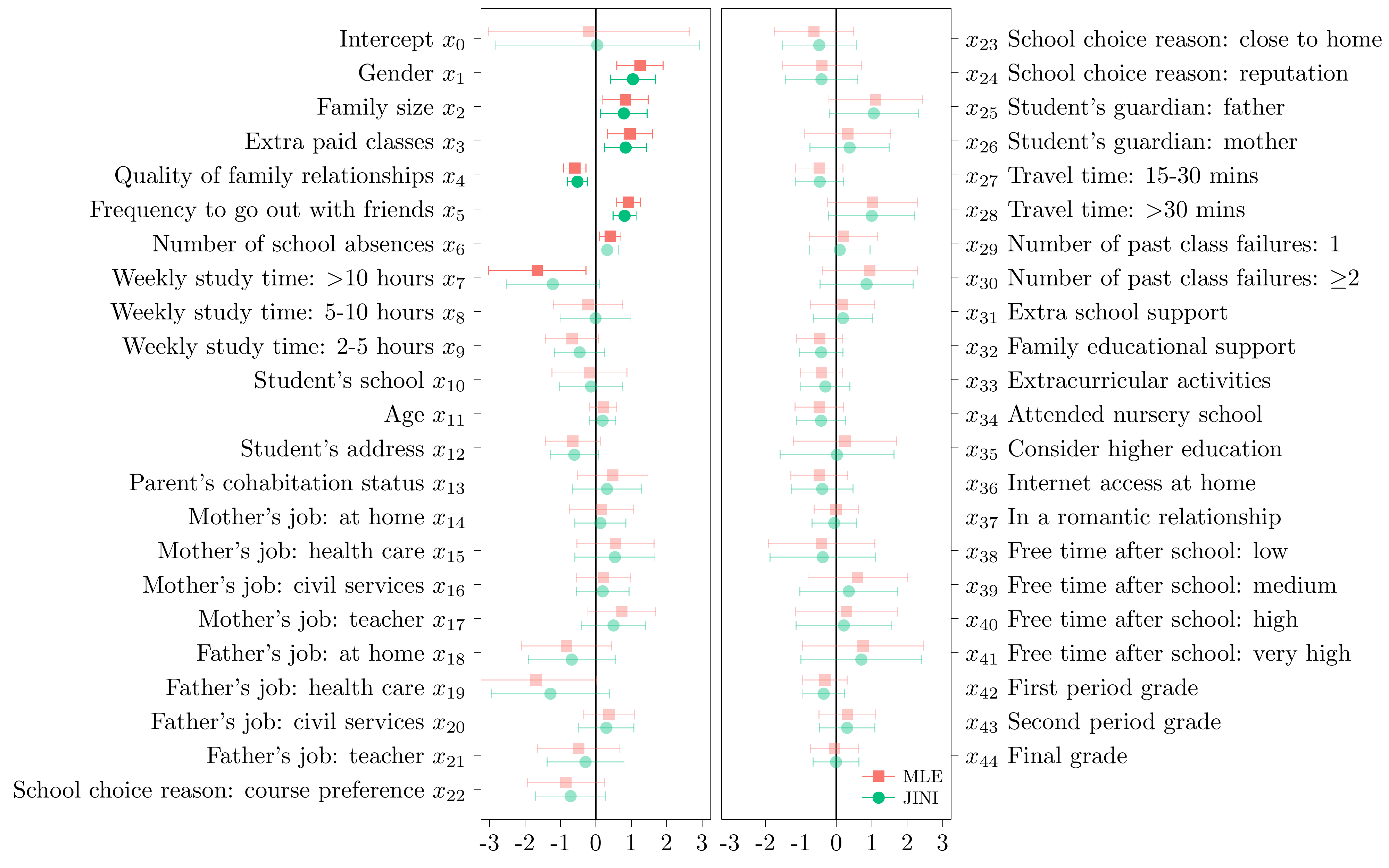}
    \caption{Point estimates and $95\%$ CIs for the real alcohol consumption data analysis presented in Section~\ref{sec:alcohol}. All parameters used in the analysis are included. When the CI does not cover zero, its color is solid. When the CI covers zero, its color is transparent.}  
    \label{fig:alcohol_real_data_all}
\end{figure}

\newpage



\bibliographystyle{apalike} 
\bibliography{biblio}       

\begin{thebibliography}{}

\bibitem[Albright et~al., 2021]{AlcoholFN:21}
Albright, D.~L., Holmes, L., Lawson, M., McDaniel, J., and Godfrey, K. (2021).
\newblock False negative {AUDIT} screening results among patients in rural
  primary care settings.
\newblock {\em Journal of Evidence-Based Social Work}, 18:585--595.

\bibitem[Arvanitis and Demos, 2014]{arvanitis2014valid}
Arvanitis, S. and Demos, A. (2014).
\newblock Valid locally uniform edgeworth expansions for a class of weakly
  dependent processes or sequences of smooth transformations.
\newblock {\em Journal of Time Series Econometrics}, 6(2):183--235.

\bibitem[Arvanitis and Demos, 2015]{arvanitis2015class}
Arvanitis, S. and Demos, A. (2015).
\newblock A class of indirect inference estimators: higher-order asymptotics
  and approximate bias correction.
\newblock {\em The Econometrics Journal}, 18(2):200--241.

\bibitem[Bai et~al., 2009]{bai2009statistical}
Bai, Z., Zheng, S., Zhang, B., and Hu, G. (2009).
\newblock Statistical analysis for rounded data.
\newblock {\em Journal of Statistical Planning and Inference},
  139(8):2526--2542.

\bibitem[Beaton and Tukey, 1974]{beaton1974fitting}
Beaton, A.~E. and Tukey, J.~W. (1974).
\newblock The fitting of power series, meaning polynomials, illustrated on
  band-spectroscopic data.
\newblock {\em Technometrics}, 16(2):147--185.

\bibitem[Beirlant and Goegebeur, 2003]{BeGo:03}
Beirlant, J. and Goegebeur, Y. (2003).
\newblock Regression with response distributions of {Pareto}-type.
\newblock {\em Computational Statistics \& Data Analysis}, 42:595--619.

\bibitem[Bianco and Yohai, 1996]{Bianco1996}
Bianco, A.~M. and Yohai, V.~J. (1996).
\newblock {\em Robust estimation in the logistic regression model}, pages
  17--34.
\newblock Springer New York, New York, NY.

\bibitem[Brillinger, 1983]{brillinger1983generalized}
Brillinger, D.~R. (1983).
\newblock A generalized linear model with ``gaussian'' regressor variables.
\newblock {\em A Festschrift for Erich L. Lehmann (P. J. Bickel, K. A. Doksum
  and J. L. Hodges, Jr., eds.)}, pages 97--114.

\bibitem[Bush et~al., 1998]{AUDIT-C:98}
Bush, K., Kivlahan, D.~R., McDonell, M.~B., Fihn, S.~D., Bradley, K.~A.,
  {Ambulatory Care Quality Improvement Project (ACQUIP)}, et~al. (1998).
\newblock The audit alcohol consumption questions (audit-c): an effective brief
  screening test for problem drinking.
\newblock {\em Archives of Internal Medicine}, 158(16):1789--1795.

\bibitem[Cantoni and Ronchetti, 2001]{CaRo:01b}
Cantoni, E. and Ronchetti, E. (2001).
\newblock Robust inference for generalized linear models.
\newblock {\em Journal of the American Statistical Association}, 96:1022--1030.

\bibitem[Cizek, 2008]{Cize:08}
Cizek, P. (2008).
\newblock Robust and efficient adaptive estimation of binary-choice regression
  models.
\newblock {\em Journal of the American Statistical Association}, 103:687--696.

\bibitem[Cordeiro and Barroso, 2007]{cordeiro2007third}
Cordeiro, G.~M. and Barroso, L.~P. (2007).
\newblock A third-order bias corrected estimate in generalized linear models.
\newblock {\em Test}, 16(1):76.

\bibitem[Cortez and Silva, 2008]{Cortez2008UsingDM}
Cortez, P. and Silva, A. M.~G. (2008).
\newblock Using data mining to predict secondary school student performance.

\bibitem[Cox, 1972]{CoxDR:72}
Cox, D.~R. (1972).
\newblock Regression models and life-tables.
\newblock {\em Journal of the Royal Statistical Society: Series B
  (Methodological)}, 34(2):187--202.

\bibitem[Cox and Hinkley, 1979]{cox1979theoretical}
Cox, D.~R. and Hinkley, D.~V. (1979).
\newblock {\em Theoretical Statistics}.
\newblock CRC Press.

\bibitem[Cox and Snell, 1968]{cox1968general}
Cox, D.~R. and Snell, E.~J. (1968).
\newblock A general definition of residuals.
\newblock {\em Journal of the Royal Statistical Society: Series B
  (Methodological)}, 30(2):248--265.

\bibitem[Cribari-Neto and Zeileis, 2010]{cribari2010beta}
Cribari-Neto, F. and Zeileis, A. (2010).
\newblock Beta regression in {R}.
\newblock {\em Journal of Statistical Software}, 34:1--24.

\bibitem[Czado and Santner, 1992]{czado1992effect}
Czado, C. and Santner, T.~J. (1992).
\newblock The effect of link misspecification on binary regression inference.
\newblock {\em Journal of Statistical Planning and Inference}, 33(2):213--231.

\bibitem[Davis et~al., 2010]{SocialDesire:10}
Davis, C.~G., Thake, J., and Vilhena, N. (2010).
\newblock Social desirability biases in self-reported alcohol consumption and
  harms.
\newblock {\em Addictive Behaviors}, 35:302--311.

\bibitem[Davison and Smith, 1990]{DaSm:90}
Davison, A.~C. and Smith, R.~L. (1990).
\newblock Models for exceedances over high thresholds.
\newblock {\em Journal of the Royal Statistical Society: Series B
  (Methodological)}, 52(3):393--425.

\bibitem[Dempster et~al., 1977]{dempster1977maximum}
Dempster, A.~P., Laird, N.~M., and Rubin, D.~B. (1977).
\newblock Maximum likelihood from incomplete data via the em algorithm.
\newblock {\em Journal of the Royal Statistical Society: Series B
  (Methodological)}, 39(1):1--22.

\bibitem[Dong, 2015]{dong2015regression}
Dong, Y. (2015).
\newblock Regression discontinuity applications with rounding errors in the
  running variable.
\newblock {\em Journal of Applied Econometrics}, 30(3):422--446.

\bibitem[Dupuis and Morgenthaler, 2002]{dupuis2002robust}
Dupuis, D.~J. and Morgenthaler, S. (2002).
\newblock Robust weighted likelihood estimators with an application to
  bivariate extreme value problems.
\newblock {\em Canadian Journal of Statistics}, 30(1):17--36.

\bibitem[Efron, 1975]{efron1975}
Efron, B. (1975).
\newblock Defining the curvature of a statistical problem (with applications to
  second order efficiency).
\newblock {\em The Annals of Statistics}, 3:1189--1242.

\bibitem[Efron and Tibshirani, 1994]{efron1994introduction}
Efron, B. and Tibshirani, R.~J. (1994).
\newblock {\em {An Introduction to the Bootstrap}}.
\newblock CRC press.

\bibitem[Ferrari and Cribari-Neto, 2004]{ferrari2004beta}
Ferrari, S. and Cribari-Neto, F. (2004).
\newblock Beta regression for modelling rates and proportions.
\newblock {\em Journal of Applied Statistics}, 31(7):799--815.

\bibitem[Field and Smith, 1994]{FiSm:94}
Field, C. and Smith, B. (1994).
\newblock Robust estimation: A weighted maximum likelihood approach.
\newblock {\em International Statistical Review/Revue Internationale de
  Statistique}, pages 405--424.

\bibitem[Firth, 1993]{Firt:93}
Firth, D. (1993).
\newblock Bias reduction of maximum likelihood estimates.
\newblock {\em Biometrika}, 80:27--38.

\bibitem[Frank et~al., 2008]{AlcoholFN:08}
Frank, D., DeBenedetti, A., Volk, R., Williams, E.~C., Kivlahan, D.~R., and
  Bradley, K.~A. (2008).
\newblock Effectiveness of the {AUDIT-C} as a screening test for alcohol misuse
  in three race/ethnic groups.
\newblock {\em Journal of General Internal Medicine}, 23:781--787.

\bibitem[Gallant and Tauchen, 1996]{gallant1996moments}
Gallant, A.~R. and Tauchen, G. (1996).
\newblock Which moments to match?
\newblock {\em Econometric Theory}, 12(4):657--681.

\bibitem[Gouri{\'e}roux et~al., 1993]{gourieroux1993indirect}
Gouri{\'e}roux, C., Monfort, A., and Renault, E. (1993).
\newblock Indirect inference.
\newblock {\em Journal of Applied Econometrics}, 8(1):85--118.

\bibitem[Gouri{\'e}roux et~al., 2000]{gourieroux2000}
Gouri{\'e}roux, C., Renault, E., and Touzi, N. (2000).
\newblock Calibration by simulation for small sample bias correction.
\newblock {\em Simulation-based Inference in Econometrics: Methods and
  Applications}, pages 328--358.

\bibitem[Greene, 1981]{greene1981asymptotic}
Greene, W.~H. (1981).
\newblock On the asymptotic bias of the ordinary least squares estimator of the
  tobit model.
\newblock {\em Econometrica: Journal of the Econometric Society}, pages
  505--513.

\bibitem[Guerrier et~al., 2019]{guerrier2018simulation}
Guerrier, S., Dupuis-Lozeron, E., Ma, Y., and Victoria-Feser, M.-P. (2019).
\newblock Simulation-based bias correction methods for complex models.
\newblock {\em Journal of the American Statistical Association (Theory and
  Methods)}, 114(525):146--157.

\bibitem[Hambuckers et~al., 2018]{HaGrKn:18}
Hambuckers, J., Groll, A., and Kneib, T. (2018).
\newblock Understanding the economic determinants of the severity of
  operational losses: A regularized generalized {Pareto} regression approach.
\newblock {\em Journal of Applied Econometrics}, 33:898--935.

\bibitem[Hampel et~al., 1986]{hampel1986robust}
Hampel, F.~R., Ronchetti, E.~M., Rousseeuw, P., and Stahel, W.~A. (1986).
\newblock {\em Robust Statistics: the Approach based on Influence Functions}.
\newblock Wiley-Interscience.

\bibitem[Hausman et~al., 1998]{HAUSMAN1998}
Hausman, J.~A., Abrevaya, J., and Scott-Morton, F.~M. (1998).
\newblock Misclassification of the dependent variable in a discrete-response
  setting.
\newblock {\em Journal of Econometrics}, 87:239--269.

\bibitem[He and Shao, 2000]{he2000parameters}
He, X. and Shao, Q.-M. (2000).
\newblock On parameters of increasing dimensions.
\newblock {\em Journal of Multivariate Analysis}, 73(1):120--135.

\bibitem[Heritier et~al., 2009]{HeCaCoVF:09}
Heritier, S., Cantoni, E., Copt, S., and Victoria-Feser, M.-P. (2009).
\newblock {\em Robust Methods in Biostatistics}.
\newblock John Wiley \& Sons.

\bibitem[Huber, 1967]{huber1967behavior}
Huber, P.~J. (1967).
\newblock The behavior of maximum likelihood estimates under nonstandard
  conditions.
\newblock In {\em Proceedings of the fifth Berkeley symposium on mathematical
  statistics and probability}, volume~1, pages 221--233.

\bibitem[Huber, 1981]{Hube:81}
Huber, P.~J. (1981).
\newblock {\em Robust Statistics}.
\newblock John Wiley, New York.

\bibitem[Imbens, 2002]{imbens2002generalized}
Imbens, G.~W. (2002).
\newblock Generalized method of moments and empirical likelihood.
\newblock {\em Journal of Business \& Economic Statistics}, 20(4):493--506.

\bibitem[Kendall, 1954]{kendall1954note}
Kendall, M.~G. (1954).
\newblock Note on bias in the estimation of autocorrelation.
\newblock {\em Biometrika}, 41(3-4):403--404.

\bibitem[Komunjer, 2012]{komunjer2012global}
Komunjer, I. (2012).
\newblock Global identification in nonlinear models with moment restrictions.
\newblock {\em Econometric Theory}, 28(4):719--729.

\bibitem[Kosmidis, 2014a]{kosmidis2014bias}
Kosmidis, I. (2014a).
\newblock Bias in parametric estimation: reduction and useful side-effects.
\newblock {\em Wiley Interdisciplinary Reviews: Computational Statistics},
  6(3):185--196.

\bibitem[Kosmidis, 2014b]{Kosm:14}
Kosmidis, I. (2014b).
\newblock Improved estimation in cumulative link models.
\newblock {\em Journal of the Royal Statistical Society: Series B
  (Methodological)}, 76:169--196.

\bibitem[Kosmidis and Firth, 2009]{KoFi:09}
Kosmidis, I. and Firth, D. (2009).
\newblock Bias reduction in exponential family nonlinear models.
\newblock {\em Biometrika}, 96:793--804.

\bibitem[Kuk, 1995]{kuk1995asymptotically}
Kuk, A. Y.~C. (1995).
\newblock Asymptotically unbiased estimation in generalized linear models with
  random effects.
\newblock {\em Journal of the Royal Statistical Society: Series B
  (Methodological)}, 57(2):395--407.

\bibitem[Lai and Robbins, 1979]{lai1979adaptive}
Lai, T.~L. and Robbins, H. (1979).
\newblock Adaptive design and stochastic approximation.
\newblock {\em The Annals of Statistics}, pages 1196--1221.

\bibitem[Li and Duan, 1989]{li1989regression}
Li, K.-C. and Duan, N. (1989).
\newblock Regression analysis under link violation.
\newblock {\em The Annals of Statistics}, 17(3):1009--1052.

\bibitem[Liu and Zhang, 2017]{LiZh:17}
Liu, H. and Zhang, Z. (2017).
\newblock Logistic regression with misclassification in binary outcome
  variables: A method and software.
\newblock {\em Behaviormetrika}, 44:447--476.

\bibitem[MacKinnon and Smith, 1998]{mackinnon1998approximate}
MacKinnon, J.~G. and Smith, A.~A. (1998).
\newblock Approximate bias correction in econometrics.
\newblock {\em Journal of Econometrics}, 85(2):205--230.

\bibitem[Mardia et~al., 1999]{mardia1999bias}
Mardia, K., Southworth, H., and Taylor, C. (1999).
\newblock On bias in maximum likelihood estimators.
\newblock {\em Journal of Statistical Planning and Inference}, 76(1-2):31--39.

\bibitem[Marriott and Pope, 1954]{marriott1954bias}
Marriott, F. and Pope, J. (1954).
\newblock Bias in the estimation of autocorrelations.
\newblock {\em Biometrika}, 41(3/4):390--402.

\bibitem[Mittelhammer et~al., 2005]{mittelhammer2005empirical}
Mittelhammer, R.~C., Judge, G.~G., and Schoenberg, R. (2005).
\newblock Empirical evidence concerning the finite sample performance of
  el-type structural equation estimation and inference methods.
\newblock {\em Chapter}, 12:282--305.

\bibitem[Moustaki and Victoria-Feser, 2006]{moustaki2006bounded}
Moustaki, I. and Victoria-Feser, M.-P. (2006).
\newblock Bounded-influence robust estimation in generalized linear latent
  variable models.
\newblock {\em Journal of the American Statistical Association},
  101(474):644--653.

\bibitem[Nelson, 1981]{nelson1981test}
Nelson, F.~D. (1981).
\newblock A test for misspecification in the censored normal model.
\newblock {\em Econometrica: Journal of the Econometric Society}, pages
  1317--1329.

\bibitem[Newey and McFadden, 1994]{newey1994large}
Newey, W.~K. and McFadden, D. (1994).
\newblock Large sample estimation and hypothesis testing.
\newblock {\em Handbook of Econometrics}, 4:2111--2245.

\bibitem[{Pickands III}, 1975]{Pick:75}
{Pickands III}, J. (1975).
\newblock Statistical inference using extreme order statistics.
\newblock {\em The Annals of Statistics}, 3:119 -- 131.

\bibitem[Polyak and Juditsky, 1992]{polyak1992acceleration}
Polyak, B.~T. and Juditsky, A.~B. (1992).
\newblock Acceleration of stochastic approximation by averaging.
\newblock {\em SIAM journal on control and optimization}, 30(4):838--855.

\bibitem[Rabe-Hesketh et~al., 2002]{rabe2002reliable}
Rabe-Hesketh, S., Skrondal, A., and Pickles, A. (2002).
\newblock Reliable estimation of generalized linear mixed models using adaptive
  quadrature.
\newblock {\em The Stata Journal}, 2(1):1--21.

\bibitem[Robbins and Monro, 1951]{robbins1951stochastic}
Robbins, H. and Monro, S. (1951).
\newblock A stochastic approximation method.
\newblock {\em The Annals of Mathematical Statistics}, pages 400--407.

\bibitem[Smith, 1985]{smith1985maximum}
Smith, R.~L. (1985).
\newblock Maximum likelihood estimation in a class of nonregular cases.
\newblock {\em Biometrika}, 72(1):67--90.

\bibitem[Sur and Cand{\`e}s, 2019]{sur2019modern}
Sur, P. and Cand{\`e}s, E. (2019).
\newblock A modern maximum-likelihood theory for high-dimensional logistic
  regression.
\newblock {\em Proceedings of the National Academy of Sciences},
  116(29):14516--14525.

\bibitem[Tibshirani and Efron, 1993]{tibshirani1993introduction}
Tibshirani, R.~J. and Efron, B. (1993).
\newblock An introduction to the bootstrap.
\newblock {\em Monographs on Statistics and Applied Probability}, 57:1--436.

\bibitem[Victoria-Feser, 2002]{MPVF:psycho02}
Victoria-Feser, M.-P. (2002).
\newblock Robust inference with binary data.
\newblock {\em Psychometrika}, 67:21--32.

\bibitem[Wang and Basu, 1999]{wang1999bias}
Wang, J. and Basu, S. (1999).
\newblock Bias-corrected confidence intervals for the concentration parameter
  in a dilution assay.
\newblock {\em Biometrics}, 55(1):111--116.

\bibitem[Wang, 2011]{wang2011gee}
Wang, L. (2011).
\newblock Gee analysis of clustered binary data with diverging number of
  covariates.
\newblock {\em The Annals of Statistics}, 39(1):389--417.

\bibitem[White, 1982]{white1982maximum}
White, H. (1982).
\newblock Maximum likelihood estimation of misspecified models.
\newblock {\em Econometrica: Journal of the Econometric Society}, pages 1--25.

\end{thebibliography}
\end{document}